\documentclass[reprint,floatfix,amsmath,amssymb,superscriptaddress]{revtex4-1}

\usepackage{graphicx}
\usepackage{bm}
\usepackage{epstopdf}
\usepackage{xr}
\usepackage{float}
\usepackage{lineno}
\usepackage{siunitx}
\DeclareSIUnit\Oersted{Oe}
\DeclareSIUnit\electronvolt{eV}
\DeclareSIUnit\emu{emu}
\usepackage{xcolor}
\usepackage[normalem]{ulem}
\usepackage{upgreek}
\usepackage{comment}

\newcommand{\kB}{k_{\mathrm{B}}}

\newcommand{\bk}{\bm{k}}

\newcommand{\zi}{i}
\newcommand{\dd}[1]{\mathrm{d} #1\,}
\newcommand{\tr}{\mathrm{tr}\,}
\newcommand{\T}{{}^{\mathrm{t}}}
\begin{document}

\title{Spin Hall effect driven by the spin magnetic moment current in Dirac materials}


\author{Zhendong Chi}
\thanks{These authors contributed equally to this work.}
\affiliation{Department of Physics, The University of Tokyo, Bunkyo-ku, Tokyo 113-0033, Japan}
\affiliation{National Institute for Materials Science, Tsukuba, Ibaraki 305-0047, Japan}

\author{Guanxiong Qu}
\thanks{These authors contributed equally to this work.}
\affiliation{Department of Physics, The University of Tokyo, Bunkyo-ku, Tokyo 113-0033, Japan}
\affiliation{RIKEN Center for Emergent Matter Science (CEMS), Wako 351-0198, Japan}

\author{Yong-Chang Lau}
\email{lau.yong.chang.d8@tohoku.ac.jp}
\affiliation{Institute for Materials Research (IMR), Tohoku University, Sendai, Miyagi 980-8577, Japan}
\affiliation{Center for Spintronics Research Network, Tohoku University, Sendai, Miyagi 980-8577, Japan}

\author{Masashi Kawaguchi}
\affiliation{Department of Physics, The University of Tokyo, Bunkyo-ku, Tokyo 113-0033, Japan}

\author{Junji Fujimoto}
\affiliation{Department of Physics, The University of Tokyo, Bunkyo-ku, Tokyo 113-0033, Japan}


\author{Koki Takanashi}
\affiliation{Institute for Materials Research (IMR), Tohoku University, Sendai, Miyagi 980-8577, Japan}
\affiliation{Center for Spintronics Research Network, Tohoku University, Sendai, Miyagi 980-8577, Japan}
\affiliation{Center for Science and Innovation in Spintronics, Core Research Cluster, Tohoku University, Sendai 980-8577, Japan}

\author{Masao Ogata}
\affiliation{Department of Physics, The University of Tokyo, Bunkyo-ku, Tokyo 113-0033, Japan}
\affiliation{Trans-scale quantum science institute (TSQS), The University of Tokyo, Bunkyo-ku, Tokyo 113-0033, Japan}

\author{Masamitsu Hayashi}
\email{hayashi@phys.s.u-tokyo.ac.jp}
\affiliation{Department of Physics, The University of Tokyo, Bunkyo-ku, Tokyo 113-0033, Japan}
\affiliation{Trans-scale quantum science institute (TSQS), The University of Tokyo, Bunkyo-ku, Tokyo 113-0033, Japan}
\affiliation{National Institute for Materials Science, Tsukuba, Ibaraki 305-0047, Japan}

\date{\today}

\begin{abstract}

The spin Hall effect of a Dirac Hamiltonian system is studied using semiclassical analyses and the Kubo formula.
In this system, the spin Hall conductivity is dependent on the definition of spin current.
All components of the spin Hall conductivity vanish when spin current is defined as the flow of spin angular momentum.
In contrast, the off-diagonal components of the spin Hall conductivity are non-zero and scale with the carrier velocity (and the effective $g$-factor) when spin current consists of the flow of spin magnetic moment.
We derive analytical formula of the conductivity, carrier mobility and the spin Hall conductivity to compare with experiments.
In experiments, we use Bi as a model system that can be characterized by the Dirac Hamiltonian. 
Te and Sn are doped into Bi to vary the electron and hole concentration, respectively. 
We find the spin Hall conductivity ($\sigma_\mathrm{SH}$) takes a maximum near the Dirac point and decreases with increasing carrier density ($n$).
The sign of $\sigma_\mathrm{SH}$ is the same regardless of the majority carrier type.  
The spin Hall mobility, proportional to $\sigma_\mathrm{SH}/n$, increases with increasing carrier mobility with a scaling coefficient of $\sim 1.4$.
These features can be accounted for quantitatively using the derived analytical formula.
The results demonstrate that the giant spin magnetic moment, with an effective $g$-factor that approaches 100, is responsible for the spin Hall effect in Bi.

\end{abstract}


\maketitle

\section{Introduction}
Spin current is defined as a flow of carriers with opposite spins moving in opposite directions.
There are a number of well established approaches to generate spin current in non-magnetic materials. 
Among them, the spin Hall effect (SHE)\cite{dyakonov1971jetp,hirsch1999prl,zhang2000prl,murakami2003science} allows electrical generation of spin current: application of current induces spin current that flows perpendicular to the current flow\cite{kato2004prl,wunderlich2005prl,valenzuela2006nature,kimura2007prl,liu2012science}.

The degree of which a material can generate spin current via the spin Hall effect is characterized by the spin Hall conductivity.
The spin Hall conductivity in non-magnetic materials can be calculated using the Kubo formula in a similar way the anomalous Hall conductivity is estimated in magnetic materials.
The form of the latter is equivalent to the Berry curvature of the electronic state of the host material\cite{nagaosa2010rmp,karplus1954pr}.
The anomalous Hall conductivity thus results from a purely geometrical property of the Bloch wave function.
(Here we do not discuss contributions from extrinsic effects, \textit{e.g.}, skew scattering and side jump.)
In contrast, there is no equivalent geometrical property, as far as we know of, that corresponds to the spin Hall conductivity.
The lack of such correspondence is intimately related to the definition of spin current: The problem stems from the fact that spin density is generally not a conserved quantity\cite{murakami2004prb,shi2006prl,gradhand2012jpcm}.

Spin current is commonly defined as the flow of spin angular momentum, \textit{i.e.}, the product of carrier velocity and the spin angular momentum\cite{sinova2004prl,guo2005prl}.
Although such spin current does not conserve spin\cite{murakami2004prb,shi2006prl,gradhand2012jpcm}, it has been widely used to characterize spin transport in metallic systems.
The difficulty in defining the spin current is particularly apparent in systems where the electronic states can be described by the Dirac Hamiltonian{\cite{crepieux2001prb,vernes2007prb,lowitzer2010prb,fuseya2012jpsj1,sun2017prl,fukazawa2017jpsj}.
We refer to such systems as Dirac materials hereafter.
Dirac semimetal\cite{armitage2018rmp} is a notable material class that belongs to this system.
As we show below, all components of the spin Hall conductivity in Dirac materials vanish when spin current is defined as the flow of spin angular momentum.

Here we present a systematic study of spin current in Dirac materials.
Two different definitions of the spin current, the flow of spin angular momentum and the flow of spin magnetic moment, are used to calculate the spin Hall conductivity.
We show that the spin current in Dirac materials is zero when flow of spin angular momentum is used as the spin current.
In contrast, the spin Hall conductivity associated with the flow of spin magnetic moment possesses non-zero off-diagonal components.
We derive analytical formulas of the carrier density, mobility, conductivity and spin Hall conductivity.

To compare with the model developed, we study the SHE of bismuth (Bi), a model Dirac material.
Te and Sn are used as dopants to add, respectively, electrons and holes in Bi and tune the position of the Fermi level.
We find the spin Hall conductivity shows significant dependence on the carrier density, taking a maximum when the Fermi level is close to the Dirac point.
The sign of the spin Hall conductivity is found to be independent of the majority carrier type.
The scaling relations between the spin Hall conductivity, carrier density and mobility are studied and compared to the model calculations. 
The results clearly show that the observed spin current is consistent with the flow of spin magnetic moment, which is defined by the effective $g$-factor of the host material.

\section{\label{sec:model}Model analyses}

\subsection{System description}
In this article, we use bismuth (Bi) as a model Dirac Hamiltonian system. 
Bi is a non-magnetic material that possesses large spin orbit coupling.
It is a semi-metal with electron and hole pockets crossing the Fermi level\cite{liu1995prb,fuseya2015jpsj}.
Both electrons and holes contribute to the transport properties.
The electron pockets preside near the three $L$-points of the reciprocal space, whereas the hole pocket is located around the $T$-point.
Owing to the large spin orbit coupling of Bi, tight binding and $k \cdot p$ model calculations show that the electronic structure of the states near the $L$-points can be described using the Dirac Hamiltonian\cite{cohen1960philmag,wolff1964jpcs,fuseya2012jpsj1}.
Here we assume that the spin transport properties of the system is effectively defined by the carriers presiding near the $L$-point.

In the following, bold fonts represent vectors.
The repeated indices imply summation.
The superscript $\dagger$ indicates transpose conjugate.
$e$ is the elementary charge ($e > 0$) and $\hbar$ is the reduced Planck constant.

\subsection{Electronic structure}

The Dirac Hamiltonian is defined as
\begin{equation}
\begin{aligned}
\label{eq:H}
\mathcal{H} &=
- \hbar v k_i \tau_2 \otimes \sigma_i + \Delta \tau_3 \otimes \sigma_0\\
&=
\begin{bmatrix}
\Delta &0 &i \hbar v k_z &i  \hbar v k_-\\
0 &\Delta &i \hbar v k_+ &-i \hbar  v k_z \\
-i \hbar v k_z &-i \hbar v k_- &-\Delta &0\\
-i \hbar v k_+ &i \hbar v k_z &0 &-\Delta \\
\end{bmatrix},
\end{aligned}
\end{equation}
where $k_{\pm} = k_x \pm i k_y$, and $k_i$ is the $i$-th component of the wave vector.
$\tau_i$ and $\sigma_i$ ($i = 1,2,3$) are the Pauli matrices that represent the orbital and spin degree of freedom, respectively.
($i = 1,2,3$ correspond to $x,y,z$ in real space coordinate, respectively.)
$\tau_0$ and $\sigma_0$ are $2 \times 2$ identity matrices.
$v$ is the carrier velocity, and $2 \Delta$ is the energy gap.
Previous studies have shown that $\mathcal{H}$ describes the electronic states near the $L$-point of Bi\cite{cohen1960philmag,wolff1964jpcs,fuseya2012jpsj1}

The eigenvalues ($E_{\tau}$) and the corresponding eigenfunctions ($\psi_{\tau,\sigma}$) of $\mathcal{H}$ are
\begin{equation}
\begin{aligned}
\label{eq:eigen}
E_\tau =& \tau \varepsilon,\\
\psi_{\tau,\sigma} =&
\frac{1}{\sqrt{2}}
\begin{bmatrix}
\tau \sqrt{1+ \frac{\tau \Delta}{\varepsilon}} \frac{\bm{k} \cdot \bm{\sigma}}{k} \chi_\sigma\\
-i \sqrt{1- \frac{\tau \Delta}{\varepsilon}} \chi_\sigma
\end{bmatrix}
\end{aligned}
\end{equation}
where $\varepsilon = \sqrt{\Delta^2 + (\hbar v k)^2}$, $k = |\bm{k}|$, $\tau(=\pm1)$ and $\sigma(=\pm1)$ represent the state index of the band and the spin, respectively.
Note that $\tau$ and $\sigma$ are integers representing orbital and spin states whereas $\tau_i$ and $\sigma_i$ ($i=0,1,2,3$) are matrices.
$\chi_\sigma$ is the spinor of the spin basis function ($\chi_{+1} = \big[ 1, 0 \big]^\dagger$, $\chi_{-1} = \big[ 0, 1 \big]^\dagger$).
The density of states $D(E)$ is obtained as the following:
\begin{equation}
\begin{aligned}
\label{eq:dos}
D(E) &=  \sum_{\tau,\sigma} \int \frac{d\bm{k}}{(2 \pi)^3} \delta(E - E_\tau)\\
 &=
\begin{cases}
\displaystyle \frac{|E| \sqrt{E^2 - \Delta^2}}{\pi^2 \hbar^3 v^3}, &|E| > \Delta,\\
\displaystyle 0, &|E| < \Delta,
\end{cases}
\end{aligned}
\end{equation}
where the spin degree of freedom has been included.
$E=0$ corresponds to the Dirac point.

The calculated band dispersion is displayed in Fig.~\ref{fig:dispersion}(a).
In the calculations, we use $2 \Delta = 15.4$ meV\cite{vecchi1974prb,fuseya2015jpsj}, which represents the band gap of Bi $L$-point.
(Due to a non-zero $\Delta$, a gap forms between the upper and lower branches of the dispersion.)
The solid lines with different color represent the dispersion relation calculated using different values of $v$. 
The corresponding density of states $D(E)$ is shown in Fig.~\ref{fig:dispersion}(b).
As evident, $D(E)$ asymptotically increases with $E^2$, which is one of the characteristics of the linear band dispersion of three-dimensional Dirac Hamiltonian system.


\begin{figure}
\begin{center}
  \includegraphics[width=1.0\columnwidth]{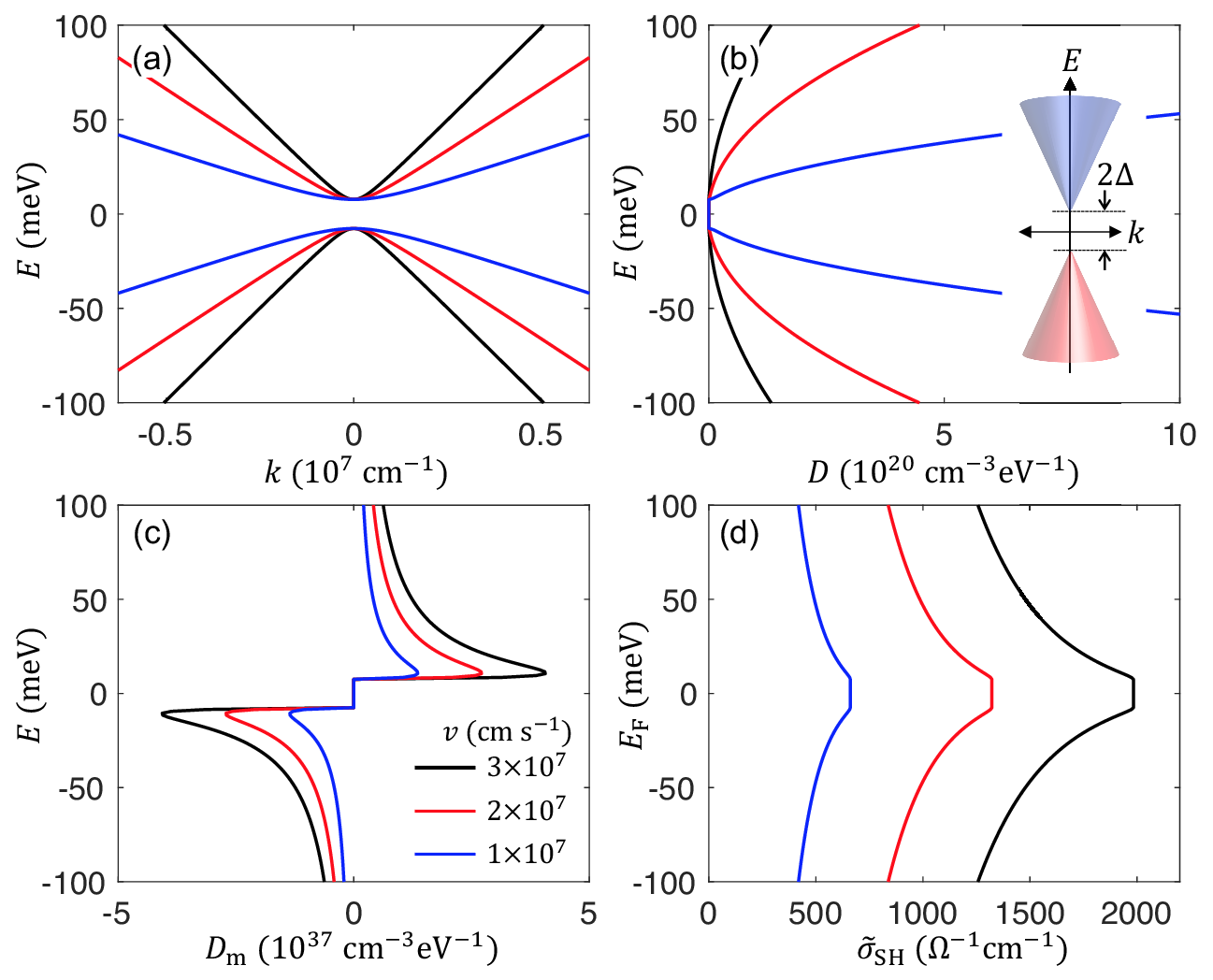}
  \caption{(a) Energy dispersion relation ($E$ vs. wave vector $k$) of the system. (b,c) The density of states $D(E)$ (b) and the density of spin Berry curvature $D_\mathrm{m}$ (c) plotted as a function of position of energy $E$. (d) The Fermi level $E_\mathrm{F}$ dependence of the spin Hall conductivity $\tilde{\sigma}_\mathrm{SH}$. (a-d). Parameters used in the calculations are: $a = 3.5 \times 10^{-7}$ s cm$^{-1}$, $\alpha = - \frac{1}{3}$, $\Delta = 7.7$ meV and $E_\mathrm{c} = -5 $ eV. Line colors represent calculation results using different values of carrier velocity $v$: see the legend of (c). Inset to (b) shows an illustration of the Bi band structure near the $L$-point.}
  \label{fig:dispersion}
\end{center}
\end{figure}

\subsection{Transport properties}

The carrier density $n$ is calculated by integrating $D(E)$ over the occupied (electron) states or the unoccupied (hole) states.
The result is
\begin{equation}
\begin{aligned}
\label{eq:density}
n &=\int_\Delta^{E_\mathrm{F}} dE D(E)
= \int_{E_\mathrm{F}}^{-\Delta} dE D(E)
= \frac{(E_\mathrm{F}^2 - \Delta^2)^{\frac{3}{2}}}{3 \pi^2 (\hbar v)^3},
\end{aligned}
\end{equation}
where $E_\mathrm{F}$ is the Fermi level.
In Fig.~\ref{fig:transport}(a), we show the Fermi level ($E_\mathrm{F}$) dependence of $n$ calculated using different values of $v$.
Note that negative (positive) $n$ indicates carriers with negative (positive) charge.
As typical of Dirac Hamiltonian systems, $|n|$ increases when $E_\mathrm{F}$ moves away from the Dirac point.

The electrical conductivity ($\sigma_{xx}$) can be obtained from the Kubo formula (see Appendix Sec.~\ref{sec:conductivity})
\begin{equation}
\begin{aligned}
\label{eq:conductivity}
\sigma_{xx}= e^2D(E_\mathrm{F}) v^2\frac{E_\mathrm{F}^2-\Delta^2}{3E_\mathrm{F}^2}\tau_\mathrm{eff}.
\end{aligned}
\end{equation}
$\tau_\mathrm{eff}$ is the relaxation time, which we assume takes the following form\cite{dassarma2013prb,dassarma2015prb}:
\begin{equation}
\begin{aligned}
\label{eq:taueff}
\tau_\mathrm{eff} = a n^{\alpha}.
\end{aligned}
\end{equation}
$a$ and $\alpha$ are constants.
The exponent $\alpha$ defines the characteristic relaxation time of the carriers involved in transport\cite{dassarma2013prb}.
Assuming carrier transport based on Drude's model, the carrier mobility $\mu_\mathrm{c}$ can be estimated from the following relation:
\begin{equation}
\begin{aligned}
\label{eq:nemu}
\sigma_{xx} = n e \mu_\mathrm{c}.
\end{aligned}
\end{equation}
From Eqs.~(\ref{eq:conductivity}), (\ref{eq:taueff}) and (\ref{eq:nemu}), $\mu_\mathrm{c}$ reads
\begin{equation}
\begin{aligned}
\label{eq:mobility}
\mu_\mathrm{c} = e D(E_\mathrm{F}) v^2\frac{E_\mathrm{F}^2-\Delta^2}{3E_\mathrm{F}^2} a n^{\alpha - 1}.
\end{aligned}
\end{equation}

The $n$ dependence of $\mu_\mathrm{c}$ is shown in Fig.~\ref{fig:transport}(b) with $a = 3.5 \times 10^{-7}$ s cm$^{-1}$ and $\alpha = - \frac{1}{3}$.
$\mu_\mathrm{c}$ takes a sharp maximum as the Fermi level approaches the Dirac point, a characteristics of systems with Dirac Hamiltonian\cite{novoselov2005nature,zhang2005nature,liang2015nmat,shekhar2015nphys}.
The sharp increase is largely due to the reduction in $|n|$ near the Dirac point.
We will see later that simple scaling relations hold for $\sigma_{xx}$ and $\mu_c$ as a function of $|n|$, and the coefficients of such relations can be compared with experiments.


\begin{figure}
\begin{center}
  \includegraphics[width=1.0\columnwidth]{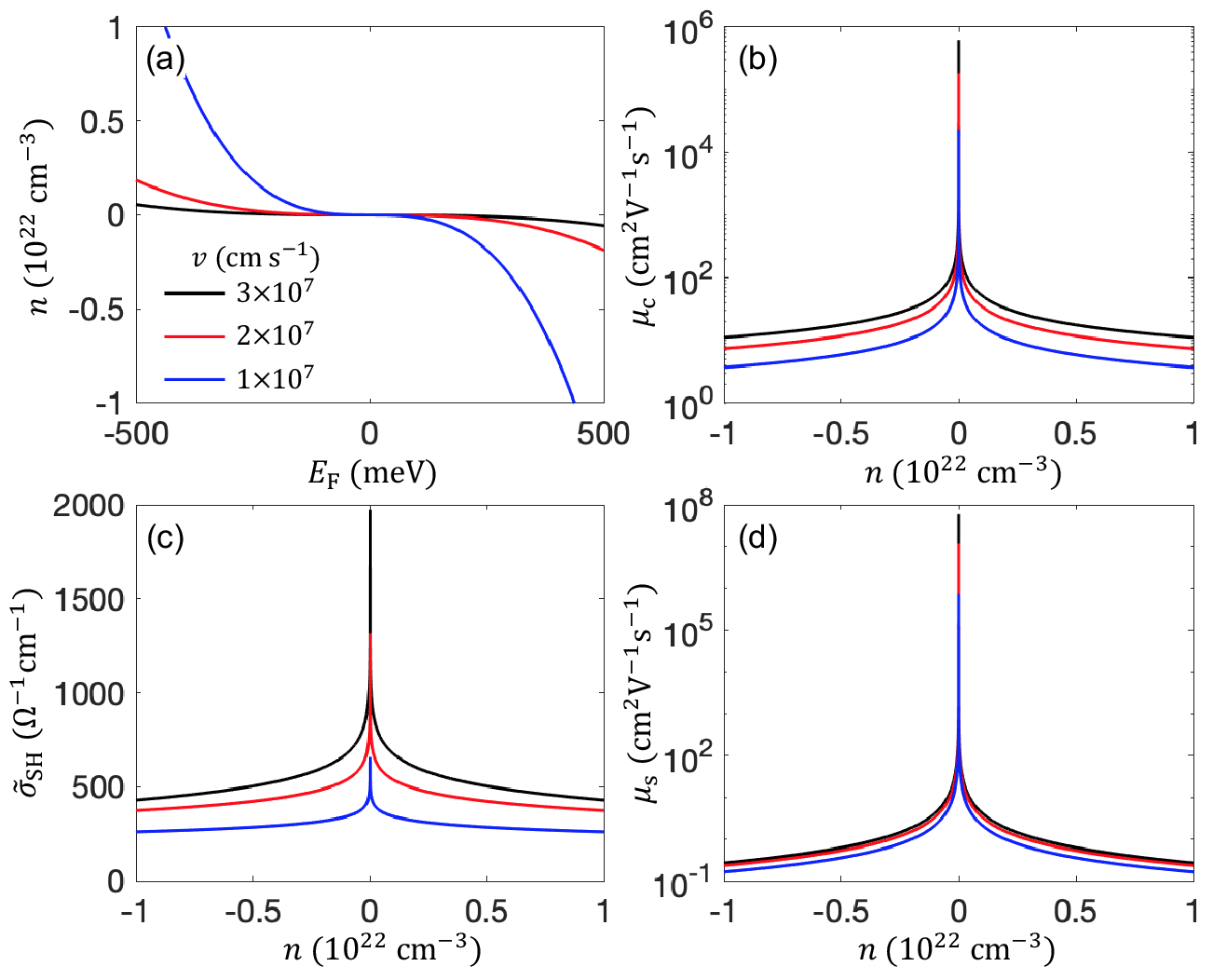}
  \caption{(a) Fermi energy ($E_\mathrm{F}$) dependence of the carrier concentration ($n$). (b-d) Carrier mobility $\mu_\mathrm{c}$ (b), spin Hall conductivity $\tilde{\sigma}_\mathrm{SH}$ (c) and spin Hall mobility $\mu_\mathrm{s}$ (d) plotted against $n$. (a-d) Line colors indicate calculation results using different values of carrier velocity $v$: see the legend of (a). Since $\mu_\mathrm{c}$ and $\mu_\mathrm{s}$ scale with the inverse of $|n|$, here we limit the plotting to a range of $|n|$ that satisfies $|E_\mathrm{F}| \geq 0.5$ meV. Parameters used in the calculations are: $a = 3.5 \times 10^{-7}$ s cm$^{-1}$, $\alpha = - \frac{1}{3}$, $\Delta = 7.7$ meV and $E_\mathrm{c} = -5 $ eV.}
  \label{fig:transport}
\end{center}
\end{figure}

\subsection{Spin angular momentum and spin magnetic moment}

In the following, we discuss the form of spin Hall conductivity.
First, the $j$-th component of the velocity operator $\bm{v}$ ($v_j$) is given by
\begin{equation}
\begin{aligned}
\label{eq:vel}
v_j = \frac{1}{\hbar} \frac{\partial \mathcal{H}}{\partial k_j}
= - v \tau_2 \otimes \sigma_j
= v
\begin{bmatrix}
0 & i \sigma_j\\
-i \sigma_j &0
\end{bmatrix}.
\end{aligned}
\end{equation}
Next, we introduce the operators representing the spin magnetic moment ($\bm{m}$) and the spin angular momentum ($\bm{s}$) of the carriers.
The spin magnetic moment is defined from the Zeeman energy term that appears in the Hamiltonian when a magnetic field is applied to the system\cite{fuseya2012jpsj1,fuseya2015jpsj}.
The $i$-th component of $\bm{m}$ ($m_i$) reads
\begin{equation}
\begin{aligned}
\label{eq:spinmoment}
m_i = -\frac{e \hbar v^2}{2 \Delta} (\tau_3 \otimes \sigma_i) = -\frac{e \hbar v^2}{2 \Delta}
\begin{bmatrix}
\sigma_i &0\\
0 &-\sigma_i
\end{bmatrix}.
\end{aligned}
\end{equation}
$m_i$ can be rewritten using the Bohr magneton $\mu_\mathrm{B}$ ($\mu_\mathrm{B} = \frac{e \hbar}{2 m_e}$ and $m_e$ is the free electron mass) as the following:
\begin{equation}
\begin{aligned}
\label{eq:spinmoment2}
m_i =& -\frac{1}{2} g^* \mu_\mathrm{B} (\tau_3 \otimes \sigma_i),
\end{aligned}
\end{equation}
where $g^*$ is the effective $g$-factor.
For the system under study (Eq.~(\ref{eq:H})), $g^*$ is given by\cite{fuseya2012jpsj1}
\begin{equation}
\begin{aligned}
\label{eq:gfactor}
g^*=\frac{2 m_e v^2}{\Delta}.
\end{aligned}
\end{equation}
The spin angular momentum ($\bm{s}$) is defined by the Lorentz invariance of the Dirac equation \cite{peskin1995book}.
The $i$-th component of $\bm{s}$ ($s_i$) reads
\begin{equation}
\begin{aligned}
\label{eq:spinang}
s_i = \frac{\hbar}{2} (\tau_0 \otimes \sigma_i) = \frac{\hbar}{2}
\begin{bmatrix}
\sigma_i &0\\
0 &\sigma_i
\end{bmatrix}.
\end{aligned}
\end{equation}

To find the form of spin current, we study the time evolution of $\langle s_i \rangle$ and $\langle m_i \rangle$ using the Heisenberg equation of motion.
We define $\langle x \rangle = \phi^\dagger x \phi $: $x$ is an operator, $\phi$ and $\phi^\dagger$ are the field operators.
$\phi$ ($\phi^\dagger$) is expressed using the electron annihilation operator $c_{\bm{k},\tau,\sigma}$ (creation operator $c^\dagger_{\bm{k},\tau,\sigma}$) and the eigenfunction $\psi_{\tau,\sigma}$ (see Eq.~(\ref{eq:eigen})) as the following: 
\begin{align}
\phi (\bm{r},t) = \sum_{\bm{k},\tau,\sigma} \psi_{\tau,\sigma} e^{i \bm{k} \cdot \bm{r} - i E_{\tau} t} c_{\bm{k},\tau,\sigma}.
\label{eq:phi}
\end{align}
$\phi (\bm{r},t)$ is the eigenfunction of the Hamiltonian (Eq.~(\ref{eq:H})) in the momentum representation: $\mathcal{H}_p =-v p_i \tau_2 \otimes \sigma_i + \Delta \tau_3 \otimes \sigma_0$, where $p_i = - i \hbar \partial_i$ is the momentum operator.
The equation of motion for $\langle s_i \rangle$ reads
\begin{equation}
\begin{aligned}
\label{eq:dirac:spin}
\frac{d \langle s_i \rangle}{dt} &= \frac{1}{i \hbar}  \langle \big[ s_i ,\mathcal{H}_p  \big]  \rangle
= - \partial_j \langle j_{\mathrm{s},j}^i \rangle + \tau_{\mathrm{s},i},\\
&\langle j_{\mathrm{s},j}^i \rangle = - \frac{\hbar}{2} v \delta_{ij} \int \phi^\dagger (\tau_2 \otimes \sigma_0) \phi d\bm{r},\\
&\tau_{\mathrm{s},i} = i \frac{\hbar}{2} v \varepsilon_{ijl} \int \big( \phi^\dagger (\tau_2 \otimes \sigma_l) (\partial_i \phi)\\
& \ \ \ \ \ \ \ \ \ \ \ \ \ \ \ \ \ \ \ \ \ \ - (\partial_i \phi^\dagger) (\tau_2 \otimes \sigma_l) \phi \big) d\bm{r},
\end{aligned}
\end{equation}
where $\delta_{ij}$ is the delta function, $\epsilon_{ijl}$ is the Levi-Civita symbol.
The equation of motion for $\langle m_i \rangle$ is given as
\begin{equation}
\begin{aligned}
\label{eq:dirac:magnetic}
\frac{d \langle m_i \rangle}{dt} &= \frac{1}{i \hbar}\langle  \big[ m_i , \mathcal{H}_p  \big] \rangle
= - \partial_j \langle j_{\mathrm{m},j}^i \rangle + \tau_{\mathrm{m},i},\\
&\langle j_{\mathrm{m},j}^i \rangle = - \mu_\mathrm{B} v \varepsilon_{ijl} \int \phi^\dagger (\tau_1 \otimes \sigma_l) \phi d\bm{r},\\
&\tau_{\mathrm{m},i} = - i \mu_\mathrm{B} v \delta_{ij} \int \big( \phi^\dagger (\tau_1 \otimes \sigma_0) (\partial_j \phi)\\
& \ \ \ \ \ \ \ \ \ \ \ \ \ \ \ \ \ \ \ \ \ \ \ \ \ - (\partial_j \phi^\dagger) (\tau_1 \otimes \sigma_0) \phi \big) d\bm{r}.
\end{aligned}
\end{equation}
Equations~(\ref{eq:dirac:spin}) and (\ref{eq:dirac:magnetic}) represent the continuity equation of the spin angular momentum and the spin magnetic moment, respectively.
$\langle j_{\mathrm{s},j}^i \rangle$ and $\langle j_{\mathrm{m},j}^i \rangle$ are the associated spin current.
With regard to spin current, the subscript $j$ represents the flow direction and the superscript $i$ indicates the spin direction.
Interestingly, $\langle j_{\mathrm{m},j}^i \rangle$ is non-zero when the flow and spin directions are orthogonal (\textit{i.e.}, $i \neq j$) whereas $\langle j_{\mathrm{s},j}^i \rangle$ vanishes under the same condition. That is,
$\langle j_{\mathrm{s},j}^i \rangle$ is non-zero only when the flow and spin directions are parallel.
In the continuity equation, if $\tau_{\mathrm{s},i}$ ($\tau_{\mathrm{m},i}$) is zero, the spin angular momentum $\langle s_i \rangle$ (the spin magnetic moment $\langle m_i \rangle$) conserves.
In general, however, $\tau_{\mathrm{s},i}$ ($\tau_{\mathrm{m},i}$) is not zero and $\langle s_i \rangle$ ($\langle m_i \rangle$) is not a conserved quantity.
One may define the spin current such that $\langle s_i \rangle$ or $\langle m_i \rangle$ conserves\cite{vernes2007prb,lowitzer2010prb}.
Since our focus here is on systems with strong spin orbit coupling (\textit{e.g.}, Bi), where spin ($\langle s_i \rangle$ or $\langle m_i \rangle$) is generally not a conserved quantity, we employ the intuitive definitions shown below (see Eqs.~(\ref{eq:spincurrent:ang}) and (\ref{eq:spincurrent:mag})).

From the second line of Eqs.~(\ref{eq:dirac:spin}) and (\ref{eq:dirac:magnetic}), we define the spin current operator as the following:
\begin{equation}
\begin{aligned}
\label{eq:spincurrent:ang}
\hat{j}_{\mathrm{s},j}^i = - \frac{\hbar}{2} v (\tau_2 \otimes \sigma_0) 
= \frac{1}{2} \{v_j , s_i \},
\end{aligned}
\end{equation}
\begin{equation}
\begin{aligned}
\label{eq:spincurrent:mag}
\hat{j}_{\mathrm{m},j}^i
= -\frac{m_e}{e} \mu_\mathrm{B} v \varepsilon_{ijl} (\tau_1 \otimes \sigma_l)
= \frac{m_e}{e} \frac{1}{2} \{v_j , m_i \},
\end{aligned}
\end{equation}
where the curly bracket indicates an anti-commutator.
Note that the definition of $\hat{j}_{\mathrm{m},j}^i$ as a spin current was employed in Ref.~\cite{fuseya2012jpsj1}.
We set the unit of the spin current to be equal to the product of energy and length, \textit{i.e.}, [erg cm] in cgs units.
The prefactor $\frac{m_e}{e}$ in Eq.~(\ref{eq:spincurrent:mag}) is a normalization constant that converts the flow of spin magnetic moment to [erg cm].
For $\hat{j}_{\mathrm{s},i}^j$, we have left out the Delta function ($\delta_{ij}$) that appears in Eq.~(\ref{eq:dirac:spin}) to allow spin current with different symmetries.
The right hand side of Eqs.~(\ref{eq:spincurrent:ang}) and (\ref{eq:spincurrent:mag}) shows that the spin current represents the product of the carrier velocity and the spin angular momentum (spin magnetic moment) for $\hat{j}_{\mathrm{s},i}^j$ ($\hat{j}_{\mathrm{m},i}^j$).

\subsection{Spin Hall conductivity}

We are now in a position to calculate the spin Hall conductivity ($\sigma_\mathrm{SH}$), 
which is obtained by integrating the spin Berry curvature ($\Omega_{\mathrm{s} (\mathrm{m}), ij,\tau,\sigma}^{l}$) of the occupied states with band $\tau$ with spin $\sigma$.
$\Omega_{\mathrm{s} (\mathrm{m}), ij,\tau,\sigma}^{l}$ associated with the spin angular momentum flow (spin magnetic momentum flow) is expressed as
\begin{equation}
\begin{aligned}
\label{eq:sbc}
&\Omega_{\mathrm{s (m)}, ij,\tau,\sigma}^{l} =\\
& - \hbar \sum_{\tau^\prime,\sigma^\prime \neq \tau,\sigma} 2 \mathrm{Im} \frac{\langle \psi_{\tau,\sigma} | \hat{j}_{\mathrm{s (m)},j}^l | \psi_{\tau^\prime,\sigma^\prime} \rangle \langle \psi_{\tau^\prime,\sigma^\prime} | v_i | \psi_{\tau,\sigma} \rangle}{(E_\tau - E_{\tau^\prime})^2}.
\end{aligned}
\end{equation}
$\hbar$ is multiplied to the right hand side of Eq.~(\ref{eq:sbc}) so that the unit of $\Omega_{ij,\tau,\sigma}^{l}$ becomes length square ([cm$^2$]).
The spin Berry curvature associated with the flow of spin angular momentum ($\Omega_{\mathrm{s}, ij,\tau,\sigma}^{l}$) is obtained by substituting Eqs.~(\ref{eq:eigen}), (\ref{eq:vel}) and (\ref{eq:spincurrent:ang}) into Eq.~(\ref{eq:sbc}), which leads to
\begin{equation}
\begin{aligned}
\label{eq:sbc:ang}
\Omega_{\mathrm{s}, ij,\tau,\sigma}^{l} = 0.
\end{aligned}
\end{equation}
This is in significant contrast to conventional Pauli-Schr\"{o}dinger Hamiltonian systems, in which $\Omega_{\mathrm{s}, ij,\tau,\sigma}^{l}$ does not vanish.
Since $\Omega_{\mathrm{s}, ij,\tau,\sigma}^{l}=0$, all components of $\sigma_\mathrm{SH}$ vanish for the flow of spin angular momentum.
This is one of the unique characters of Dirac materials.
See Sec. VI and the Appendix (Sec.~\ref{sec:vanishment}) for the implication of the vanishing $\sigma_\mathrm{SH}$ associated with the spin angular momentum flow.

In contrast, the spin Berry curvature associated with the flow of spin magnetic moment ($\Omega_{\mathrm{m}, ij,\tau,\sigma}^{l}$) reads 
\begin{equation}
\begin{aligned}
\label{eq:sbc:mag}
\Omega_{\mathrm{m}, ij,\tau,\sigma}^{l} = \tau \varepsilon_{ijl} \frac{\hbar^2 g^* v^2 \Delta}{8 \varepsilon^3}.
\end{aligned}
\end{equation}
For calculating the spin Hall conductivity $\sigma_\mathrm{SH}$, it is convenient to introduce the density of spin Berry curvature ($D_{\mathrm{m}, \tau, \sigma}$), defined as
\begin{equation}
\begin{aligned}
\label{eq:sbcdensity}
D_{\mathrm{m}, \tau, \sigma} &= \int \frac{d\bm{k}}{(2 \pi)^3} \Omega_{\mathrm{m}, ij,\tau,\sigma}^{l} \delta(E - E_\tau).
\end{aligned}
\end{equation}
Substituting $\Omega_{\mathrm{m}, ij,\tau,\sigma}^{l}$ into Eq.~(\ref{eq:sbcdensity}), we find
\begin{equation}
\begin{aligned}
\label{eq:sbcdensity:ang:mag}
D_{\mathrm{m}, \tau, \sigma} &= \frac{g^* \tau \Delta}{8 \pi^2 \hbar \hbar v} \frac{\sqrt{E^2 - \Delta^2}}{E^2}.
\end{aligned}
\end{equation}
In contrary to the density of states ($D(E)$), $D_{\mathrm{m}, \tau, \sigma}$ can be negative as its sign determines the sign of spin Hall conductivity.
Since $D_{\mathrm{m},\tau,\sigma}$ depends on $\tau$, the sign of $D_{\mathrm{m},\tau,\sigma}$ is opposite for the bands below or above the gap.

For the flow of spin magnetic moment, we obtain
\begin{equation}
\begin{aligned}
\label{eq:shc}
\sigma_\mathrm{SH}(E_\mathrm{F}) = -e \sum_{\tau,\sigma} \int_{E_\mathrm{c}}^{E_\mathrm{F}} dE f_\mathrm{FD} D_{\mathrm{m},\tau,\sigma}
\end{aligned}
\end{equation}
where $f_\mathrm{FD}$ is the Fermi-Dirac distribution function.
For the energy integration, a cutoff energy $E_\mathrm{c}$ is introduced.
$E_\mathrm{c}$ is determined by the band structure.
From tight binding calculations, the valence band near the Dirac point lies in the energy range of a few electron volts.
We thus set $E_\mathrm{c}$ to -5 eV in the calculations.
At 0 K, the integration results in the following expression for $\sigma_\mathrm{SH}$\cite{fuseya2012jpsj1,fuseya2015jpsj}:
\begin{equation}
\begin{aligned}
\label{eq:shc}
\sigma_\mathrm{SH}&(E_\mathrm{F}) = -\frac{e m_e v}{4 \pi^2 \hbar} \big( F(E_\mathrm{F}) - F(E_\mathrm{c}) \big),\\
&\begin{cases}
F(E) = \ln{\big( \frac{|E| + \sqrt{E^2 - \Delta^2}}{\Delta} \big)} - \frac{\sqrt{E^2 - \Delta^2}}{|E|}, \textrm{for } |E| > \Delta,\\
F(E) = 0, \textrm{for } |E| \leq \Delta.
\end{cases}
\end{aligned}
\end{equation}
This result is equal to that reported in Refs.~\cite{fuseya2012jpsj1,fuseya2015jpsj}.
The present approach to obtain the form of $\sigma_\mathrm{SH}$ using spin Berry curvature is thus equivalent to that using thermal Green's functions.
The unit of $\sigma_\mathrm{SH}$ is [C cm$^{-1}$]. We define $\tilde{\sigma}_\mathrm{SH} \equiv \frac{2e}{\hbar} \sigma_\mathrm{SH}$, which has the same unit with conductivity, \textit{i.e.}, [$\Omega^{-1}$ cm$^{-1}$].
In analogy to the carrier mobility (Eq.~(\ref{eq:nemu})), the spin Hall mobility is defined as
\begin{equation}
\begin{aligned}
\label{eq:shmobility}
\mu_\mathrm{s} = \frac{\tilde{\sigma}_\mathrm{SH}}{n e}.
\end{aligned}
\end{equation}
The unit of $\mu_\mathrm{s}$ is the same with that of carrier mobility, \textit{i.e.}, [cm$^2$ V$^{-1}$ s$^{-1}$]. $\mu_\mathrm{s}$ can be considered the mobility associated with the spin current\cite{chi2020sciadv}.

Figure~\ref{fig:dispersion}(c) shows the energy dependence of the density of spin Berry curvature $D_{\mathrm{m},\tau,\sigma}$ calculated using different values of $v$.
$D_{\mathrm{m},\tau,\sigma}$ takes an extremum near $|E| \sim \Delta$ and the sign of $D_{\mathrm{m},\tau,\sigma}$ changes across the band gap.
The sign change of $D_{\mathrm{m},\tau,\sigma}$ across the band gap results in a plateau of $\tilde{\sigma}_\textrm{SH}$ around the Dirac point ($E=0$): See Fig.~\ref{fig:dispersion}(d), which shows the Fermi level dependence of $\tilde{\sigma}_\textrm{SH}$.
The constant $\tilde{\sigma}_\textrm{SH}$ within the band gap is a characteristic feature of systems with gapped Dirac Hamiltonian\cite{fuseya2012jpsj1,fukazawa2017jpsj}.
Note that the sign of $\sigma_\mathrm{SH}$ does not change when the Fermi level moves across the gap.
In Figs.~\ref{fig:transport}(c) and \ref{fig:transport}(d), we show the calculated $\tilde{\sigma}_\mathrm{SH}$ and $\mu_\mathrm{s}$ plotted as a function of carrier density.
Both $\tilde{\sigma}_\mathrm{SH}$ and $\mu_\mathrm{s}$ show a sharp peak at small $|n|$.


\begin{figure}
\begin{center}
  \includegraphics[width=1.0\columnwidth]{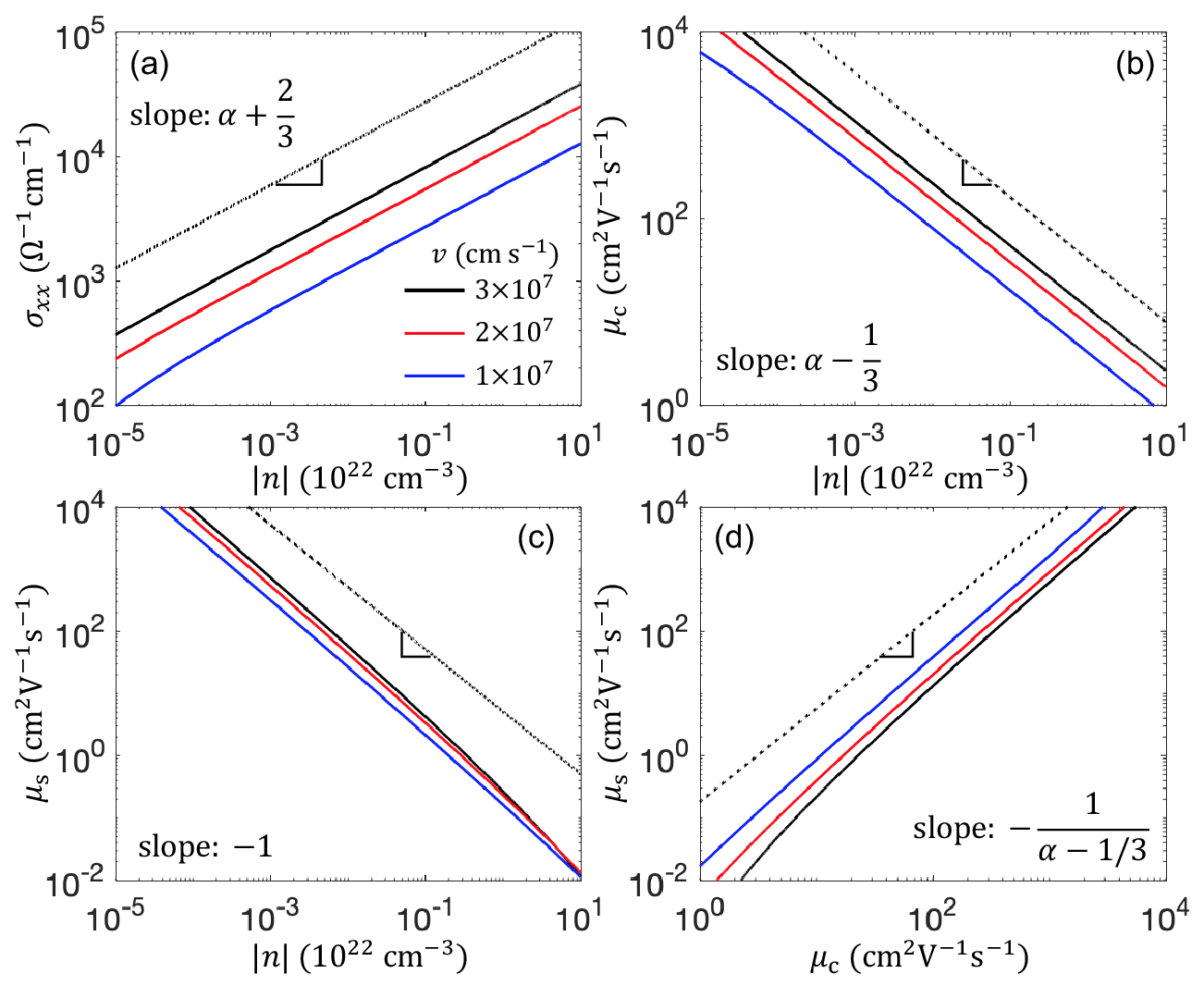}
  \caption{(a-c) Carrier density $|n|$ dependence of the conductivity $\sigma_{xx}$ (a), carrier mobility $\mu_\mathrm{c}$ (b) and spin Hall mobility $\mu_\mathrm{s}$ (c). (d) $\mu_\mathrm{s}$ vs. $\mu_\mathrm{c}$. (a-d) All plots are in log-log scale.
Line colors indicate calculation results using different values of carrier velocity $v$: see the legend of (a). Parameters used in the calculations are: $a = 3.5 \times 10^{-7}$ s cm$^{-1}$, $\alpha = - \frac{1}{3}$, $\Delta = 7.7$ meV and $E_\mathrm{c} = -5 $ eV. The dotted line in (a-d) show the corresponding relation based on Eq.~(\ref{eq:scaling:conductivity}) (a),  Eq.~(\ref{eq:scaling:mobility}) (b), Eq.~(\ref{eq:scaling:spinHallmobility}) (c) and Eq.~(\ref{eq:scaling:musmuc}) (d). Lines are shifted vertically. The slope of the lines are denoted in each panel. 
}
  \label{fig:scaling}
\end{center}
\end{figure}

\subsection{\label{sec:model:scaling}Scaling relations}

Since many of the transport properties depend on the carrier density, it is useful to study the scaling relation among the parameters.
In Figs.~\ref{fig:scaling}(a-c), the carrier density ($|n|$) dependence of the conductivity ($\sigma_{xx}$), the carrier mobility ($\mu_\mathrm{c}$) and the spin Hall mobility ($\mu_\mathrm{s}$) are presented by the solid lines in log-log scale.
Similarly, the relation between $\mu_\mathrm{s}$ and $\mu_\mathrm{c}$ is shown in Fig.~\ref{fig:scaling}(d). 
Different colors represent results when different values of $v$ are used.
All plots show a clear power law.

When $|E_\mathrm{F}| \gg \Delta$, $\sigma_{xx}$ in Eq.~(\ref{eq:conductivity}), $\mu_\mathrm{c}$ in Eq.~(\ref{eq:mobility}) and  $\mu_\mathrm{s}$ in Eq.~(\ref{eq:shmobility}) can be approximated as
\begin{equation}
\label{eq:scaling:conductivity}
\sigma_{xx} \sim \frac{2ae^2v}{3(3\pi^2)^{1/3}\hbar} n^{\alpha + \frac{2}{3}},
\end{equation}
\begin{equation}
\label{eq:scaling:mobility}
\mu_\mathrm{c} \sim \frac{2aev}{3(3\pi^2)^{1/3}\hbar} n^{\alpha - \frac{1}{3}}.
\end{equation}
\begin{equation}
\begin{aligned}
\label{eq:scaling:spinHallmobility}
\mu_\mathrm{s} = \frac{em_e v}{2 \pi^2 \hbar^2} n^{-1} \ln \Big(\frac{E_\mathrm{c}}{(3 \pi^2 n)^{\frac{1}{3}} \hbar v} \Big).
\end{aligned}
\end{equation}
In the last equation, $\mu_\mathrm{s}$ obeys a power law with a scaling coefficient of $-1$ if the ln$(n^{-1})$ dependence can be neglected.
To study the scaling coefficient, we fit the lines in Fig.~\ref{fig:scaling}(c) with a linear function in the appropriate $|n|$ range ($10^{18} \leq |n| \leq 10^{21}$ cm$^{-3}$): 
\begin{equation}
\label{eq:scaling}
\log_{10}(\mu_\mathrm{s}) = a_s \log_{10}(|n|) + b_s
\end{equation}
We find the scaling coefficient, $a_s$, is close to -1: see Fig.~\ref{fig:exponents} in which $a_s$ is plotted as a function of $v$ using different values of $\alpha$.
Although $a_s$ slightly decreases with increasing $v$, its variation is small.
$a_s$ does not depend on $\alpha$ since $\tilde{\sigma}_\textrm{SH}$ is independent of the relaxation time $\tau_\mathrm{eff}$.
These results show that the scaling of $\mu_\mathrm{s}$ can be approximated as $|n|^{-1}$, with a 10-20\% variation, when $v$ is in the range shown in Fig.~\ref{fig:exponents}.
\begin{figure}
\begin{center}
  \includegraphics[width=0.5\columnwidth]{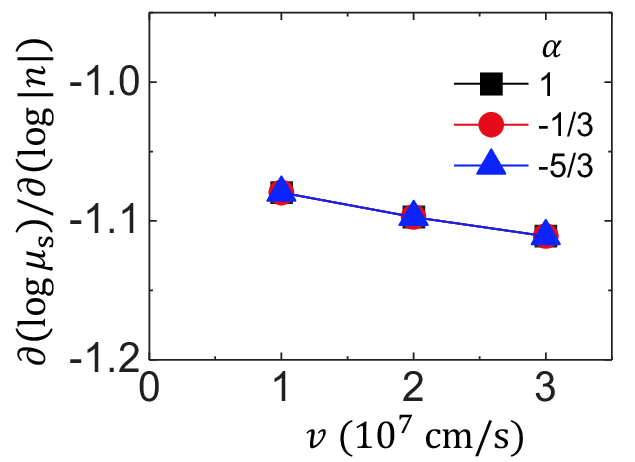}
  \caption{Carrier velocity dependence of the scaling coefficient $a_s$ obtained by linear fitting the spin Hall mobility $\mu_\mathrm{s}$ in log-log scale as a function of carrier density $|n|$, see Fig.~\ref{fig:scaling}(c). Symbols represent calculation results using different values of $\alpha$ denoted in the legend.}
  \label{fig:exponents}
\end{center}
\end{figure}
Assuming that $\mu_\mathrm{s}$ scales as $|n|^{-1}$, the relation between $\mu_\mathrm{s}$ and $\mu_\mathrm{c}$ reads:
\begin{equation}
\label{eq:scaling:musmuc}
\mu_\mathrm{s} \propto \mu_\mathrm{c}^{-\frac{1}{\alpha - \frac{1}{3}}}.
\end{equation}
The scaling relations derived in Eqs.~(\ref{eq:scaling:conductivity}), (\ref{eq:scaling:mobility}), (\ref{eq:scaling:spinHallmobility}) and (\ref{eq:scaling:musmuc}) are part of the main results of this paper.
The scaling coefficients are summarized in Table~\ref{table:scaling}.
The dotted lines in Fig.~\ref{fig:scaling} show the expected scaling relations from Eqs.~(\ref{eq:scaling:conductivity}), (\ref{eq:scaling:mobility}), (\ref{eq:scaling:spinHallmobility}) and (\ref{eq:scaling:musmuc}).
The lines are shifted vertically for better visibility.
As evident, the scaling coefficient are consistent with the slope of the linear lines plotted in Figs.~\ref{fig:scaling}.
In the next section, we compare the scaling coefficients obtained in the experiments with those predicted by the model.



\begingroup
\setlength{\tabcolsep}{10pt} 
\renewcommand{\arraystretch}{1.3} 
\begin{table*}[t]
 \caption{
 Comparison of the scaling coefficients obtained in the experiments and model calculations. For the latter, the following parameters are used: $v = 2.5 \times 10^7$ cm s$^{-1}$, $\alpha = - \frac{1}{3}$, $a = 3.5 \times 10^{-7}$ s cm$^{-1}$, $\Delta = 7.7$ meV and $E_\textrm{c} = -5$ eV.}
 \label{table:scaling}
 \centering
   \vspace{3pt}
  \begin{tabular}{c c c c c}
   \hline \hline
   Relation & Experiments & Model\footnote{Scaling coefficients when $E_\mathrm{F} \gg \Delta$ is assumed. $\sigma_{xx}$ vs. $|n|$: Eq.~(\ref{eq:scaling:conductivity}). $\mu_\textrm{c}$ vs. $|n|$: Eq.~(\ref{eq:scaling:mobility}). $\mu_\textrm{s}$ vs. $|n|$: Eq.~(\ref{eq:scaling:spinHallmobility}) and the $\log(n^{-\frac{1}{3}})$ dependence is neglected. $\mu_\textrm{s}$ vs. $\mu_\textrm{c}$: Eq.~(\ref{eq:scaling:musmuc}).} & $(\alpha = -\frac{1}{3})$ & Fitting\footnote{Slope of the linear line fitted to the calculations presented in Fig.~\ref{fig:scaling2}.}\\
   \hline 
   $\sigma_{xx}$ vs. $|n|$ & $0.22 \pm 0.08$ & $\alpha + \frac{2}{3}$ & 0.33 & 0.33\\
   $\mu_\textrm{c}$ vs. $|n|$ & $-0.68 \pm 0.08$ & $\alpha - \frac{1}{3}$ & -0.67 & -0.66\\
   $\mu_\textrm{s}$ vs. $|n|$ & $-1.0 \pm 0.05$ & -1 & -1 & -1.2\\
   $\mu_\textrm{s}$ vs. $\mu_\textrm{c}$ & $1.4 \pm 0.2$ & $-\frac{1}{\alpha - 1/3}$ & 1.5 & 1.8\\
         \hline
  \end{tabular}
\end{table*}
\endgroup

\section{Experimental results}
\subsection{Sample preparation and structural characterization}
We use bismuth (Bi) as a model system to study the spin Hall effect under the Dirac Hamiltonian.
Pristine Bi (no doping), Te-doped Bi (Bi$_{1-x}$Te$_x$) and Sn-doped Bi (Bi$_{1-y}$Sn$_y$) thin films were grown by magnetron sputtering on thermally oxidized Si substrates with or without a 0.5 nm thick Ta seed layer. 
$x$ and $y$ represent the nominal doping level of Te and Sn in Bi, respectively. 
The nominal thickness of the doped and pristine Bi films is $\sim$10 to $\sim$15 nm.
See the Appendix (Sec.~\ref{sec:Sample}) for the details of sample preparation and device fabrication. 
Figure~\ref{fig:structure}(a) shows the $x$ and $y$ dependence of $\theta-2\theta$ X-ray diffraction (XRD) spectra for pristine and doped Bi films.
The films are polycrystalline with Bi(012) and Bi(104) being the two primary planes that grow along the film normal for films with $x$ and $y$ up to $\sim$0.4.
The appearance of $\beta$-Sn (200) peak for $y\geq0.3$ may indicate agglomeration of $\beta$-Sn clusters in Bi.
Note that all XRD peaks presented in Fig.~\ref{fig:structure}(a), except for the weak $\beta$-Sn peak at $2\theta \sim 31^o$, can be indexed by the diffraction peaks for rhombohedral Bi, suggesting that the crystal structure of Bi is maintained upon Te- and Sn-doping. 
Focusing on the (012) diffraction peak, we plot in Fig.~\ref{fig:structure}(b) the corresponding interatomic plane distance $d_{\rm{012}}$ deduced from the Bragg's law as functions of $x$ and $y$.
$d_{\rm{012}}$ varies linearly with increasing $x$ and $y$, suggesting that the doped element is distributed uniformly within Bi.
The opposite trend of $d_{\rm{012}}$ for Bi$_{1-x}$Te$_x$ and Bi$_{1-y}$Sn$_y$ may arise from the different atomic radii $r$ of the elements satisfying the relation $r_{\rm{Sn}} \geq r_{\rm{Bi}} > r_{\rm{Te}}$\cite{clementi1967jcp}.
\begin{figure}[b]
\begin{center}
\includegraphics[width=1\columnwidth]{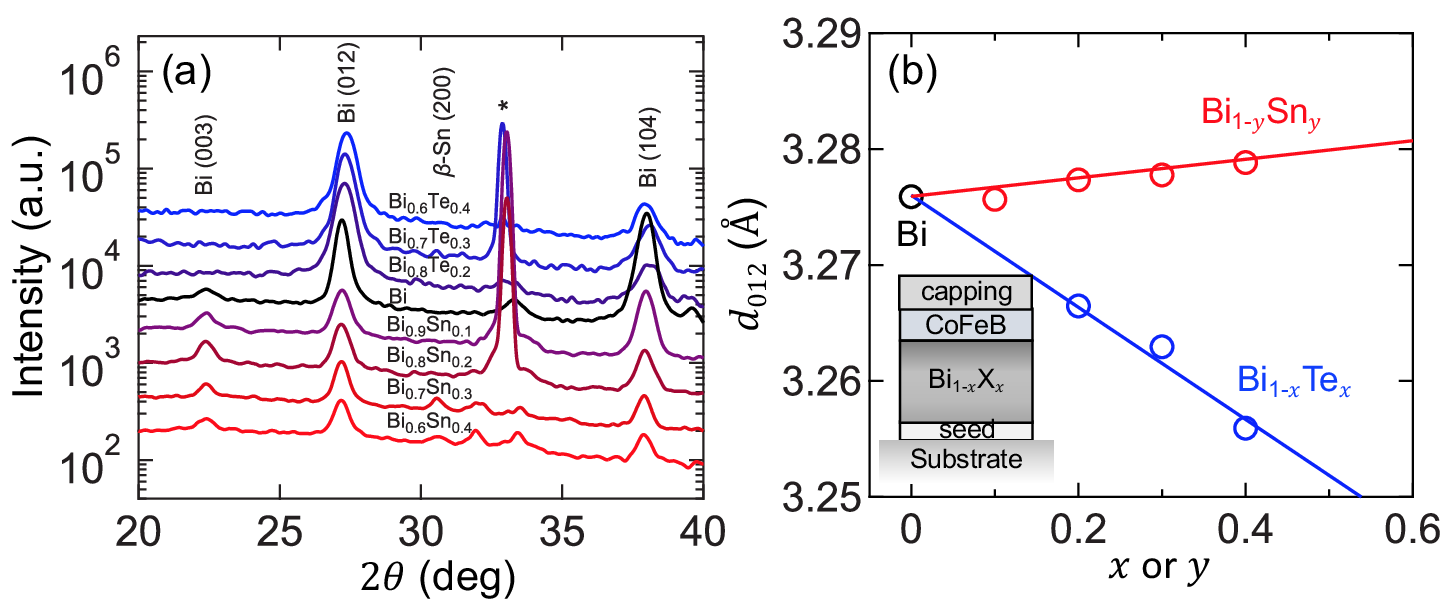}
\caption{(a) XRD spectra of carrier-doped Bi thin films grown on the substrates with the 0.5 nm Ta seed layer. Data are shifted vertically for clarity. The alloy composition is denoted for each spectra. (b) Nominal doping concentration ($x$ and $y$) dependence of (012) atomic plane distance, $d_{\rm{012}}$. Inset shows a sketch of the film structure. (The CoFeB layer sketched here is inserted only for the films used in the spin-orbit torque measurements.)} 
\label{fig:structure}
\end{center}
\end{figure}

\subsection{Transport properties}
We first investigate the transport properties of pristine and doped Bi thin films.
The carrier density $n$ and mobility $\mu_\mathrm{c}$ of the films are estimated using ordinary Hall coefficient and resistivity measurements.
For films with small doping where the single-carrier model is not valid, longitudinal magnetoresistance is measured and analyzed using a two-carrier model\cite{chi2020sciadv}.
Figure~\ref{fig:exp:transport}(a) shows $n$ as a function of $x$ and $y$.
Positive (negative) $n$ indicates that the majority carrier is hole (electron).
We find the smallest $|n|$ in pristine Bi among the films studied: The minimum value is $|n| \sim 10^{19}$ cm$^{-3}$.
Note however that such $|n|$ is nearly two orders of magnitude larger than that of single crystal Bi\cite{behnia2007science}.
We infer that disorder (defects, impurities) induced by the deposition process causes the extra carriers.
The difference in $n$ with respect to nominal concentration of Te and Sn dopants may be attributed to their different solubility in Bi.

The change in $n$ with respect to the dopant concentration can be understood using a rigid-band model.
A simplified image of the band structure near the $L$-point of Bi is illustrated in the inset to Fig.~\ref{fig:dispersion}(b).
The Fermi level of pristine Bi lies close to ($\sim$30 meV above) the Dirac point\cite{liu1995prb}.
Electron and hole doping shift the Fermi level ($E_\mathrm{F}$) away from the Dirac point, leading to dramatic enhancement of $|n|$ due to the increased density of states.
This is consistent with the Fermi level dependence of the density of states ($D$) shown in Fig.~\ref{fig:dispersion}(b).
Note that it is difficult to identify the exact position of the Dirac point with respect to doping concentration since the measurements are carried out at room temperature and the precision of the doping concentration is not determined.
Hereafter, we assume $n$ as an indicator of $E_\mathrm{F}$ relative to the Dirac point. 

\begin{figure}
\begin{center}
  \includegraphics[width=1\columnwidth]{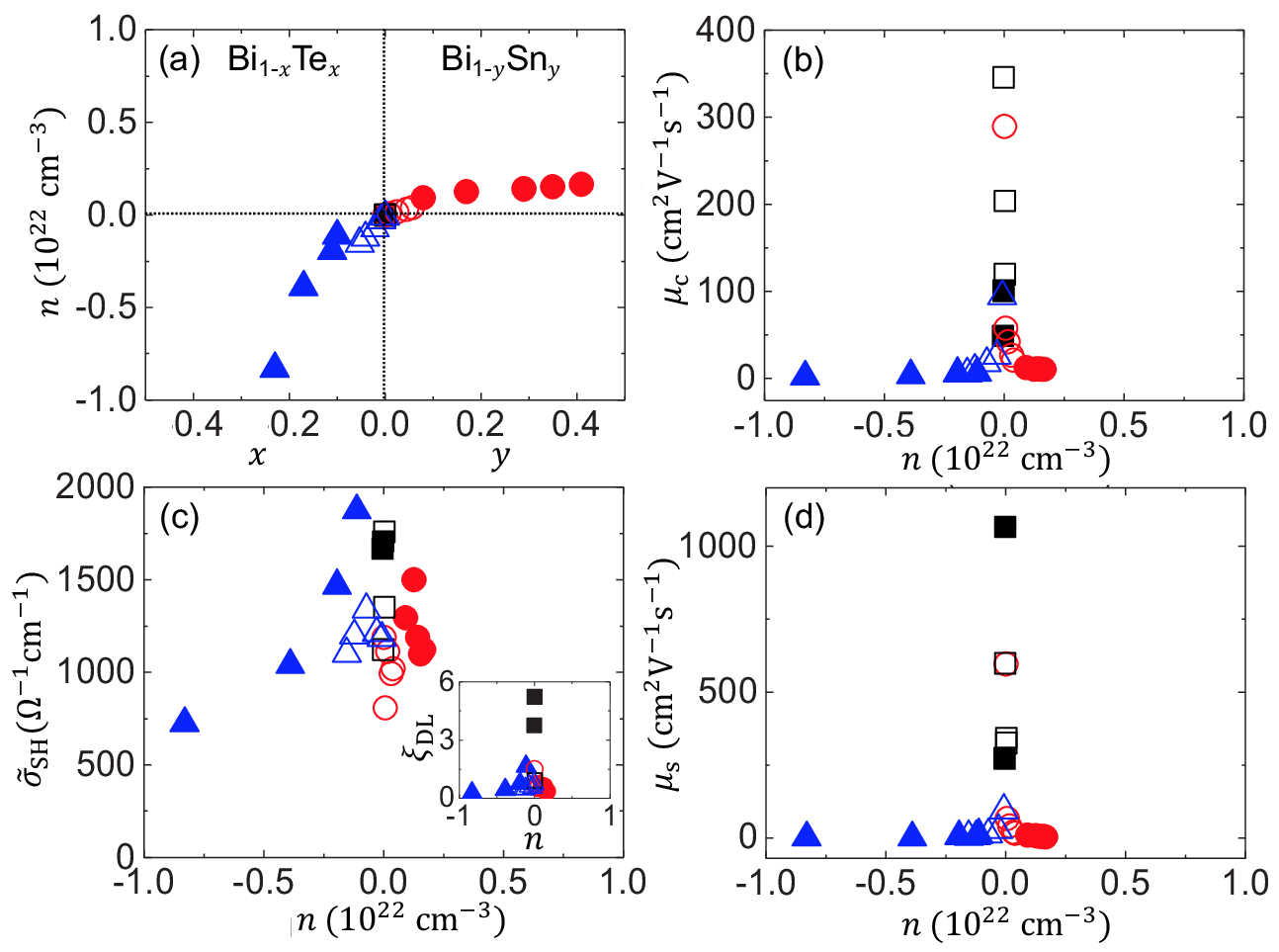}
  \caption{(a) Carrier concentration $n$ plotted as a function of nominal doping concentration $x$ for Bi$_{1-x}$Te$_x$ alloy and $y$ for Bi$_{1-y}$Sn$_y$ alloy. $x=y=0$ corresponds to pristine Bi. Positive (negative) $n$ indicates the majority carrier is hole (electron). (b-d) $n$ dependence of mobility $\mu_\mathrm{c}$ (b), spin Hall conductivity $\tilde{\sigma}_\textrm{SH}$ (c) and spin Hall mobility $\mu_\mathrm{s}$ (d). Inset to (c) shows the spin torque efficiency $\xi_\mathrm{DL}$ vs. $n$. (a-d) Red circles: Sn-doped Bi, blue triangles: Te-doped Bi, black squares: pristine Bi. Solid (open) symbols show results from films with (without) the 0.5 nm Ta seed layer.}
  \label{fig:exp:transport}
\end{center}
\end{figure}

The $n$ dependence of mobility $\mu_\mathrm{c}$ is presented in Fig.~\ref{fig:exp:transport}(b).
As evident, $\mu_\mathrm{c}$ takes a maximum at small $|n|$ and abruptly decays as the number of carriers increases.
Such $n$ dependence of $\mu_\mathrm{c}$ has been observed in materials with Dirac cones (\textit{e.g.}, in graphene), in which the mobility takes an extrema when the Fermi level is close to the Dirac point.
The maximum $\mu_\mathrm{c}$ at small $|n|$ is $\sim 350$ (cm$^2$V$^{-1}$s$^{-1}$).
Note that the longitudinal resistivity $\rho_{xx}$ exhibits a sharp peak at small $|n|$ and decreases monotonically with increasing $x$ and $y$ (see Fig.~\ref{fig:resistivity} in the Appendix (Sec.~\ref{sec:supps:sample})).
Such a reduction of the resistivity upon doping is primarily due to the rapid increase of $n$: The slower decrease of the mobility is unable to counter balance such trend.

\subsection{Spin torque efficiency}
The harmonic Hall voltage measurement\cite{kim2013nmat,garello2013nnano,hayashi2014prb,avci2014prb,roschewsky2019prb} is used to quantify the spin torque efficiency via measurements of the spin-orbit torques.
A 2-nm-thick CoFeB layer is deposited on top of the pristine or carrier doped Bi layer.
The anomalous Hall and the planar Hall effects of the CoFeB layer are used to detect the change in the magnetization direction as current is supplied to the films.
From the measurements (see Appendix (Sec.~\ref{sec:supps:sot}) for the details) the dampinglike (fieldlike) spin-orbit effective field $h_\mathrm{DL(FL)}$ is obtained.
Contributions from thermoelectric voltages, in particular, the large ordinary Nernst effect of Bi, are taken into account in the measurements\cite{avci2014prb,roschewsky2019prb,chi2020sciadv}.
The dampinglike (fieldlike) spin torque efficiency $\xi_\mathrm{DL(FL)}$ is estimated from $h_\mathrm{DL(FL)}$ using the following relation:
\begin{equation}
\label{eq:xi_DL}
\xi_\mathrm{DL(FL)}=\frac{2e}{\hbar}\frac{h_\mathrm{DL(FL)}{M_\mathrm{s}}{t_\mathrm{eff}}}{j}
\end{equation}
where $e$ is the elementary charge, $\hbar$ is the reduced Planck constant, $j$ is the current density that flows in the pristine or carrier doped Bi layer, $M_\mathrm{s}$ and $t_{\textrm{eff}}$ are the saturation magnetization and the effective thickness of the CoFeB layer, respectively.
The product of $M_\mathrm{s} t_{\textrm{eff}}$ is measured using vibrating sample magnetometry; see Appendix (Sec.~\ref{sec:Sample}) for the details.
$\xi_\mathrm{DL}$ is related to the spin Hall angle ($\theta_\mathrm{SH}$) via the relation $\xi_\mathrm{DL} = T \tan \theta_\mathrm{SH}$, where $T$ is the interface spin transmission coefficient\cite{pai2015prb,zhang2015nphys,rojassanchez2014prl}.
We assume $T=1$: $\xi_\mathrm{DL}$ thus provides a lower limit of $\theta_\mathrm{SH}$ (as $\theta_\mathrm{SH}$ varies from $-\frac{\pi}{2}$ to $\frac{\pi}{2}$, the range of $\xi_\mathrm{DL}$ is, in general, unlimited).

$\xi_\mathrm{DL}$ of all structures are plotted as a function of $n$ in the inset to Fig.~\ref{fig:exp:transport}(c).
The $n$ dependence of $\xi_\mathrm{FL}$ is presented in the Appendix, Fig.~\ref{fig:FL-SOT} (Sec.~\ref{sec:supps:sot}). 
Positive $\xi_\mathrm{DL}$ corresponds to generation of spin current that has the same sign with that of Pt\cite{sagasta2016prb} and BiSb\cite{chi2020sciadv}.
Similar to $\mu_\mathrm{c}$, $\xi_\mathrm{DL}$ reaches a maximum at small $|n|$.
For pristine Bi, we obtain $\xi_\mathrm{DL}$ that exceeds 5.
Note that such large spin torque efficiency is in contrast to previous studies on the spin Hall effect of Bi\cite{hou2012apl,emoto2016prb,yue2018prl}.
We infer the structure of Bi at the interface with the ferromagnetic layer plays a crucial role for the spin torque efficiency.
For example, reversing the stacking order (placing the ferromagnetic layer below Bi) results in significant reduction in the spin torque efficiency, likely caused by the difference in film growth\cite{hirose2021prb}.
Here we consider the rhombohedral crystal of Bi, in which the electronic structure at the $L$-point can be described by the Dirac Hamiltonian, is essential in obtaining the large spin Hall conductivity.
Reversing the stacking order may cause the rhombohedral crystal structure to be altered by, for example, strain and/or film growth.
Interestingly, the sign of $\xi_\mathrm{DL}$ is the same for Te- and Sn-doped Bi in which the majority carrier is, respectively, electrons and holes.
This is consistent with theoretical predictions based on the Dirac Hamiltonian\cite{fuseya2012jpsj1,fukazawa2017jpsj}: see also Sec.~\ref{sec:model} E.
\begin{figure}[t]
\begin{center}
  \includegraphics[width=1\columnwidth]{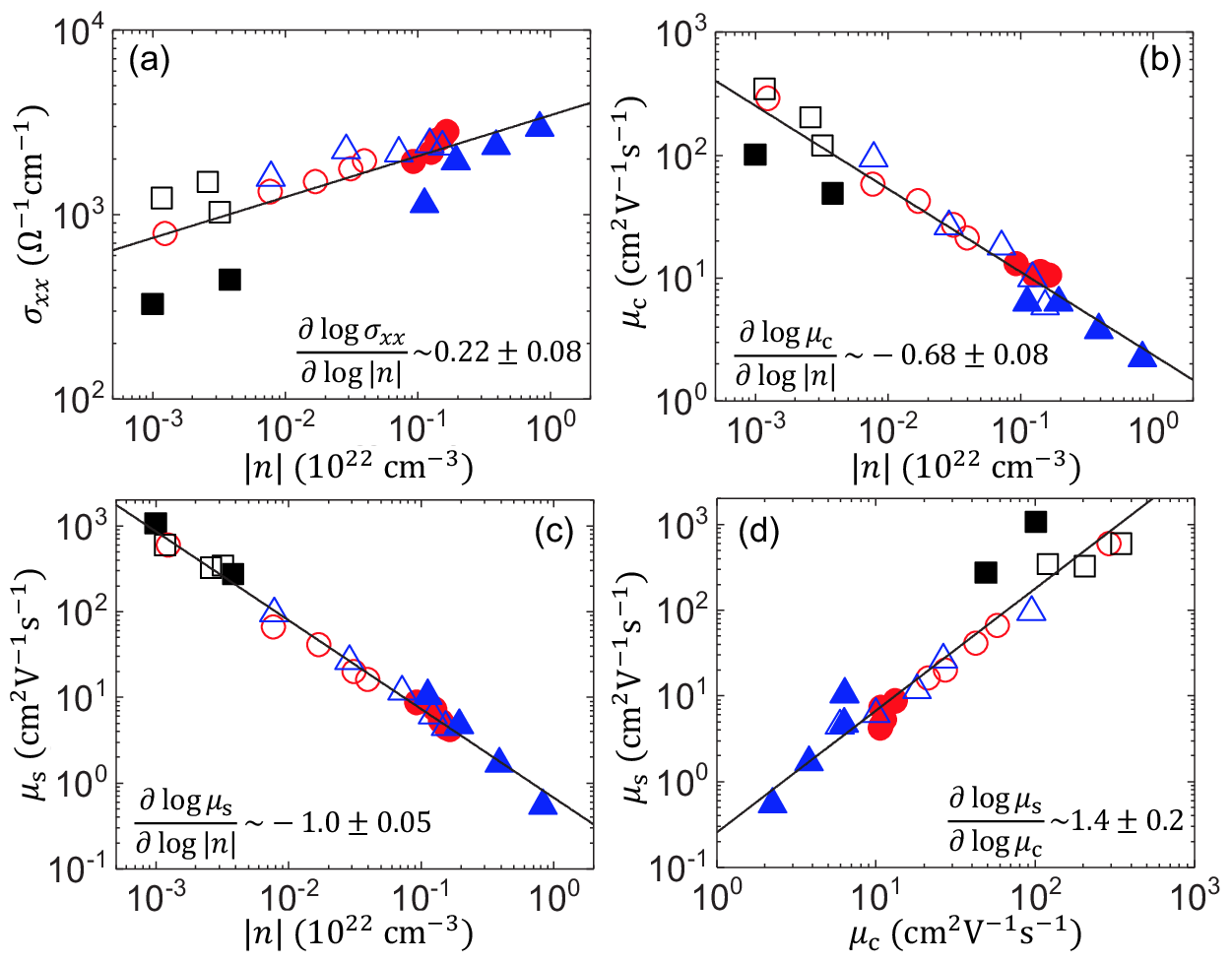}
  \caption{(a-c) Conductivity $\sigma_{xx}$ (a), mobility $\mu_\mathrm{c}$ (b) and spin Hall mobility $\mu_\mathrm{s}$ (c) plotted as a function of carrier concentration $|n|$. (d) Dependence of spin Hall mobility $\mu_\mathrm{s}$ on $\mu_\mathrm{c}$. (a-d) Solid lines are linear fit to the data plotted in log-log scale. The slope of the linear fit is denoted in each panel. The error range represents 95\% confidence interval. Red circles: Sn-doped Bi, blue triangles: Te-doped Bi, black squares: pristine Bi. Solid (open) symbols show results from films with (without) the 0.5 nm Ta seed layer.}
  \label{fig:exp:scaling}
\end{center}
\end{figure}

The spin Hall conductivity $\tilde{\sigma}_\textrm{SH}$ is calculated using the relation $\tilde{\sigma}_\textrm{SH} =  \xi_\mathrm{DL} / \rho_{xx}$ (assuming $T=1$).
The $n$ dependence of $\tilde{\sigma}_\textrm{SH}$ is presented in Fig.~\ref{fig:exp:transport}(c).
$\tilde{\sigma}_\textrm{SH}$ shows a broad maximum at small $|n|$ and decay with increasing $|n|$. 
For samples with small $|n|$, $\sigma_\mathrm{SH}$ ranges from $\sim 1000$ to 2000 $(\frac{\hbar}{2e} \Omega^{-1}$cm$^{-1}$), which is comparable to that of Pt\cite{sagasta2016prb} and BiSb alloys\cite{chi2020sciadv}.
As an independent verification, we have also quantified the spin torque efficiency of pristine Bi using the spin-torque ferromagnetic resonance technique.
We obtain $\sigma_\mathrm{SH} \sim 1220$ ($\frac{\hbar}{2e}$ $\Omega^{-1}$ cm$^{-1}$), which is consistent with the results obtained using the harmonic Hall measurements (see Fig.~\ref{fig:ST-FMR} in the Appendix (Sec.~\ref{sec:supps:st-fmr})).
The spin Hall mobility ($\mu_\mathrm{s}$) is plotted against $|n|$ in Fig.~\ref{fig:exp:transport}(d). The $n$ dependence of $\mu_\mathrm{s}$ shows similar characteristics with $\mu_\mathrm{c}$: it takes a sharp peak at small $|n|$ and abruptly decreases with increasing $|n|$.

\subsection{Scaling relations}
These results show that the SHE of Bi strongly depends on the carrier density.
To quantify the scaling, we show in Figs.~\ref{fig:exp:scaling}(a-c) the $|n|$ dependence of $\sigma_{xx}$, $\mu_\mathrm{c}$ and $\mu_\mathrm{s}$ in log scale.
The relation of $\mu_\mathrm{s}$ and $\mu_\mathrm{c}$ is presented in Fig.~\ref{fig:exp:scaling}(d).
The log-log plots can be fitted with a linear function: the fitted curve is shown by the solid line.
The slope of the log-log plot, presented in each panel, gives the scaling coefficient.

\subsection{Comparison to model calculations}
We use the formulas described in Sec.\ref{sec:model} to study their applicability.
To find a parameter set that best describes the experimental results, we adjust $v$, $a$ and $\alpha$ in the model calculations.
$2 \Delta$ is set to the band gap of Bi ($\Delta \sim 7.7$ meV)\cite{smith1964pr,vecchi1974prb} and the cutoff energy $E_\textrm{c}$ is fixed to -5 eV, which roughly represents the band width near the Fermi level. 
Since $\mu_\mathrm{c}$ and $\mu_\mathrm{s}$ scale with the inverse of $|n|$, here we introduce impurity induced carrier density $n_0$, which is set to the minimum carrier density found in experiments, \textit{i.e.}, $n_0 \sim 1 \times 10^{19}$ cm$^{-3}$. 

The calculated Fermi level dependence of $n$ and the $n$ dependence of $\mu_\mathrm{c}$, $\tilde{\sigma}_\mathrm{SH}$ and $\mu_\mathrm{s}$ are shown in Fig.~\ref{fig:transport2}.
We set $v$ so that the maximum value of $\tilde{\sigma}_\mathrm{SH}$ at small $|n|$ agrees with that of the experiments.
$\alpha$ is determined by the scaling relations shown in Fig.~\ref{fig:scaling2} (discussed in the following).
$a$ is adjusted to match the maximum value of $\mu_\mathrm{c}$ and $\mu_\mathrm{s}$ with that of the experiments.
We thus obtain $v = 2.5 \times 10^7$ cm s$^{-1}$, $\alpha  \sim - \frac{1}{3}$ and $a \sim 3.5 \times 10^{-7}$ s cm$^{-1}$.
The effective $g$-factor obtained from $\Delta$ and $v$ is $g^* \sim 92$: studies have reported values similar in order of magnitude\cite{fuseya2015prl}.
The value of $\alpha$ is in accordance with theoretical calculations using the Dirac Hamiltonian\cite{dassarma2015prb}.
Note that $\mu_\mathrm{c}$ drops to zero as $|E_\mathrm{F}|$ approaches $\Delta$ owing to the introduction of $n_0$ (see Eq.~(\ref{eq:mobility})).
\begin{figure}[t]
\begin{center}
  \includegraphics[width=1.0\columnwidth]{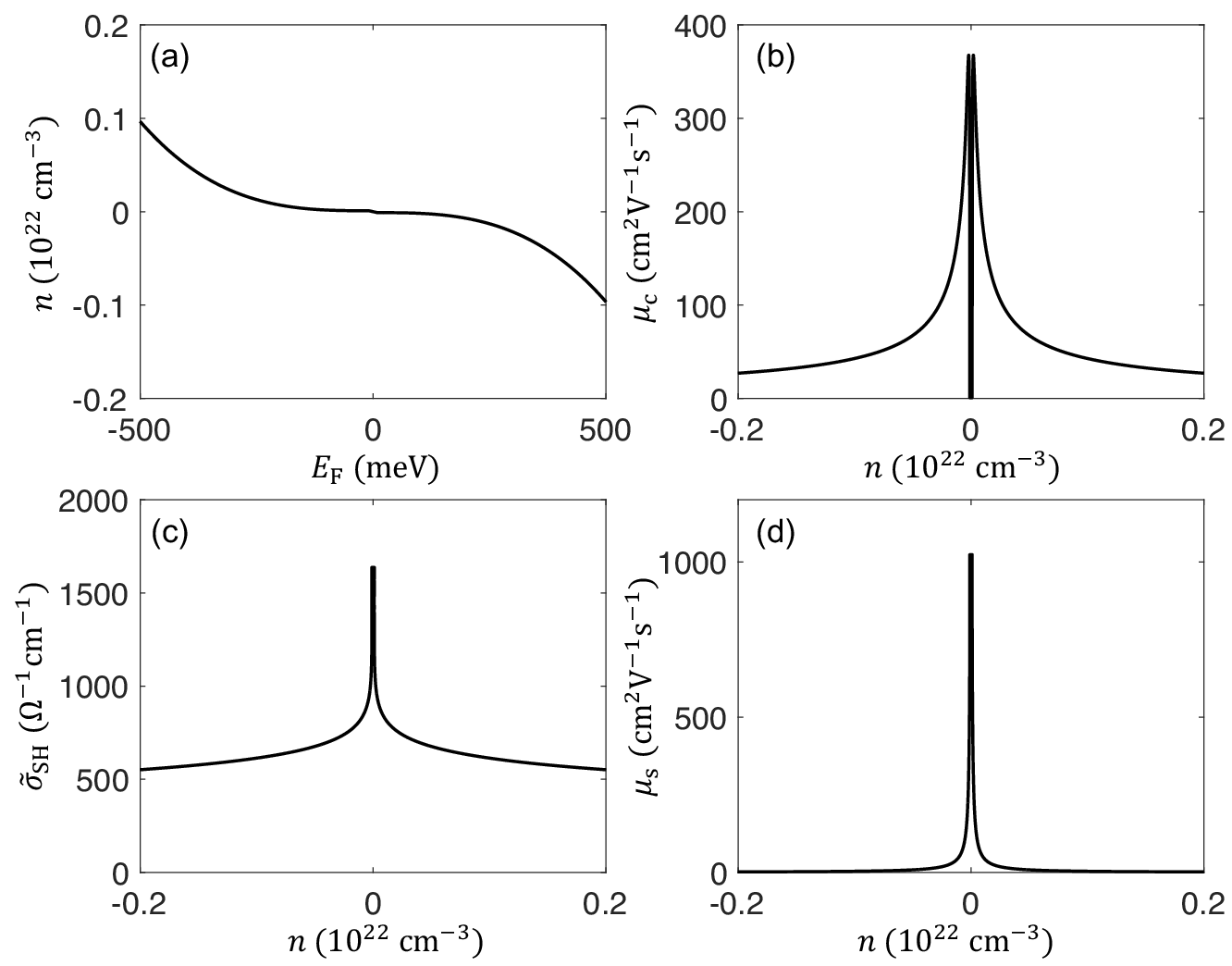}
  \caption{(a) Fermi energy ($E_\mathrm{F}$) dependence of the carrier concentration ($n$). (b-d) Carrier mobility $\mu_\mathrm{c}$ (b), spin Hall conductivity $\tilde{\sigma}_\mathrm{SH}$ (c) and spin Hall mobility $\mu_\mathrm{s}$ (d) plotted against $n$. 
  Parameters used in the calculations are: $v = 2.5 \times 10^{7}$ cm s$^{-1}$, $a = 3.5 \times 10^{-7}$ s cm$^{-1}$, $\alpha = - \frac{1}{3}$, $\Delta = 7.7$ meV, $E_\mathrm{c} = -5 $ eV and $n_0 \sim 1 \times 10^{19}$ cm$^{-3}$.}
  \label{fig:transport2}
\end{center}
\end{figure}

The calculated scaling relations are presented in Fig.~\ref{fig:scaling2}.
The ranges of the $x$- and $y$- axes are set to the same as those in Fig.~\ref{fig:exp:scaling} to allow direct comparison.
As $\mu_\mathrm{c}$ approaches zero when $|E_\mathrm{F}| \rightarrow \Delta$, Fig.~\ref{fig:scaling2}(b) reflects such trend.
To verify that introduction of $n_0$ does not influence the scaling relation, we fit the log-log plots with a linear function in the appropriate range ($10^{20} \leq |n| \leq 10^{22}$ cm$^{-3}$).
The red solid lines show the fitting results.
The slope of the linear lines is denoted in each panel and Table~\ref{table:scaling}.
\begin{figure}[h]
\begin{center}
  \includegraphics[width=1.0\columnwidth]{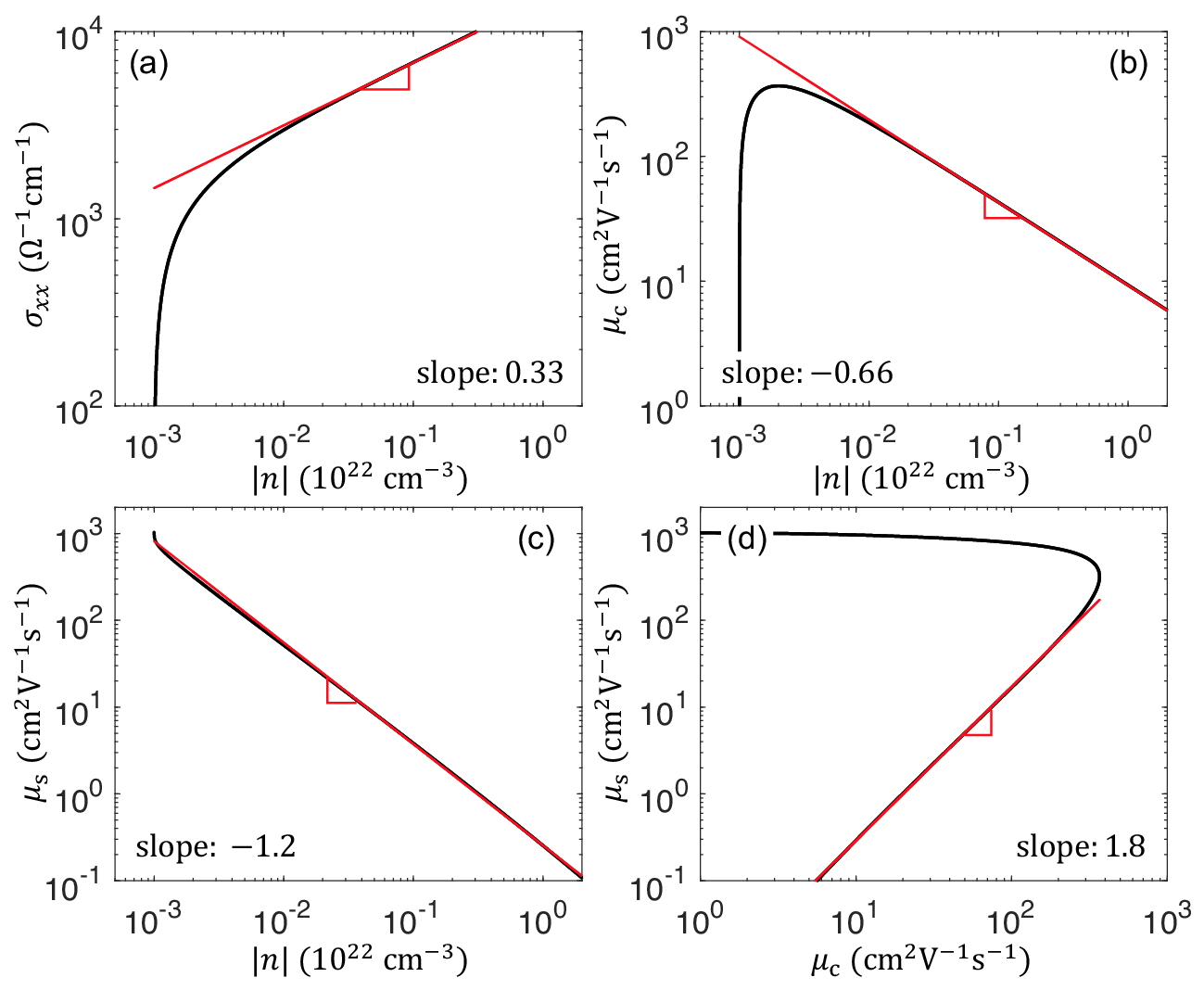}
  \caption{(a-c) Carrier density $|n|$ dependence of the conductivity $\sigma_{xx}$ (a), mobility $\mu_\mathrm{c}$ (b) and spin Hall mobility $\mu_\mathrm{s}$ (c). (d) $\mu_\mathrm{s}$ vs. $\mu_\mathrm{c}$. (a-d) All plots are in log-log scale. The red solid lines represent linear fitting to the calculated results in the appropriate range ($10^{20} \leq |n| \leq 10^{22}$ cm$^{-3}$). Slope of the linear line is denoted in each panel.
Parameters used in the calculations are: $v = 2.5 \times 10^{7}$ cm s$^{-1}$, $a = 3.5 \times 10^{-7}$ s cm$^{-1}$, $\alpha = - \frac{1}{3}$, $\Delta = 7.7$ meV, $E_\mathrm{c} = -5 $ eV and $n_0 \sim 1 \times 10^{19}$ cm$^{-3}$.
}
  \label{fig:scaling2}
\end{center}
\end{figure}

Table~\ref{table:scaling} summarizes the scaling relations obtained from the model analyses and experiments.
For the former, the analytical formulas of the scaling coefficient (Eqs.~(\ref{eq:scaling:conductivity}), (\ref{eq:scaling:mobility}), (\ref{eq:scaling:spinHallmobility}) and (\ref{eq:scaling:musmuc})) and the corresponding values with $\alpha = -\frac{1}{3}$ are shown.
Note that $E_\mathrm{F} \gg \Delta$ is assumed to derive these formulas.
The scaling coefficient is set to -1 for the relation between $\mu_\textrm{s}$ and $|n|$ (see the discussion in Sec.~\ref{sec:model:scaling}).
The slope of linear lines displayed in Fig.~\ref{fig:scaling2}, which describe the scaling relations when $n_0$ is non-zero, are shown in the column "Fitting".
As evident, introduction of $n_0$ does not significantly alter the scaling coefficients.
The deviation of the coefficient for the $\mu_\textrm{s}$ vs. $|n|$ scaling (and consequently the $\mu_\textrm{s}$ vs. $\mu_\textrm{c}$ scaling) between the analytical formula and the fitting is largely caused by the omission of the $\log(n^{-\frac{1}{3}})$ factor in Eq.~(\ref{eq:scaling:spinHallmobility}).
We find the model calculations roughly reproduce the coefficients obtained by the experiments.
As the experimental data exhibit relatively large scattering, the differences between the experiments and the calculations are within the experimental error.

\section{Discussion}
Here we discuss potential alternative interpretations of the experimental results.
First, the model described in Sec.~\ref{sec:model} neglects contribution, if any, from the holes presiding at the $T$ point.
From tight binding calculations, however, we find contribution from the holes on the spin Hall conductivity is rather small due to the large effective mass.
The extrinsic carriers induced by Te and Sn doping may also contribute to the generation of spin current via the extrinsic SHE (e.g., skew scattering and side jump mechanisms).
As shown in Fig.~\ref{fig:exp:transport}(c), the spin Hall conductivity drops drastically upon alloying Bi with Te (with one additional valence electron than Bi) or Sn (with one less valence electron than Bi).
It is interesting to compare these results with those when Bi is alloyed with Sb.
Sb is placed at the same row with Sn and Te in the periodic table and has the same number of valence electrons with Bi.
Reference~\cite{chi2020sciadv} showed that alloying Bi with Sb leads to a spin Hall conductivity plateau for composition up to $\sim40$\% Sb. 
We therefore consider the observed modulation of spin Hall conductivity with doping is not a simple function of the extrinsic carrier density but rather reflects the position of the Fermi level. 
Moreover, a theoretical study\cite{fukazawa2017jpsj} predicted that the $n$ dependence of $\sigma_\mathrm{SH}$ becomes an odd function of $n$ when skew scattering is taken into account.
The study also showed that $\sigma_\mathrm{SH}$ increases with $|n|$ when side jump mechanism is considered.
Such features are not found in the experiments, suggesting that contributions from the extrinsic SHE, if any, are negligible.
Further studies are required to experimentally identify the origin of the SHE in Bi.

The relatively good agreement between the experiments and calculations suggests the validity of the model used.
The flow of spin magnetic moment as a spin current in Bi, however, poses a question on whether it itself can exert spin torque on the magnetization of the ferromagnetic layer. 
The dampinglike spin-orbit torque is, in general, based on conservation of spin angular momentum.
Since the flow of spin angular momentum vanishes in Bi, one expects zero spin torque on the CoFeB magnetization.
To account for the observed spin-orbit torque at the Bi/CoFeB interface, we infer the spin current due to the flow of spin magnetic moment in Bi is converted to spin angular momentum in CoFeB at the interface.
Note that the wave function of Bi is composed of a four-component spinor, whereas that of CoFeB, a typical transition metal, is formed from a two-component spinor.
Thus a $2 \times 4$ spin transmission matrix must be introduced to match the wave function at the interface, which may promote the conversion.
Further modeling is required to identify the form of the $2 \times 4$ spin transmission matrix.

\section{Conclusion}
In summary, we have studied the spin transport properties of Dirac Hamiltonian systems.
The Hamiltonian represents the electronic structure of gapped Dirac semimetals that includes, for example, states near the $L$-point in the reciprocal space of Bi.
The spin Hall conductivity is calculated using the Kubo formula.
If the spin current is defined as a flow of spin angular momentum, all components of the spin Hall conductivity vanish.
In contrast, when the spin current is defined as the flow of spin magnetic moment, the off-diagonal components of the spin Hall conductivity are non-zero.
The spin Hall conductivity associated with the flow of spin magnetic moment scales with the carrier velocity (and the effective $g$-factor).
Analytical formulas for the carrier density, mobility, conductivity and spin Hall conductivity and their mutual scaling relations are derived.
The scaling coefficients, which depend on the carrier velocity and relaxation time, can be used to characterize the spin transport properties of the system.

The model analyses are compared to experimental results on the spin Hall effect of Bi.
Te or Sn is substituted into Bi to control the Fermi level via carrier doping. 
The spin torque efficiency and the spin Hall conductivity take a maximum near the Dirac point.
The sign of the spin torque efficiency is the same regardless of the majority carrier type (\textit{i.e.}, the Fermi level position).
A clear power law is found for the carrier density dependence of the conductivity, mobility and the spin Hall mobility, suggesting that strong scaling relations exist.
We find these experimental results, including the scaling coefficients, are consistent with the model calculations.
These results thus demonstrate what spin current stands for in Dirac materials.
Further studies are required to identify how the flow of spin magnetic moment, with a large effective $g$-factor, is converted to spin-orbit torque at the interface with magnetic materials.

\section*{Acknowledgements}
We thank G. Tatara for fruitful discussion. This work was partly supported by JST CREST (Grant No. JPMJCR19T3), JSPS Grant-in-Aid for Early-Career Scientists (Grant No. JP20K15156), and the Center for Spintronics Research Network (CSRN). Z.C. is supported by Materials Education program for the future leaders in Research, Industry, and Technology (MERIT), the University of Tokyo and the JSR Fellowship.

\section{Appendix}

\subsection{\label{sec:conductivity}Conductivity by Kubo formula}
\setcounter{equation}{0}
\numberwithin{equation}{subsection}
In the present Hamiltonian (\ref{eq:H}), the thermal Green's function is 
\begin{equation}
\label{eq:ThermalG}
\mathcal{G}({\bm k}, i\varepsilon_n) =\frac{i\varepsilon_n 
+ \hbar v k_i \tau_2 \otimes \sigma_i - \Delta \tau_3 \otimes \sigma_0}{\mathcal {D}}, 
\end{equation}
where $\varepsilon_n=(2n+1)\pi k_\mathrm{B}T$, with an integer $n$, is Matsubara frequency.
$\mathcal{D}:=(i\varepsilon_n)^2-\varepsilon^2$ and $\varepsilon=\sqrt{\Delta^2+(\hbar v k)^2}$.
Using the velocity in Eq.~(\ref{eq:vel}), the electric conductivity is obtained from the current-current correlation function
\begin{equation}
\Phi_{xx}(i\omega_\lambda) = -\frac{k_\mathrm{B}Te^2}{V}\sum_{n,{\bm k}} {\rm Tr} 
\mathcal{G}_+ {\hat v}_x \mathcal{G} {\hat v}_x,
\end{equation}
where $\omega_\lambda=2\pi \lambda k_\mathrm{B}T$ ($\lambda$ is an integer) represents the frequency of the applied electric field.
Here the vertex corrections are neglected.
Tr is the trace over the $4\times 4$ matrix, and ${\mathcal G}_+:={\mathcal G}({\bm k}, i\varepsilon_n+i\omega_\lambda)$.
After taking the trace, $\Phi_{xx}$ becomes
\begin{equation}
\Phi_{xx}(i\omega_\lambda) = -\frac{4k_\mathrm{B}Te^2v^2}{V}\sum_{n,{\bm k}} 
\frac{i\varepsilon_n(i\varepsilon_n+i\omega_\lambda)-\varepsilon^2+2K_x^2}{{\mathcal D}{\mathcal D}_+},
\end{equation}
with $K_x=\hbar v k_x$ and $\mathcal{D}_+:=(i\varepsilon_n+i\omega_\lambda)^2-\varepsilon^2$. The summation over $n$ can be taken by the standard technique using contour integral.
We assume the relaxation rate is given by $\Gamma=\hbar/2\tau_\mathrm{eff}$.
Taking the analytic continuation $i\omega_\lambda \rightarrow \hbar \omega+i\delta$, we obtain the conductivity, up to linear order of $\omega$, as
\begin{equation}
\begin{aligned}
\label{eq:KuboXX}
\sigma_{xx} =& \lim_{\omega\rightarrow 0} \frac{\Phi_{xx}(\hbar\omega+i\delta)-\Phi_{xx}(0)}{i\omega} \\
=&\frac{4e^2v^2\hbar}{V} \sum_{\bm k} \int_{-\infty}^\infty \frac{dz}{2\pi} (-f'(z)) \biggl[ 
\frac{z^2+\Gamma^2-\varepsilon^2+2K_x^2}{D^R D^A} \\
&-\frac{D^R+2K_x^2}{2(D^R)^2} -\frac{D^A+2K_x^2}{2(D^A)^2} \biggr] \\
=&\frac{e^2v^2\hbar}{V} \sum_{\tau, \tau', {\bm k}} \int \frac{dz}{2\pi} (-f'(z)) 
\left\{ 1+\tau\tau'\left(\frac{2K_x^2}{\varepsilon^2} -1\right) \right\} \\
&\times \biggl[\frac{1}{(z+i\Gamma -\tau\varepsilon)(z-i\Gamma-\tau' \varepsilon)} \\
&-{\rm Re} \frac{1}{(z+i\Gamma -\tau\varepsilon)(z+i\Gamma-\tau' \varepsilon)} \biggr],
\end{aligned}
\end{equation}
with
\begin{equation}
D^R = (z+i\Gamma)^2-\varepsilon^2, \quad D^A = (z-i\Gamma)^2-\varepsilon^2.
\end{equation}

If $\Gamma$ is small compared to $\Delta$ or $E_\mathrm{F}$, the intraband conbribution with $\tau=\tau'$ becomes dominant and $\sigma_{xx}$ is approximated as 
\begin{equation}
\sigma_{xx}= 
\frac{e^2v^2\hbar}{V} \sum_{\tau, {\bm k}} \int dz \left( -\frac{f'(z)}{\Gamma}\right) \frac{K_x^2}{\varepsilon^2} \delta(z-\tau \varepsilon).
\end{equation}
At $T=0$, this becomes
\begin{equation}
\sigma_{xx}= e^2D(E_\mathrm{F}) v^2\frac{E_\mathrm{F}^2-\Delta^2}{3E_\mathrm{F}^2}\tau_\mathrm{eff}.
\end{equation}
This result is consistent with that in the Boltzmann transport theory with relaxation time approximation, 
$\sigma_{xx}= e^2D(E_\mathrm{F}) \langle v_{x}^2 \rangle_{\rm FS} \tau_\mathrm{eff}=e^2D(E_\mathrm{F})v_\mathrm{F}^2 \tau_\mathrm{eff}/3$, since the Fermi velocity in the present case is
\begin{equation}
v_\mathrm{F} = \frac{1}{\hbar} \frac{\partial \varepsilon}{\partial k} \biggl|_{\varepsilon=E_\mathrm{F}} = v\frac{\sqrt{E_\mathrm{F}^2-\Delta^2}}{E_\mathrm{F}}.
\end{equation}
See Ref.~\cite{fujimoto2021condmat} for the derivation of the conductivity using the Born approximation, where the carrier relaxation rate, obtained from the self energy due to impurity potentials, depends on the energy of the carriers.

\subsection{\label{sec:vanishment}Zero spin Hall conductivity of spin angular momentum flow}
We discuss the reason behind the zero spin Hall conductivity when the flow of the spin angular momentum is used as the spin current.
First, we show that the Dirac Hamiltonian (Eq.~(\ref{eq:H})) can be decomposed into two subspaces.
The basis state $c^{}_{\bk} = \T(c_1, c_2, c_3, c_4)$ is transformed using an unitary transformation $U_{\bk} = \tau_0 \otimes u_{\bk}$, where $u_{\bk}$ satisfies $u_{\bk} (\bk \cdot \bm{\sigma}) u_{\bk} = |\bk| \sigma^z$; that is, $\tilde{c}_{\bk} = U^{}_{\bk} c^{}_{\bk}$.
The Hamiltonian therefore transforms as 
\begin{align}
\sum_{\bk} \tilde{c}^{\dagger}_{\bk} H \tilde{c}^{}_{\bk}
	& = \sum_{\bk} c^{\dagger}_{\bk} U^{\dagger}_{\bk} H U^{}_{\bk} c^{}_{\bk}
\notag \\
	& = \sum_{\bk} c^{\dagger}_{\bk} \begin{pmatrix}
		\Delta
	&	0
	&	\zi \hbar v |\bk|
	&	0
	\\	0
	&	\Delta
	&	0
	&	- \zi \hbar v |\bk|
	\\	- \zi \hbar v |\bk|
	&	0
	&	- \Delta
	&	0
	\\	0
	&	\zi \hbar v |\bk|
	&	0
	&	- \Delta
	\end{pmatrix}
	c^{}_{\bk}.
\end{align}
By interchanging rows and columns, we may block-diagonalize the Hamiltonian as follows:
\begin{align}
\sum_{\bk} \tilde{c}^{\dagger}_{\bk} H \tilde{c}^{}_{\bk}
	& = \sum_{\bk} \bar{c}_{\bk}^{\dagger}
	\begin{pmatrix}
		\Delta
	&	\zi \hbar v |\bk|
	&	0
	&	0
	\\	- \zi \hbar v |\bk|
	&	- \Delta
	&	0
	&	0
	\\	0
	&	0
	&	- \Delta
	&	\zi \hbar v |\bk|
	\\	0
	&	0
	&	- \zi \hbar v |\bk|
	&	\Delta
	\end{pmatrix}
	\bar{c}^{}_{\bk}
,\end{align}
where $\bar{c}_{\bk} = \T (c_1, c_3, c_4, c_2)$.

Next, we define two-component spinors $a_{\bk}$ and $b_{\bk}$ as the projections of the four component spinor $\bar{c}_{\bk}$; $a_{\bk} = \T(c_{1}, c_{3})$ and $b_{\bk} = \T(c_{4}, c_{2})$.
The Hamiltonian can thus be expressed as
\begin{align}
\sum_{\bk} \tilde{c}^{\dagger}_{\bk} H \tilde{c}^{}_{\bk}
	& = \sum_{\bk} a^{\dagger}_{\bk} h_{\bk}^{\Delta} a^{}_{\bk}
		+ \sum_{\bk} b^{\dagger}_{\bk} h_{\bk}^{\bar{\Delta}} b^{}_{\bk}
,\end{align}
where $\bar{\Delta} = - \Delta$ and
\begin{align}
h_{\bk}^{\Delta}
	& = \begin{pmatrix}
		\Delta
	&	\zi \hbar v |\bk|
	\\	- \zi \hbar v |\bk|
	&	- \Delta	
	\end{pmatrix}
.\end{align}
We refer to the states described by $a_{\bk}$ and $b_{\bk}$ as the $a$- and $b$-carriers, respectively.
Note that the Hamiltonians of $a$- and $b$-carriers have the same matrix elements except for the sign of the diagonal term, \textit{i.e.}, the mass gap ($\Delta$).

Now we show that the magnitude of the $a$- and $b$-carriers' Berry curvature-like contributions to spin Hall conductivity are the same but their signs are opposite.
Since the Hall conductivity in general has a direct correspondence with the Berry curvature, we look for its form.
First we calculate the Hall conductivity in the absence of magnetic field.
The electric current operator is given as
\begin{align}
J_i
	& = - e \sum_{\bk} \tilde{c}_{\bk}^{\dagger} \left( \frac{1}{\hbar} \frac{\partial H}{\partial k_i} \right) \tilde{c}_{\bk}^{}
.\end{align}
Using $a$- and $b$-carriers' representation, $J_i$ can be expressed as
\begin{align}
J_i
	& = e v \sum_{\bk} \left[
		\frac{k_i}{|\bk|} a_{\bk}^{\dagger}
		\begin{pmatrix}
			0
		&	- \zi
		\\	\zi
		&	0
		\end{pmatrix}
		a^{}_{\bk}
		+ \frac{k_i}{|\bk|} b_{\bk}^{\dagger}
		\begin{pmatrix}
			0
		&	- \zi
		\\	\zi
		&	0
		\end{pmatrix}
		b^{}_{\bk}
\right. \notag \\ & \hspace{1em} \left.
		+ |\bk| A^{+}_{\bk, i} a^{\dagger}_{\bk}
		\begin{pmatrix}
			1
		&	0
		\\	0
		&	- 1
		\end{pmatrix}
		b_{\bk}^{}
		+ |\bk| A^{-}_{\bk, i} b^{\dagger}_{\bk}
		\begin{pmatrix}
			1
		&	0
		\\	0
		&	- 1
		\end{pmatrix}
		a_{\bk}^{}
	\right]
,\end{align}
where the gauge field in the momentum space $A_{\bk, i}^{\alpha}$ is defined as
\begin{align}
A_{\bk, i}^{\alpha}
	& = \frac{- \zi}{4} \tr \left[
	U_{\bk}^{\dagger} \frac{\partial U_{\bk}}{\partial k_i}
	\sigma^{\alpha}
	\right]
\qquad (\alpha = x, y, z)
\label{eq:gauge}
,\end{align}
and $A^{\pm}_{\bk, i} = A_{\bk, i}^x \pm \zi A_{\bk, i}^y$.
To calculate $\sigma_{x y}$, we introduce the correlation function of charge currents
\begin{align}
\langle J_x; J_y \rangle (\zi \omega_{\lambda})
	& = \int_0^{\beta} \dd{\tau} e^{\zi \omega_{\lambda} \tau} \langle \mathrm{T}_{\tau} J_x (\tau) J_y \rangle_{H}
,\end{align}
where $\beta = 1 / \kB T$, $\omega_{\lambda}$ is the Matsubara frequency (of boson), $\mathrm{T}_{\tau}$ is the time-ordering operator, $J_i (\tau)$ is the Heisenberg representation of $J_i$, and $\langle \cdots \rangle_{H}$ represents the thermal average of $H$.
Using the correlation function, $\sigma_{x y}$ reads
\begin{align}
\label{eq:sigmaxy}
\sigma_{x y}
	& = \lim_{\omega \to 0} \frac{\langle J_x; J_y \rangle (\omega) - \langle J_x; J_y \rangle (0)}{\zi \omega}
\notag \\
	& = \frac{e^2}{\hbar} \frac{1}{V} \sum_{\bk} \sum_{\eta = \pm} f(\epsilon_{\bk \eta}) \big( \Omega_{\eta, x y}^a (\bk) + \Omega_{\eta, x y}^b (\bk) \big)
,\end{align}
where $\langle J_x; J_y \rangle (\omega)$ is the retarded correlation function obtained by taking the analytic continuation $\zi \omega_{\lambda} \to \hbar \omega + \zi 0$ in $\langle J_x; J_y \rangle (\zi \omega_{\lambda})$, and $\epsilon_{\bk \eta} = \eta \sqrt{\hbar^2 v^2 k^2 + \Delta^2}$ is the eigenenergy.
$\Omega_{\eta, x y}^a$ and $\Omega_{\eta, x y}^b$ are defined as
\begin{align}
\label{eq:berry}
\Omega_{\eta, x y}^a &= - \Omega_{\eta, x y}^b \\
	 &= - \frac{\zi \hbar^2 v^2 k^2}{4 \epsilon_{\bk \eta}^2}
		\left( A_{\bk, x}^{+} A_{\bk, y}^{-} - A_{\bk, y}^{+} A_{\bk, x}^{-} \right).
\end{align}
$\Omega_{\eta, x y}^{a}$ and $\Omega_{\eta, x y}^{b}$ are $a$ and $b$-carriers' Berry curvature-like contributions to the Hall conductivity.
As their signs are opposite, but with equal magnitude, the Hall conductivity ($\sigma_{x y}$) vanishes.

The spin Hall conductivity can also be expressed using $\Omega_{\eta, x y}^{a}$ and $\Omega_{\eta, x y}^{b}$.
The spin Hall conductivity based on the Kubo formula is obtained via calculation of the correlation function between spin current and charge current.
The spin current operator for the flow of spin angular momentum is given by
\begin{align}
J_{\mathrm{s}, i}^{\alpha}
	& = \frac{\hbar}{2} \sum_{\bk} \tilde{c}^{\dagger}_{\bk}
		\frac{1}{2} \left\{ \frac{1}{\hbar} \frac{\partial H}{\partial k_i}, s^{\alpha} \right\}
	\tilde{c}^{}_{\bk}
,\end{align}
where $\{A, B\} = A B + B A$ is the anticommutator and $s^{\alpha} = \tau_0 \otimes \sigma^{\alpha}$.
Using $a$ and $b$-carriers' representation, $J_{\mathrm{s}, i}^{\alpha}$ can be expressed as
\begin{align}
J_{\mathrm{s}, i}^{\alpha}
	& = \sum_{\bk} \frac{- k_i}{|\bk|} S_{a a}^{\alpha}
		a_{\bk}^{\dagger}
		\begin{pmatrix}
			0
		&	- \zi
		\\	\zi
		&	0
		\end{pmatrix}
		a_{\bk}^{}
		+ \sum_{\bk} \frac{- k_i}{|\bk|} S_{b b}^{\alpha}
		b_{\bk}^{\dagger}
		\begin{pmatrix}
			0
		&	- \zi
		\\	\zi
		&	0
		\end{pmatrix}
		b_{\bk}^{}
\notag \\ & \hspace{1em}
		- \sum_{\bk} |\bk| A_{\bk, i}^{+} \left( \frac{S_{a a}^{\alpha} + S_{b b}^{\alpha}}{2} \right)
		a_{\bk}^{\dagger}
		\begin{pmatrix}
			1
		&	0
		\\	0
		&	- 1
		\end{pmatrix}
		b_{\bk}^{}
\notag \\ & \hspace{1em}
		- \sum_{\bk} |\bk| A_{\bk, i}^{-} \left( \frac{S_{a a}^{\alpha} + S_{b b}^{\alpha}}{2} \right)
		b_{\bk}^{\dagger}
		\begin{pmatrix}
			1
		&	0
		\\	0
		&	- 1
		\end{pmatrix}
		a_{\bk}^{}
\label{eq:Js}
,\end{align}
where $S_{a a}^{\alpha}$ and $S_{b b}^{\alpha}$ are the matrix elements of $s^\alpha$ with [see Eq.~(\ref{eq:spinang})] $a$- and $b$-carriers, respectively.
It turns out $S_{b b}^{\alpha} = - S_{a a}^{\alpha}$ and thus the third and forth terms of the right hand side of Eq.~(\ref{eq:Js}), \textit{i.e.}, the hybridization terms, are zero.
We keep these terms hereafter to show how the spin Hall conductivity vanishes.

The correlation function of spin current and charge current is given as
\begin{align}
\langle J_{\mathrm{s},s i}^{\alpha}; J_j \rangle (\zi \omega_{\lambda}) = \frac{1}{V} \int_0^{\beta} \dd{\tau} e^{\zi \omega_{\lambda} \tau} \langle \mathrm{T}_{\tau} J_{\mathrm{s}, i}^{\alpha} (\tau) J_j \rangle
 .\end{align}
The spin Hall conductivity is defined as
\begin{align}
\sigma_{i j}^{\alpha}
	& = \lim_{\omega \to 0} \frac{\langle J_{\mathrm{s}, i}^{\alpha}; J_j \rangle (\omega) - \langle J_{\mathrm{s}, i}^{\alpha}; J_j \rangle (0)}{\zi \omega}
,\end{align}
where $\langle J_{\mathrm{s},s i}^{\alpha}; J_j \rangle (\omega)$ is obtained by taking the analytic continuation $\zi \omega_{\lambda} \to \hbar \omega + \zi 0$ in $\langle J_{\mathrm{s}, i}^{\alpha}; J_j \rangle (\zi \omega_{\lambda})$.
After some straight forward calculations, we obtain
\begin{align}
\sigma_{i j}^{\alpha}
	& = \frac{- e}{\hbar} \frac{1}{V} \sum_{\bk} \sum_{\eta = \pm} f(\epsilon_{\bk \eta}) \big( \Omega_{\eta, x y}^a (\bk) + \Omega_{\eta, x y}^b (\bk) \big)
		( S_{a a}^{\alpha} + S_{b b}^{\alpha} )
.\end{align}
We thus find that $\sigma_{i j}^{\alpha}$ can also be expressed using $\Omega_{\eta, x y}^{a}$ and $\Omega_{\eta, x y}^{b}$ and the two cancel out each other. 
Note that the term $S_{a a}^{\alpha} + S_{b b}^{\alpha}$ also sets $\sigma_{i j}^{\alpha} = 0$.
This puts a strong constraint on $\sigma_{i j}^{\alpha}$, which must be zero even if $\Omega_{\eta, x y}^{a} + \Omega_{\eta, x y}^{b} \neq 0$. However, with the current model, the relation $\Omega_{\eta, x y}^{a} + \Omega_{\eta, x y}^{b} = 0$ always hold due to the symmetry of the $a$- and $b$-carriers and thus $\sigma_{i j}^{\alpha} = 0$.

\subsection{\label{sec:supps:sample}Sample preparation and characterization}
\label{sec:Sample}
Bi$_{1-x}$Te$_x$ and Bi$_{1-y}$Sn$_y$ alloy thin films were grown by radio frequency (RF) magnetron sputtering on $10 \times 10$ mm$^2$ Si substrates covered with 100 nm-thick Si oxide. The film structures are sub./seed/[$t_\mathrm{Bi}$ Bi$|$$t_\mathrm{Sn}$ Sn]$_N$/0.3 Bi/FM/2 MgO/1 Ta and sub./seed/[$t_\mathrm{Bi}$ Bi$|$$t_\mathrm{Te}$ Te]$_N$/0.3 Bi/FM/2 MgO/1 Ta (thicknesses in nanometer).
Bi, Te, and Sn layers were sputtered from elemental targets.
$t_\mathrm{Bi(Te, Sn)}$ is the thickness of each Bi (Te, Sn) layer.
$N$ represents the repeat number of [Bi$|$Sn] and [Bi$|$Te] bilayers.
(Here $N=16$ is used.)
The thickness of a bilayer unit is set to meet $t_\mathrm{Bi}+t_\mathrm{Te(Sn)} \sim$ 0.65 nm so that the two elements intermix.
We define the nominal Te (Sn) concentration, $x$ ($y$), as $x(y) \equiv t_\mathrm{Te(Sn)}/(t_\mathrm{Bi}+t_\mathrm{Te(Sn)})$.
The total thickness of the carrier doped Bi (including the Bi termination layer) is denoted by $t \equiv (t_\mathrm{Bi} + t_\mathrm{Te(Sn)}) N + 0.3$ (thickness in nanometer).
For $N=16$, $t \sim 10.7$ nm.
Films without carrier doping (pristine Bi) are also made: sub./seed/$t$ Bi/FM/2 MgO/1 Ta.
The seed layer is either 0.5 nm Ta or no insertion layer.
The FM layer is 2 nm CoFeB.
The nominal composition of CoFeB is Co:Fe:B = 20:60:20 at\%.


The transport properties and the spin-orbit torques (SOT) of the heterostructures were measured using devices with $t_\mathrm{CoFeB} = 0$ and 2 nm, respectively.
Hall bar devices were fabricated using standard optical lithography and Ar ion etching.
The width of the Hall bar $w$ and the distance between two longitudinal voltage probes $L$ are 10 and 25 $\mu\rm{m}$, respectively.
5 Ta/60 Cu/5 Pt contact pads were formed using standard lift-off processes onto the Hall bars.
The effective saturation magnetization, i.e., the product of saturation magnetization $M_\mathrm{s}$ and the effective thickness $t_\mathrm{eff}$ of the CoFeB layer, of unpatterned heterostructures with $t_\mathrm{CoFeB} = 2$ nm were determined using a vibrating sample magnetometer. $M_\mathrm{s} t_\mathrm{eff}$ is 182 $\upmu$emu cm$^{-1}$ for Bi, 201, 188 $\upmu$emu cm$^{-1}$ for $x = 0.3, 0.5$ and 186, 154 $\upmu$emu cm$^{-1}$ for $y = 0.2, 0.5$. The data is interpolated to obtain the value for all samples studied.
The magnetic easy axis of the CoFeB layer points along the film plane for all samples.
The resistivity of pristine and doped Bi thin films is presented in Fig.~\ref{fig:resistivity}.
\begin{figure}[h!]
\begin{center}
  \includegraphics[width=0.5\columnwidth]{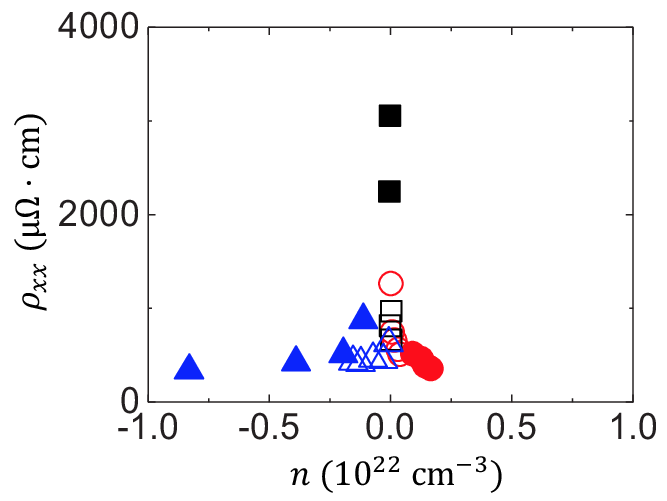}
  \caption{Resistivity $\rho$ plotted as a function of carrier concentration $n$. Red circles: Sn-doped Bi, blue triangles:
Te-doped Bi, black squares: pristine Bi. Solid (open) symbols show results from films with (without) the 0.5-nm Ta seed layer.}
  \label{fig:resistivity}
\end{center}
\end{figure}

\subsection{\label{sec:supps:sot}Spin-orbit torque measurements}
We used the harmonic Hall technique to measure the spin torque efficiency in the heterostructures with $t_\mathrm{CoFeB} = 2$ nm.
An ac current is applied to the Hall bar made of the heterostructure and the resulting Hall voltage is measured.
The r.m.s. amplitude ($I_0$) and the frequency ($\omega/2\pi$) of the ac current are 2.5 $\rm{mA}$ and 17.5 $\rm{Hz}$, respectively.
An in-plane magnetic field $H_\mathrm{ext}$ is applied and rotated within the film plane: the angle between $H_\mathrm{ext}$ and the current flow is defined as $\varphi$.
The magnetic easy axis of the CoFeB layer points along the film plane.
The in-plane anisotropy of CoFeB is negligibly small compared to $H_\mathrm{ext}$: we assume the magnetization is parallel to the direction of $H_\mathrm{ext}$.
The $\varphi$ dependence of the first harmonic Hall voltage $V_{1\omega}$ and the second harmonic Hall voltage $V_{2\omega}$ were simultaneously measured.
The first and second harmonic Hall resistances, $R_{1\omega}$ and $R_{2\omega}$, are obtained by dividing the harmonic voltages with $I_0$, i.e., $R_{1\omega(2\omega)}=V_{1\omega(2\omega)}/I_0$.
$R_{1\omega}$ and $R_{2\omega}$ can be expressed as\cite{avci2014prb,roschewsky2019prb,chi2020sciadv}:
\begin{equation}
\begin{aligned}
\label{eq:R1w}
R_{1\omega}&=R_\mathrm{PHE}\sin{2\varphi}+{\zeta}R_\mathrm{AHE}\cos{\varphi},
\end{aligned}
\end{equation}
\begin{equation}
\begin{aligned}
\label{eq:R2w}
R_{2\omega}&=\frac{1}{2} \left(R_\mathrm{AHE}\dfrac{h_\mathrm{DL}}{H_\mathrm{ext}+H_\mathrm{k}}+R_\mathrm{ONE}H_\mathrm{ext}+R_\mathrm{const}\right)\cos\varphi\\
&\ \ \ - R_\mathrm{PHE}\dfrac{h_\mathrm{FL}+h_\mathrm{Oe}}{H_\mathrm{ext}}\cos2\varphi\cos\varphi\\
& \equiv A \cos\varphi+B \cos2\varphi\cos\varphi.
\end{aligned}
\end{equation}
$R_{\rm{AHE}}$ is the anomalous Hall resistance, $R_{\rm{PHE}}$ is the planar Hall resistance, $h_\mathrm{DL}$ is the dampinglike (DL) spin-orbit effective field, $h_\mathrm{FL}$ is the fieldlike (FL) spin-orbit effective field, $h_{\rm{Oe}}$ is the Oersted field, and $H_{\rm{k}}$ is the magnetic anisotropy field.
$R_{\rm{const}}$ and $R_\textrm{ONE}H_\textrm{ext}$ represent thermoelectric contributions to $R_{2\omega}$ that are independent of and linear with $H_\mathrm{ext}$, respectively.
The former is due to the anomalous Nernst effect and/or the combined action of spin Seebeck effect and inverse spin Hall effect. The latter originates from the ordinary Nernst effect (ONE)\cite{avci2014prb,roschewsky2019prb,chi2020sciadv}.
$R_{1\omega}$ is used to extract $R_{\rm{PHE}}$. $R_{\rm{AHE}}$ and $H_{\rm{k}}$ are obtained from separate measurements of the out-of-plane field dependence of the Hall resistance.
$h_{\rm{Oe}}$ is calculated using Ampere's law\cite{hayashi2014prb}.
Data are fitted with Eq.~(\ref{eq:R2w}) to extract $h_\mathrm{DL}$, $h_\mathrm{FL}$, $R_{\rm{const}}$ and $R_\textrm{ONE}H_\textrm{ext}$.

Representative $\varphi$-dependence of $R_{2\omega}$ for Bi$_{0.89}$Te$_{0.11}$/CoFeB and Bi$_{0.92}$Sn$_{0.08}$/CoFeB bilayers measured at different $H_{\rm{ext}}$ are plotted in Fig.~\ref{fig:DL-SOT}(a) and (b). Black solid curves are the best fits using Eq.~\ref{eq:R2w}. $H_\mathrm{ext}$ dependence of the $\cos\varphi$ component ($A$) for these two structures are compared in Fig.~\ref{fig:DL-SOT}(c) and (d). The best fits are shown by the black lines. Other colored lines represent decomposition of various contributions. For both samples,  contribution from the DL spin-orbit torque (red curves) dominates in small $H_{\rm{ext}}$ regimes, whereas that of ONE (green lines) mainly contributes at larger fields.
The sign of the ONE is opposite for Bi$_{1-x}$Te$_x$/CoFeB and Bi$_{1-y}$Sn$_y$/CoFeB, which may be related to the change of the majority carriers in these two series of doped Bi. In contrast, the DL spin-orbit torque  maintains its sign upon traversing the Dirac point. 
Note that due to the distinct $H_{\rm{ext}}$ dependence, signal contamination from the ONE can be safely ruled out\cite{yue2018prl}.

\begin{figure}
\begin{center}
  \includegraphics[width=1\columnwidth]{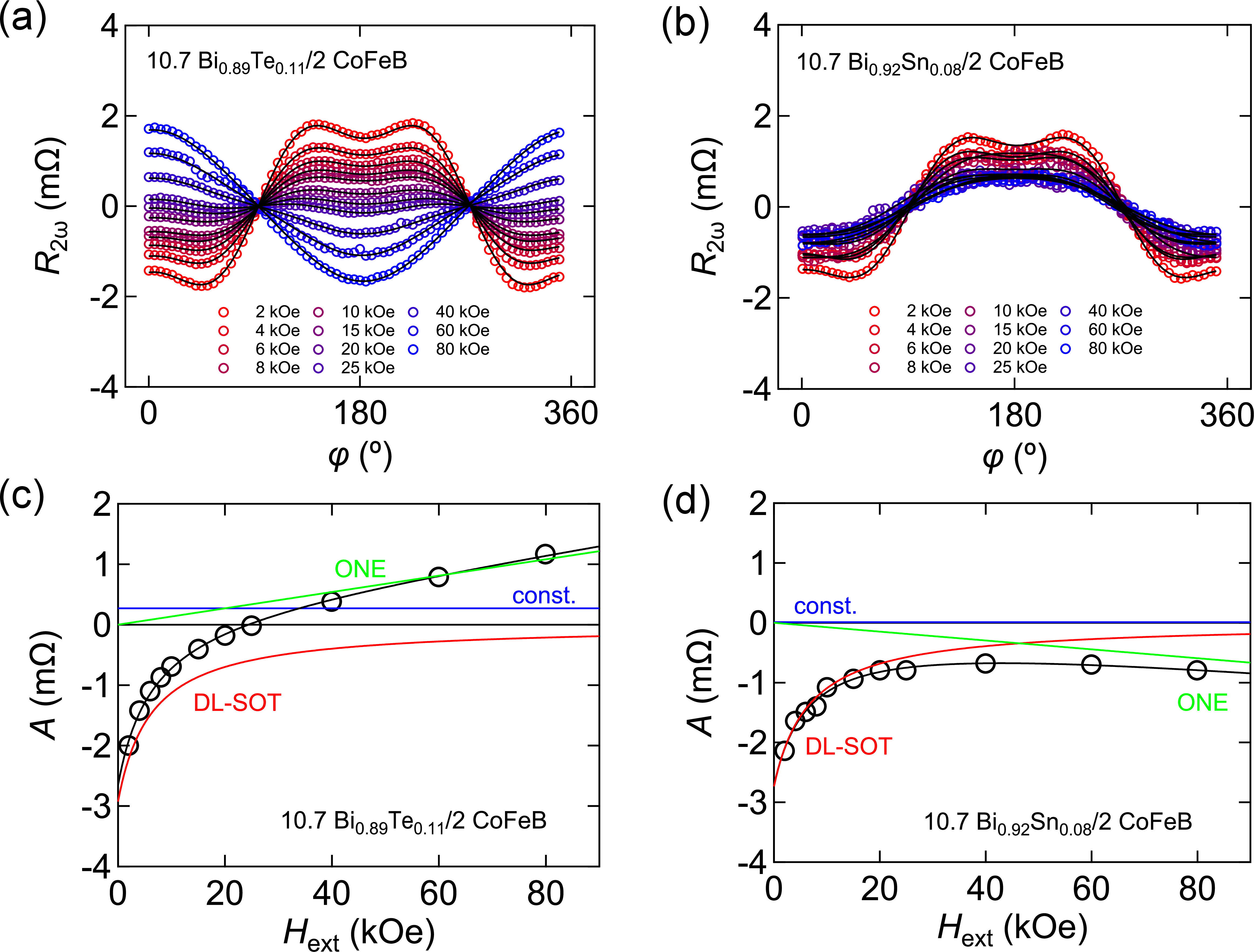}
  \caption{(a,b) Azimuthal angle, $\varphi$ dependence of second harmonic Hall resistance $R_{2\omega}$ for 10.7 Bi$_{0.89}$Te$_{0.11}$/2 CoFeB (a) and 10.7 Bi$_{0.92}$Sn$_{0.08}$/2 CoFeB (b) bilayers measured at different external magnetic fields $H_{\rm{ext}}$. (c,d) $\cos\varphi$ component ($A$) of $R_{2\omega}$ against $H_{\rm{ext}}$ for 10.7 Bi$_{0.89}$Te$_{0.11}$/2 CoFeB (c) and 10.7 Bi$_{0.92}$Sn$_{0.08}$/2 CoFeB (d), respectively. The colored curves show contributions from different effects and the black curves are the sum of all. All data were obtained at 300 K.}
  \label{fig:DL-SOT}
\end{center}
\end{figure}

The FL spin-orbit torque contribution is estimated using Eq.~(\ref{eq:R2w}).
The $\cos2\varphi\cos\varphi$ component ($B$) of $R_{2\omega}$ codes its information.
The FL spin torque efficiency ($\xi_\mathrm{FL}$) of the heterostructures, estimated using Eq.~\ref{eq:xi_DL} in the main text, is plotted as a function of the corresponding carrier density $n$ in Fig.~\ref{fig:FL-SOT}.
We find the sign of $\xi_\mathrm{FL}$ is positive (i.e. opposing the Oersted field) for all samples studied.
The sign agrees with that of Pt/Co/AlO$_x$\cite{garello2013nnano} and BiSb/CoFeB\cite{chi2020sciadv}, and is opposite to that of Bi$_2$Se$_3$/Py\cite{mellnik2014nature}.
$\xi_\mathrm{FL}$ exhibits a maximum at small $|n|$ and decreases with increasing $|n|$.
Note that the magnitude of $\xi_\mathrm{FL}$ is smaller than $\xi_\mathrm{DL}$ for all samples.
\begin{figure}[h!]
\begin{center}
  \includegraphics[width=0.5\columnwidth]{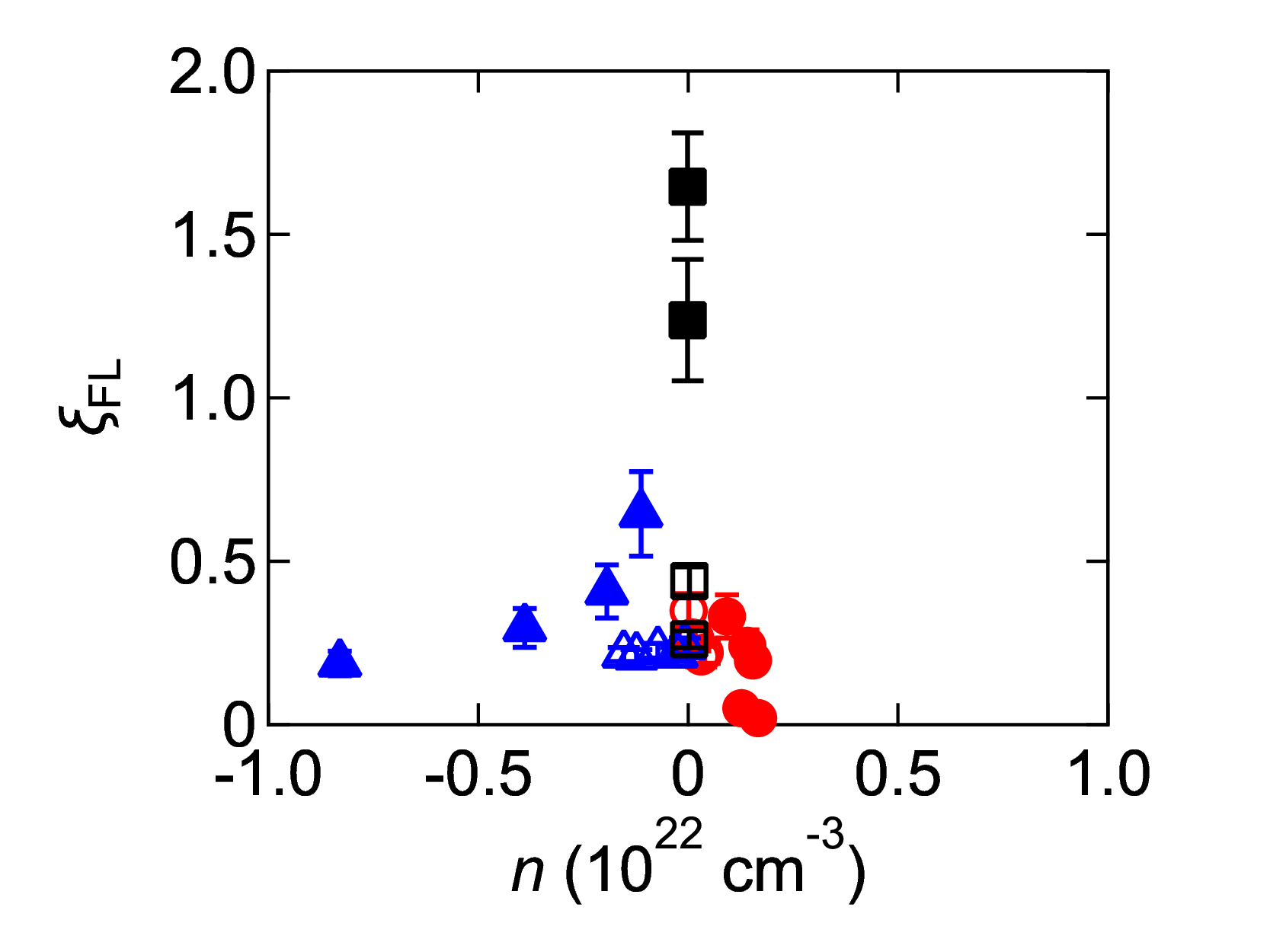}
  \caption{Carrier concentration $n$ dependence of the the FL spin torque efficiency ($\xi_\mathrm{FL}$) of heterostructures with the 0.5 nm Ta seed layer.}
  \label{fig:FL-SOT}
\end{center}
\end{figure}


\subsection{\label{sec:supps:st-fmr}Spin-torque ferromagnetic resonance of Bi/CoFeB}
\label{sec:ST-FMR}
We carried out spin-torque ferromagnetic resonance (ST-FMR) measurement\cite{liu2011prl} as an independent verification of the large spin torque efficiency in Bi/CoFeB.
We focus on pristine Bi grown directly on Si/SiO$_2$ substrate without the Ta seed layer.
The film structure is sub./11 Bi/0-10 CoFeB wedge/2 MgO/1 Ta (thicknesses in nanometer), i.e., a wedge film with the thickness of CoFeB ($t_\textrm{CoFeB}$) varying from 0 to 10 nm.
The wedge film was fabricated using standard optical lithography and Ar ion milling into microstrips with a nominal length of $L = 40$ $\upmu$m and a width of $w = 10$ $\upmu$m.
An amplitude modulated (AM) radio-frequency (RF) microwave power of 17 dBm was applied through a ground-signal-ground coplanar waveguide while an in-plane magnetic field $H_\mathrm{ext}$ was swept along the azimuthal angle $\varphi = 45$ or 225 degrees with respect to the long axis of the microstrip.
The mixing voltage $V_\mathrm{mix}$ was measured using a lock-in amplifier being synchronized to the modulation frequency of 9997 Hz.

The resistivity of pristine Bi ($\rho_\mathrm{Bi}=\SI{484}{\textit{\micro}\ohm\cdot\centi\meter}$) and CoFeB ($\rho_\mathrm{CoFeB}=\SI{158}{\textit{\micro}\ohm\cdot\centi\meter}$) are extracted from the $t_\textrm{CoFeB}$-dependence of the sheet conductance for the wedge film after microfabrication. 
We assume a magnetic dead layer thickness $t_\mathrm{D}\approx0.3\pm0.2$ nm \cite{chi2020sciadv} and define the effective CoFeB thickness as $t_\mathrm{eff} \equiv t_\mathrm{CoFeB}-t_\mathrm{D}$.

The ST-FMR spectra at various RF excitation frequencies $f$ of a 11 Bi/7.4 CoFeB device are plotted in Fig.~\ref{fig:ST-FMR}(a).
The resonance lineshape can be decomposed into the sum of a symmetric and an antisymmetric Lorentzian, which is proportional to the DL spin-orbit effective field and the RF effective field, respectively. The apparent DL spin torque efficiency $\xi_\mathrm{FMR}$ is extracted based on the lineshape analysis\cite{liu2011prl}:
\begin{equation}
\centering
\label{eq:xi_FMR}
\xi_\mathrm{FMR}=\frac{S}{A}\frac{e \mu_0 M_\mathrm{s} t_\mathrm{eff} t_\mathrm{Bi}}{\hbar}\sqrt{1+\frac{4\pi M_\mathrm{eff}}{H_\mathrm{ext}}}.
\end{equation}

A typical decomposition of the resonance spectrum at $f = \SI{10}{\giga\hertz}$ is illustrated in Fig.~\ref{fig:ST-FMR}(b). The strong symmetric component (in red) relative to the antisymmetric counterpart (in blue) indicates the remarkable spin torque efficiency in Bi/CoFeB bilayer. We further verified that $\xi_\mathrm{FMR}$ is nearly independent of $f$, within the experimental uncertainty, as shown in Fig.~\ref{fig:ST-FMR}(c).

Next, we have studied the CoFeB thickness dependence of $\xi_\mathrm{FMR}$ in order to take into account contribution of the FL spin-orbit effective field to the total RF field acting on the ferromagnetic CoFeB layer. Assuming $\xi_\mathrm{FL}$ is independent of the CoFeB thickness for reasonably thick $t_\mathrm{eff}$, $\xi_\mathrm{DL}$ and $\xi_\mathrm{FL}$ can be extracted\cite{pai2015prb} from the 1/$\xi_\mathrm{FMR}$ vs 1/$t_\mathrm{eff}$ plot in Fig.~\ref{fig:ST-FMR}(d) via the following relation:
\begin{equation}
\centering
\label{eq:xi_FMR2}
\frac{1}{\xi_\mathrm{FMR}}=\frac{1}{\xi_\mathrm{DL}}\left(1+\frac{\hbar}{e}\frac{\xi_\mathrm{FL}}{\mu_0M_\mathrm{s} t_\mathrm{eff}t_\mathrm{Bi}}\right)
\end{equation}
We obtain $\xi_\mathrm{DL} \approx +0.59$ and $\xi_\mathrm{FL} \approx +0.007$ (opposing the Oersted field), respectively. The spin Hall conductivity is estimated to be $\sim1220$ ($\frac{\hbar}{2e}$ $\Omega^{-1}\cdot\mathrm{cm}^{-1}$), which is in good agreement with that from the harmonic Hall measurements.
The size of $\xi_\mathrm{FL} \approx 0.007$ estimated here is an order of magnitude smaller than that estimated using the harmonic Hall measurements. The origin of this discrepancy is currently unknown and requires further investigation.

\begin{figure}
\begin{center}
  \includegraphics[width=1\columnwidth]{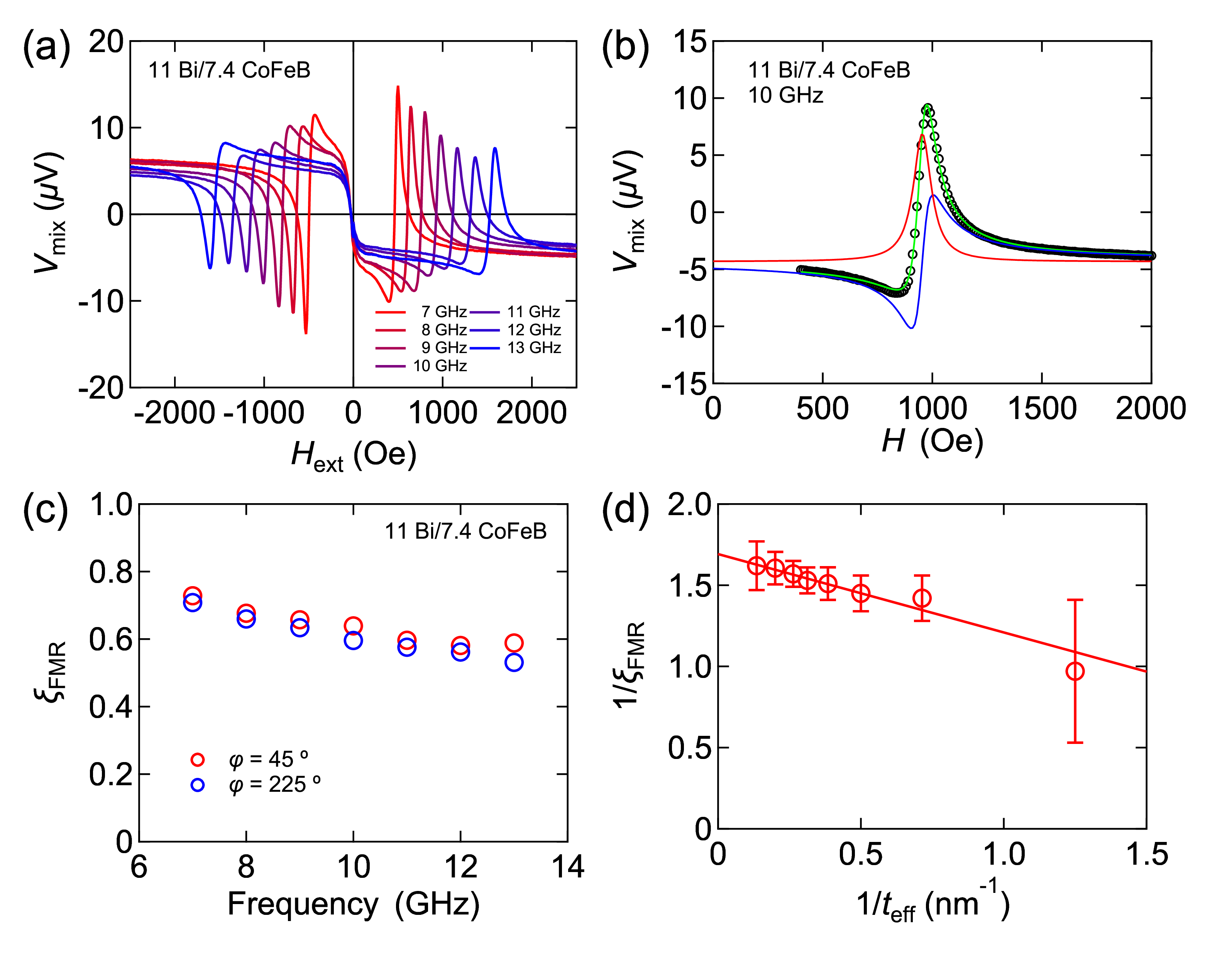}
  \caption{(a) Spin-torque ferromagnetic resonance (ST-FMR) spectra for a representative 11 Bi/7.4 CoFeB bilayer device at various RF excitation frequencies $f$. (b) A typical decomposition of the resonance spectrum for 11 Bi/7.4 CoFeB bilayer device measured at $f=$\SI{10}{\giga\hertz}. (c) $f$ dependence of apparent spin torque efficiency $\xi_\mathrm{FMR}$ with in-plane magnetic field swept along the azimuthal angle $\varphi = \SI{45}{\degree}$ or $\varphi = \SI{225}{\degree}$. (d) $1/\xi_\mathrm{FMR}$ plotted as a function of the inverse of effective CoFeB thickness $t_\mathrm{eff}$.}
  \label{fig:ST-FMR}
\end{center}
\end{figure}
\clearpage

\bibliography{ref_041822}

\begin{thebibliography}{56}%
\makeatletter
\providecommand \@ifxundefined [1]{%
 \@ifx{#1\undefined}
}%
\providecommand \@ifnum [1]{%
 \ifnum #1\expandafter \@firstoftwo
 \else \expandafter \@secondoftwo
 \fi
}%
\providecommand \@ifx [1]{%
 \ifx #1\expandafter \@firstoftwo
 \else \expandafter \@secondoftwo
 \fi
}%
\providecommand \natexlab [1]{#1}%
\providecommand \enquote  [1]{``#1''}%
\providecommand \bibnamefont  [1]{#1}%
\providecommand \bibfnamefont [1]{#1}%
\providecommand \citenamefont [1]{#1}%
\providecommand \href@noop [0]{\@secondoftwo}%
\providecommand \href [0]{\begingroup \@sanitize@url \@href}%
\providecommand \@href[1]{\@@startlink{#1}\@@href}%
\providecommand \@@href[1]{\endgroup#1\@@endlink}%
\providecommand \@sanitize@url [0]{\catcode `\\12\catcode `\$12\catcode
  `\&12\catcode `\#12\catcode `\^12\catcode `\_12\catcode `\%12\relax}%
\providecommand \@@startlink[1]{}%
\providecommand \@@endlink[0]{}%
\providecommand \url  [0]{\begingroup\@sanitize@url \@url }%
\providecommand \@url [1]{\endgroup\@href {#1}{\urlprefix }}%
\providecommand \urlprefix  [0]{URL }%
\providecommand \Eprint [0]{\href }%
\providecommand \doibase [0]{http://dx.doi.org/}%
\providecommand \selectlanguage [0]{\@gobble}%
\providecommand \bibinfo  [0]{\@secondoftwo}%
\providecommand \bibfield  [0]{\@secondoftwo}%
\providecommand \translation [1]{[#1]}%
\providecommand \BibitemOpen [0]{}%
\providecommand \bibitemStop [0]{}%
\providecommand \bibitemNoStop [0]{.\EOS\space}%
\providecommand \EOS [0]{\spacefactor3000\relax}%
\providecommand \BibitemShut  [1]{\csname bibitem#1\endcsname}%
\let\auto@bib@innerbib\@empty
\bibitem [{\citenamefont {Dyakonov}\ and\ \citenamefont
  {Perel}(1971)}]{dyakonov1971jetp}%
  \BibitemOpen
  \bibfield  {author} {\bibinfo {author} {\bibfnamefont {M.~I.}\ \bibnamefont
  {Dyakonov}}\ and\ \bibinfo {author} {\bibfnamefont {V.~I.}\ \bibnamefont
  {Perel}},\ }\href@noop {} {\bibfield  {journal} {\bibinfo  {journal} {JETP
  Lett.}\ }\textbf {\bibinfo {volume} {13}},\ \bibinfo {pages} {467} (\bibinfo
  {year} {1971})}\BibitemShut {NoStop}%
\bibitem [{\citenamefont {Hirsch}(1999)}]{hirsch1999prl}%
  \BibitemOpen
  \bibfield  {author} {\bibinfo {author} {\bibfnamefont {J.~E.}\ \bibnamefont
  {Hirsch}},\ }\href@noop {} {\bibfield  {journal} {\bibinfo  {journal} {Phys.
  Rev. Lett.}\ }\textbf {\bibinfo {volume} {83}},\ \bibinfo {pages} {1834}
  (\bibinfo {year} {1999})}\BibitemShut {NoStop}%
\bibitem [{\citenamefont {Zhang}(2000)}]{zhang2000prl}%
  \BibitemOpen
  \bibfield  {author} {\bibinfo {author} {\bibfnamefont {S.~F.}\ \bibnamefont
  {Zhang}},\ }\href@noop {} {\bibfield  {journal} {\bibinfo  {journal} {Phys.
  Rev. Lett.}\ }\textbf {\bibinfo {volume} {85}},\ \bibinfo {pages} {393}
  (\bibinfo {year} {2000})}\BibitemShut {NoStop}%
\bibitem [{\citenamefont {Murakami}\ \emph {et~al.}(2003)\citenamefont
  {Murakami}, \citenamefont {Nagaosa},\ and\ \citenamefont
  {Zhang}}]{murakami2003science}%
  \BibitemOpen
  \bibfield  {author} {\bibinfo {author} {\bibfnamefont {S.}~\bibnamefont
  {Murakami}}, \bibinfo {author} {\bibfnamefont {N.}~\bibnamefont {Nagaosa}}, \
  and\ \bibinfo {author} {\bibfnamefont {S.~C.}\ \bibnamefont {Zhang}},\
  }\href@noop {} {\bibfield  {journal} {\bibinfo  {journal} {Science}\ }\textbf
  {\bibinfo {volume} {301}},\ \bibinfo {pages} {1348} (\bibinfo {year}
  {2003})}\BibitemShut {NoStop}%
\bibitem [{\citenamefont {Kato}\ \emph {et~al.}(2004)\citenamefont {Kato},
  \citenamefont {Myers}, \citenamefont {Gossard},\ and\ \citenamefont
  {Awschalom}}]{kato2004prl}%
  \BibitemOpen
  \bibfield  {author} {\bibinfo {author} {\bibfnamefont {Y.~K.}\ \bibnamefont
  {Kato}}, \bibinfo {author} {\bibfnamefont {R.~C.}\ \bibnamefont {Myers}},
  \bibinfo {author} {\bibfnamefont {A.~C.}\ \bibnamefont {Gossard}}, \ and\
  \bibinfo {author} {\bibfnamefont {D.~D.}\ \bibnamefont {Awschalom}},\
  }\href@noop {} {\bibfield  {journal} {\bibinfo  {journal} {Phys. Rev. Lett.}\
  }\textbf {\bibinfo {volume} {93}},\ \bibinfo {pages} {176601} (\bibinfo
  {year} {2004})}\BibitemShut {NoStop}%
\bibitem [{\citenamefont {Wunderlich}\ \emph {et~al.}(2005)\citenamefont
  {Wunderlich}, \citenamefont {Kaestner}, \citenamefont {Sinova},\ and\
  \citenamefont {Jungwirth}}]{wunderlich2005prl}%
  \BibitemOpen
  \bibfield  {author} {\bibinfo {author} {\bibfnamefont {J.}~\bibnamefont
  {Wunderlich}}, \bibinfo {author} {\bibfnamefont {B.}~\bibnamefont
  {Kaestner}}, \bibinfo {author} {\bibfnamefont {J.}~\bibnamefont {Sinova}}, \
  and\ \bibinfo {author} {\bibfnamefont {T.}~\bibnamefont {Jungwirth}},\
  }\href@noop {} {\bibfield  {journal} {\bibinfo  {journal} {Phys. Rev. Lett.}\
  }\textbf {\bibinfo {volume} {94}},\ \bibinfo {pages} {047204} (\bibinfo
  {year} {2005})}\BibitemShut {NoStop}%
\bibitem [{\citenamefont {Valenzuela}\ and\ \citenamefont
  {Tinkham}(2006)}]{valenzuela2006nature}%
  \BibitemOpen
  \bibfield  {author} {\bibinfo {author} {\bibfnamefont {S.~O.}\ \bibnamefont
  {Valenzuela}}\ and\ \bibinfo {author} {\bibfnamefont {M.}~\bibnamefont
  {Tinkham}},\ }\href@noop {} {\bibfield  {journal} {\bibinfo  {journal}
  {Nature}\ }\textbf {\bibinfo {volume} {442}},\ \bibinfo {pages} {176}
  (\bibinfo {year} {2006})}\BibitemShut {NoStop}%
\bibitem [{\citenamefont {Kimura}\ \emph {et~al.}(2007)\citenamefont {Kimura},
  \citenamefont {Otani}, \citenamefont {Sato}, \citenamefont {Takahashi},\ and\
  \citenamefont {Maekawa}}]{kimura2007prl}%
  \BibitemOpen
  \bibfield  {author} {\bibinfo {author} {\bibfnamefont {T.}~\bibnamefont
  {Kimura}}, \bibinfo {author} {\bibfnamefont {Y.}~\bibnamefont {Otani}},
  \bibinfo {author} {\bibfnamefont {T.}~\bibnamefont {Sato}}, \bibinfo {author}
  {\bibfnamefont {S.}~\bibnamefont {Takahashi}}, \ and\ \bibinfo {author}
  {\bibfnamefont {S.}~\bibnamefont {Maekawa}},\ }\href@noop {} {\bibfield
  {journal} {\bibinfo  {journal} {Phys. Rev. Lett.}\ }\textbf {\bibinfo
  {volume} {98}},\ \bibinfo {pages} {156601} (\bibinfo {year}
  {2007})}\BibitemShut {NoStop}%
\bibitem [{\citenamefont {Liu}\ \emph {et~al.}(2012)\citenamefont {Liu},
  \citenamefont {Pai}, \citenamefont {Li}, \citenamefont {Tseng}, \citenamefont
  {Ralph},\ and\ \citenamefont {Buhrman}}]{liu2012science}%
  \BibitemOpen
  \bibfield  {author} {\bibinfo {author} {\bibfnamefont {L.}~\bibnamefont
  {Liu}}, \bibinfo {author} {\bibfnamefont {C.-F.}\ \bibnamefont {Pai}},
  \bibinfo {author} {\bibfnamefont {Y.}~\bibnamefont {Li}}, \bibinfo {author}
  {\bibfnamefont {H.~W.}\ \bibnamefont {Tseng}}, \bibinfo {author}
  {\bibfnamefont {D.~C.}\ \bibnamefont {Ralph}}, \ and\ \bibinfo {author}
  {\bibfnamefont {R.~A.}\ \bibnamefont {Buhrman}},\ }\href@noop {} {\bibfield
  {journal} {\bibinfo  {journal} {Science}\ }\textbf {\bibinfo {volume}
  {336}},\ \bibinfo {pages} {555} (\bibinfo {year} {2012})}\BibitemShut
  {NoStop}%
\bibitem [{\citenamefont {Nagaosa}\ \emph {et~al.}(2010)\citenamefont
  {Nagaosa}, \citenamefont {Sinova}, \citenamefont {Onoda}, \citenamefont
  {MacDonald},\ and\ \citenamefont {Ong}}]{nagaosa2010rmp}%
  \BibitemOpen
  \bibfield  {author} {\bibinfo {author} {\bibfnamefont {N.}~\bibnamefont
  {Nagaosa}}, \bibinfo {author} {\bibfnamefont {J.}~\bibnamefont {Sinova}},
  \bibinfo {author} {\bibfnamefont {S.}~\bibnamefont {Onoda}}, \bibinfo
  {author} {\bibfnamefont {A.~H.}\ \bibnamefont {MacDonald}}, \ and\ \bibinfo
  {author} {\bibfnamefont {N.~P.}\ \bibnamefont {Ong}},\ }\href@noop {}
  {\bibfield  {journal} {\bibinfo  {journal} {Rev. Mod. Phys.}\ }\textbf
  {\bibinfo {volume} {82}},\ \bibinfo {pages} {1539} (\bibinfo {year}
  {2010})}\BibitemShut {NoStop}%
\bibitem [{\citenamefont {Karplus}\ and\ \citenamefont
  {Luttinger}(1954)}]{karplus1954pr}%
  \BibitemOpen
  \bibfield  {author} {\bibinfo {author} {\bibfnamefont {R.}~\bibnamefont
  {Karplus}}\ and\ \bibinfo {author} {\bibfnamefont {J.~M.}\ \bibnamefont
  {Luttinger}},\ }\href@noop {} {\bibfield  {journal} {\bibinfo  {journal}
  {Physical Review}\ }\textbf {\bibinfo {volume} {95}},\ \bibinfo {pages}
  {1154} (\bibinfo {year} {1954})}\BibitemShut {NoStop}%
\bibitem [{\citenamefont {Murakami}\ \emph {et~al.}(2004)\citenamefont
  {Murakami}, \citenamefont {Nagaosa},\ and\ \citenamefont
  {Zhang}}]{murakami2004prb}%
  \BibitemOpen
  \bibfield  {author} {\bibinfo {author} {\bibfnamefont {S.}~\bibnamefont
  {Murakami}}, \bibinfo {author} {\bibfnamefont {N.}~\bibnamefont {Nagaosa}}, \
  and\ \bibinfo {author} {\bibfnamefont {S.~C.}\ \bibnamefont {Zhang}},\
  }\href@noop {} {\bibfield  {journal} {\bibinfo  {journal} {Phys. Rev. B}\
  }\textbf {\bibinfo {volume} {69}},\ \bibinfo {pages} {235206} (\bibinfo
  {year} {2004})}\BibitemShut {NoStop}%
\bibitem [{\citenamefont {Shi}\ \emph {et~al.}(2006)\citenamefont {Shi},
  \citenamefont {Zhang}, \citenamefont {Xiao},\ and\ \citenamefont
  {Niu}}]{shi2006prl}%
  \BibitemOpen
  \bibfield  {author} {\bibinfo {author} {\bibfnamefont {J.~R.}\ \bibnamefont
  {Shi}}, \bibinfo {author} {\bibfnamefont {P.}~\bibnamefont {Zhang}}, \bibinfo
  {author} {\bibfnamefont {D.}~\bibnamefont {Xiao}}, \ and\ \bibinfo {author}
  {\bibfnamefont {Q.}~\bibnamefont {Niu}},\ }\href@noop {} {\bibfield
  {journal} {\bibinfo  {journal} {Phys. Rev. Lett.}\ }\textbf {\bibinfo
  {volume} {96}},\ \bibinfo {pages} {076604} (\bibinfo {year}
  {2006})}\BibitemShut {NoStop}%
\bibitem [{\citenamefont {Gradhand}\ \emph {et~al.}(2012)\citenamefont
  {Gradhand}, \citenamefont {Fedorov}, \citenamefont {Pientka}, \citenamefont
  {Zahn}, \citenamefont {Mertig},\ and\ \citenamefont
  {Gyorffy}}]{gradhand2012jpcm}%
  \BibitemOpen
  \bibfield  {author} {\bibinfo {author} {\bibfnamefont {M.}~\bibnamefont
  {Gradhand}}, \bibinfo {author} {\bibfnamefont {D.~V.}\ \bibnamefont
  {Fedorov}}, \bibinfo {author} {\bibfnamefont {F.}~\bibnamefont {Pientka}},
  \bibinfo {author} {\bibfnamefont {P.}~\bibnamefont {Zahn}}, \bibinfo {author}
  {\bibfnamefont {I.}~\bibnamefont {Mertig}}, \ and\ \bibinfo {author}
  {\bibfnamefont {B.~L.}\ \bibnamefont {Gyorffy}},\ }\href@noop {} {\bibfield
  {journal} {\bibinfo  {journal} {J. Phys.-Condes. Matter}\ }\textbf {\bibinfo
  {volume} {24}},\ \bibinfo {pages} {213202} (\bibinfo {year}
  {2012})}\BibitemShut {NoStop}%
\bibitem [{\citenamefont {Sinova}\ \emph {et~al.}(2004)\citenamefont {Sinova},
  \citenamefont {Culcer}, \citenamefont {Niu}, \citenamefont {Sinitsyn},
  \citenamefont {Jungwirth},\ and\ \citenamefont {MacDonald}}]{sinova2004prl}%
  \BibitemOpen
  \bibfield  {author} {\bibinfo {author} {\bibfnamefont {J.}~\bibnamefont
  {Sinova}}, \bibinfo {author} {\bibfnamefont {D.}~\bibnamefont {Culcer}},
  \bibinfo {author} {\bibfnamefont {Q.}~\bibnamefont {Niu}}, \bibinfo {author}
  {\bibfnamefont {N.~A.}\ \bibnamefont {Sinitsyn}}, \bibinfo {author}
  {\bibfnamefont {T.}~\bibnamefont {Jungwirth}}, \ and\ \bibinfo {author}
  {\bibfnamefont {A.~H.}\ \bibnamefont {MacDonald}},\ }\href@noop {} {\bibfield
   {journal} {\bibinfo  {journal} {Phys. Rev. Lett.}\ }\textbf {\bibinfo
  {volume} {92}},\ \bibinfo {pages} {126603} (\bibinfo {year}
  {2004})}\BibitemShut {NoStop}%
\bibitem [{\citenamefont {Guo}\ \emph {et~al.}(2005)\citenamefont {Guo},
  \citenamefont {Yao},\ and\ \citenamefont {Niu}}]{guo2005prl}%
  \BibitemOpen
  \bibfield  {author} {\bibinfo {author} {\bibfnamefont {G.~Y.}\ \bibnamefont
  {Guo}}, \bibinfo {author} {\bibfnamefont {Y.}~\bibnamefont {Yao}}, \ and\
  \bibinfo {author} {\bibfnamefont {Q.}~\bibnamefont {Niu}},\ }\href@noop {}
  {\bibfield  {journal} {\bibinfo  {journal} {Phys. Rev. Lett.}\ }\textbf
  {\bibinfo {volume} {94}},\ \bibinfo {pages} {226601} (\bibinfo {year}
  {2005})}\BibitemShut {NoStop}%
\bibitem [{\citenamefont {Crepieux}\ and\ \citenamefont
  {Bruno}(2001)}]{crepieux2001prb}%
  \BibitemOpen
  \bibfield  {author} {\bibinfo {author} {\bibfnamefont {A.}~\bibnamefont
  {Crepieux}}\ and\ \bibinfo {author} {\bibfnamefont {P.}~\bibnamefont
  {Bruno}},\ }\href@noop {} {\bibfield  {journal} {\bibinfo  {journal} {Phys.
  Rev. B}\ }\textbf {\bibinfo {volume} {64}},\ \bibinfo {pages} {014416}
  (\bibinfo {year} {2001})}\BibitemShut {NoStop}%
\bibitem [{\citenamefont {Vernes}\ \emph {et~al.}(2007)\citenamefont {Vernes},
  \citenamefont {Gyorffy},\ and\ \citenamefont {Weinberger}}]{vernes2007prb}%
  \BibitemOpen
  \bibfield  {author} {\bibinfo {author} {\bibfnamefont {A.}~\bibnamefont
  {Vernes}}, \bibinfo {author} {\bibfnamefont {B.~L.}\ \bibnamefont {Gyorffy}},
  \ and\ \bibinfo {author} {\bibfnamefont {P.}~\bibnamefont {Weinberger}},\
  }\href@noop {} {\bibfield  {journal} {\bibinfo  {journal} {Phys. Rev. B}\
  }\textbf {\bibinfo {volume} {76}},\ \bibinfo {pages} {012408} (\bibinfo
  {year} {2007})}\BibitemShut {NoStop}%
\bibitem [{\citenamefont {Lowitzer}\ \emph {et~al.}(2010)\citenamefont
  {Lowitzer}, \citenamefont {Kodderitzsch},\ and\ \citenamefont
  {Ebert}}]{lowitzer2010prb}%
  \BibitemOpen
  \bibfield  {author} {\bibinfo {author} {\bibfnamefont {S.}~\bibnamefont
  {Lowitzer}}, \bibinfo {author} {\bibfnamefont {D.}~\bibnamefont
  {Kodderitzsch}}, \ and\ \bibinfo {author} {\bibfnamefont {H.}~\bibnamefont
  {Ebert}},\ }\href@noop {} {\bibfield  {journal} {\bibinfo  {journal} {Phys.
  Rev. B}\ }\textbf {\bibinfo {volume} {82}},\ \bibinfo {pages} {140402}
  (\bibinfo {year} {2010})}\BibitemShut {NoStop}%
\bibitem [{\citenamefont {Fuseya}\ \emph {et~al.}(2012)\citenamefont {Fuseya},
  \citenamefont {Ogata},\ and\ \citenamefont {Fukuyama}}]{fuseya2012jpsj1}%
  \BibitemOpen
  \bibfield  {author} {\bibinfo {author} {\bibfnamefont {Y.}~\bibnamefont
  {Fuseya}}, \bibinfo {author} {\bibfnamefont {M.}~\bibnamefont {Ogata}}, \
  and\ \bibinfo {author} {\bibfnamefont {H.}~\bibnamefont {Fukuyama}},\
  }\href@noop {} {\bibfield  {journal} {\bibinfo  {journal} {J. Phys. Soc.
  Jpn.}\ }\textbf {\bibinfo {volume} {81}},\ \bibinfo {pages} {093704}
  (\bibinfo {year} {2012})}\BibitemShut {NoStop}%
\bibitem [{\citenamefont {Sun}\ \emph {et~al.}(2016)\citenamefont {Sun},
  \citenamefont {Zhang}, \citenamefont {Felser},\ and\ \citenamefont
  {Yan}}]{sun2017prl}%
  \BibitemOpen
  \bibfield  {author} {\bibinfo {author} {\bibfnamefont {Y.}~\bibnamefont
  {Sun}}, \bibinfo {author} {\bibfnamefont {Y.}~\bibnamefont {Zhang}}, \bibinfo
  {author} {\bibfnamefont {C.}~\bibnamefont {Felser}}, \ and\ \bibinfo {author}
  {\bibfnamefont {B.~H.}\ \bibnamefont {Yan}},\ }\href@noop {} {\bibfield
  {journal} {\bibinfo  {journal} {Phys. Rev. Lett.}\ }\textbf {\bibinfo
  {volume} {117}},\ \bibinfo {pages} {146403} (\bibinfo {year}
  {2016})}\BibitemShut {NoStop}%
\bibitem [{\citenamefont {Fukazawa}\ \emph {et~al.}(2017)\citenamefont
  {Fukazawa}, \citenamefont {Kohno},\ and\ \citenamefont
  {Fujimoto}}]{fukazawa2017jpsj}%
  \BibitemOpen
  \bibfield  {author} {\bibinfo {author} {\bibfnamefont {T.}~\bibnamefont
  {Fukazawa}}, \bibinfo {author} {\bibfnamefont {H.}~\bibnamefont {Kohno}}, \
  and\ \bibinfo {author} {\bibfnamefont {J.}~\bibnamefont {Fujimoto}},\
  }\href@noop {} {\bibfield  {journal} {\bibinfo  {journal} {J. Phys. Soc.
  Jpn.}\ }\textbf {\bibinfo {volume} {86}},\ \bibinfo {pages} {094704}
  (\bibinfo {year} {2017})}\BibitemShut {NoStop}%
\bibitem [{\citenamefont {Armitage}\ \emph {et~al.}(2018)\citenamefont
  {Armitage}, \citenamefont {Mele},\ and\ \citenamefont
  {Vishwanath}}]{armitage2018rmp}%
  \BibitemOpen
  \bibfield  {author} {\bibinfo {author} {\bibfnamefont {N.~P.}\ \bibnamefont
  {Armitage}}, \bibinfo {author} {\bibfnamefont {E.~J.}\ \bibnamefont {Mele}},
  \ and\ \bibinfo {author} {\bibfnamefont {A.}~\bibnamefont {Vishwanath}},\
  }\href@noop {} {\bibfield  {journal} {\bibinfo  {journal} {Rev. Mod. Phys.}\
  }\textbf {\bibinfo {volume} {90}},\ \bibinfo {pages} {015001} (\bibinfo
  {year} {2018})}\BibitemShut {NoStop}%
\bibitem [{\citenamefont {Liu}\ and\ \citenamefont {Allen}(1995)}]{liu1995prb}%
  \BibitemOpen
  \bibfield  {author} {\bibinfo {author} {\bibfnamefont {Y.}~\bibnamefont
  {Liu}}\ and\ \bibinfo {author} {\bibfnamefont {R.~E.}\ \bibnamefont
  {Allen}},\ }\href@noop {} {\bibfield  {journal} {\bibinfo  {journal}
  {Physical Review B}\ }\textbf {\bibinfo {volume} {52}},\ \bibinfo {pages}
  {1566} (\bibinfo {year} {1995})}\BibitemShut {NoStop}%
\bibitem [{\citenamefont {Fuseya}\ \emph
  {et~al.}(2015{\natexlab{a}})\citenamefont {Fuseya}, \citenamefont {Ogata},\
  and\ \citenamefont {Fukuyama}}]{fuseya2015jpsj}%
  \BibitemOpen
  \bibfield  {author} {\bibinfo {author} {\bibfnamefont {Y.}~\bibnamefont
  {Fuseya}}, \bibinfo {author} {\bibfnamefont {M.}~\bibnamefont {Ogata}}, \
  and\ \bibinfo {author} {\bibfnamefont {H.}~\bibnamefont {Fukuyama}},\
  }\href@noop {} {\bibfield  {journal} {\bibinfo  {journal} {J. Phys. Soc.
  Jpn.}\ }\textbf {\bibinfo {volume} {84}},\ \bibinfo {pages} {012001}
  (\bibinfo {year} {2015}{\natexlab{a}})}\BibitemShut {NoStop}%
\bibitem [{\citenamefont {Cohen}\ and\ \citenamefont
  {Blount}(1960)}]{cohen1960philmag}%
  \BibitemOpen
  \bibfield  {author} {\bibinfo {author} {\bibfnamefont {M.~H.}\ \bibnamefont
  {Cohen}}\ and\ \bibinfo {author} {\bibfnamefont {E.~I.}\ \bibnamefont
  {Blount}},\ }\href@noop {} {\bibfield  {journal} {\bibinfo  {journal} {Phil.
  Mag.}\ }\textbf {\bibinfo {volume} {5}},\ \bibinfo {pages} {115} (\bibinfo
  {year} {1960})}\BibitemShut {NoStop}%
\bibitem [{\citenamefont {Wolff}(1964)}]{wolff1964jpcs}%
  \BibitemOpen
  \bibfield  {author} {\bibinfo {author} {\bibfnamefont {P.~A.}\ \bibnamefont
  {Wolff}},\ }\href@noop {} {\bibfield  {journal} {\bibinfo  {journal} {J.
  Phys. Chem. Solids}\ }\textbf {\bibinfo {volume} {25}},\ \bibinfo {pages}
  {1057} (\bibinfo {year} {1964})}\BibitemShut {NoStop}%
\bibitem [{\citenamefont {Vecchi}\ and\ \citenamefont
  {Dresselhaus}(1974)}]{vecchi1974prb}%
  \BibitemOpen
  \bibfield  {author} {\bibinfo {author} {\bibfnamefont {M.~P.}\ \bibnamefont
  {Vecchi}}\ and\ \bibinfo {author} {\bibfnamefont {M.~S.}\ \bibnamefont
  {Dresselhaus}},\ }\href@noop {} {\bibfield  {journal} {\bibinfo  {journal}
  {Phys. Rev. B}\ }\textbf {\bibinfo {volume} {10}},\ \bibinfo {pages} {771}
  (\bibinfo {year} {1974})}\BibitemShut {NoStop}%
\bibitem [{\citenamefont {Sarma}\ and\ \citenamefont
  {Hwang}(2013)}]{dassarma2013prb}%
  \BibitemOpen
  \bibfield  {author} {\bibinfo {author} {\bibfnamefont {S.~D.}\ \bibnamefont
  {Sarma}}\ and\ \bibinfo {author} {\bibfnamefont {E.~H.}\ \bibnamefont
  {Hwang}},\ }\href@noop {} {\bibfield  {journal} {\bibinfo  {journal} {Phys.
  Rev. B}\ }\textbf {\bibinfo {volume} {88}},\ \bibinfo {pages} {035439}
  (\bibinfo {year} {2013})}\BibitemShut {NoStop}%
\bibitem [{\citenamefont {Sarma}\ \emph {et~al.}(2015)\citenamefont {Sarma},
  \citenamefont {Hwang},\ and\ \citenamefont {Min}}]{dassarma2015prb}%
  \BibitemOpen
  \bibfield  {author} {\bibinfo {author} {\bibfnamefont {S.~D.}\ \bibnamefont
  {Sarma}}, \bibinfo {author} {\bibfnamefont {E.~H.}\ \bibnamefont {Hwang}}, \
  and\ \bibinfo {author} {\bibfnamefont {H.}~\bibnamefont {Min}},\ }\href@noop
  {} {\bibfield  {journal} {\bibinfo  {journal} {Phys. Rev. B}\ }\textbf
  {\bibinfo {volume} {91}},\ \bibinfo {pages} {035201} (\bibinfo {year}
  {2015})}\BibitemShut {NoStop}%
\bibitem [{\citenamefont {Novoselov}\ \emph {et~al.}(2005)\citenamefont
  {Novoselov}, \citenamefont {Geim}, \citenamefont {Morozov}, \citenamefont
  {Jiang}, \citenamefont {Katsnelson}, \citenamefont {Grigorieva},
  \citenamefont {Dubonos},\ and\ \citenamefont {Firsov}}]{novoselov2005nature}%
  \BibitemOpen
  \bibfield  {author} {\bibinfo {author} {\bibfnamefont {K.~S.}\ \bibnamefont
  {Novoselov}}, \bibinfo {author} {\bibfnamefont {A.~K.}\ \bibnamefont {Geim}},
  \bibinfo {author} {\bibfnamefont {S.~V.}\ \bibnamefont {Morozov}}, \bibinfo
  {author} {\bibfnamefont {D.}~\bibnamefont {Jiang}}, \bibinfo {author}
  {\bibfnamefont {M.~I.}\ \bibnamefont {Katsnelson}}, \bibinfo {author}
  {\bibfnamefont {I.~V.}\ \bibnamefont {Grigorieva}}, \bibinfo {author}
  {\bibfnamefont {S.~V.}\ \bibnamefont {Dubonos}}, \ and\ \bibinfo {author}
  {\bibfnamefont {A.~A.}\ \bibnamefont {Firsov}},\ }\href@noop {} {\bibfield
  {journal} {\bibinfo  {journal} {Nature}\ }\textbf {\bibinfo {volume} {438}},\
  \bibinfo {pages} {197} (\bibinfo {year} {2005})}\BibitemShut {NoStop}%
\bibitem [{\citenamefont {Zhang}\ \emph {et~al.}(2005)\citenamefont {Zhang},
  \citenamefont {Tan}, \citenamefont {Stormer},\ and\ \citenamefont
  {Kim}}]{zhang2005nature}%
  \BibitemOpen
  \bibfield  {author} {\bibinfo {author} {\bibfnamefont {Y.~B.}\ \bibnamefont
  {Zhang}}, \bibinfo {author} {\bibfnamefont {Y.~W.}\ \bibnamefont {Tan}},
  \bibinfo {author} {\bibfnamefont {H.~L.}\ \bibnamefont {Stormer}}, \ and\
  \bibinfo {author} {\bibfnamefont {P.}~\bibnamefont {Kim}},\ }\href@noop {}
  {\bibfield  {journal} {\bibinfo  {journal} {Nature}\ }\textbf {\bibinfo
  {volume} {438}},\ \bibinfo {pages} {201} (\bibinfo {year}
  {2005})}\BibitemShut {NoStop}%
\bibitem [{\citenamefont {Liang}\ \emph {et~al.}(2015)\citenamefont {Liang},
  \citenamefont {Gibson}, \citenamefont {Ali}, \citenamefont {Liu},
  \citenamefont {Cava},\ and\ \citenamefont {Ong}}]{liang2015nmat}%
  \BibitemOpen
  \bibfield  {author} {\bibinfo {author} {\bibfnamefont {T.}~\bibnamefont
  {Liang}}, \bibinfo {author} {\bibfnamefont {Q.}~\bibnamefont {Gibson}},
  \bibinfo {author} {\bibfnamefont {M.~N.}\ \bibnamefont {Ali}}, \bibinfo
  {author} {\bibfnamefont {M.~H.}\ \bibnamefont {Liu}}, \bibinfo {author}
  {\bibfnamefont {R.~J.}\ \bibnamefont {Cava}}, \ and\ \bibinfo {author}
  {\bibfnamefont {N.~P.}\ \bibnamefont {Ong}},\ }\href@noop {} {\bibfield
  {journal} {\bibinfo  {journal} {Nat. Mater.}\ }\textbf {\bibinfo {volume}
  {14}},\ \bibinfo {pages} {280} (\bibinfo {year} {2015})}\BibitemShut
  {NoStop}%
\bibitem [{\citenamefont {Shekhar}\ \emph {et~al.}(2015)\citenamefont
  {Shekhar}, \citenamefont {Nayak}, \citenamefont {Sun}, \citenamefont
  {Schmidt}, \citenamefont {Nicklas}, \citenamefont {Leermakers}, \citenamefont
  {Zeitler}, \citenamefont {Skourski}, \citenamefont {Wosnitza}, \citenamefont
  {Liu}, \citenamefont {Chen}, \citenamefont {Schnelle}, \citenamefont
  {Borrmann}, \citenamefont {Grin}, \citenamefont {Felser},\ and\ \citenamefont
  {Yan}}]{shekhar2015nphys}%
  \BibitemOpen
  \bibfield  {author} {\bibinfo {author} {\bibfnamefont {C.}~\bibnamefont
  {Shekhar}}, \bibinfo {author} {\bibfnamefont {A.~K.}\ \bibnamefont {Nayak}},
  \bibinfo {author} {\bibfnamefont {Y.}~\bibnamefont {Sun}}, \bibinfo {author}
  {\bibfnamefont {M.}~\bibnamefont {Schmidt}}, \bibinfo {author} {\bibfnamefont
  {M.}~\bibnamefont {Nicklas}}, \bibinfo {author} {\bibfnamefont
  {I.}~\bibnamefont {Leermakers}}, \bibinfo {author} {\bibfnamefont
  {U.}~\bibnamefont {Zeitler}}, \bibinfo {author} {\bibfnamefont
  {Y.}~\bibnamefont {Skourski}}, \bibinfo {author} {\bibfnamefont
  {J.}~\bibnamefont {Wosnitza}}, \bibinfo {author} {\bibfnamefont {Z.~K.}\
  \bibnamefont {Liu}}, \bibinfo {author} {\bibfnamefont {Y.~L.}\ \bibnamefont
  {Chen}}, \bibinfo {author} {\bibfnamefont {W.}~\bibnamefont {Schnelle}},
  \bibinfo {author} {\bibfnamefont {H.}~\bibnamefont {Borrmann}}, \bibinfo
  {author} {\bibfnamefont {Y.}~\bibnamefont {Grin}}, \bibinfo {author}
  {\bibfnamefont {C.}~\bibnamefont {Felser}}, \ and\ \bibinfo {author}
  {\bibfnamefont {B.~H.}\ \bibnamefont {Yan}},\ }\href@noop {} {\bibfield
  {journal} {\bibinfo  {journal} {Nat. Phys.}\ }\textbf {\bibinfo {volume}
  {11}},\ \bibinfo {pages} {645} (\bibinfo {year} {2015})}\BibitemShut
  {NoStop}%
\bibitem [{\citenamefont {Peskin}\ and\ \citenamefont
  {Schroeder}(1995)}]{peskin1995book}%
  \BibitemOpen
  \bibfield  {author} {\bibinfo {author} {\bibfnamefont {M.~E.}\ \bibnamefont
  {Peskin}}\ and\ \bibinfo {author} {\bibfnamefont {D.~V.}\ \bibnamefont
  {Schroeder}},\ }\href@noop {} {\emph {\bibinfo {title} {An Introduction To
  Quantum Field Theory}}}\ (\bibinfo  {publisher} {CRC Press},\ \bibinfo {year}
  {1995})\BibitemShut {NoStop}%
\bibitem [{\citenamefont {Chi}\ \emph {et~al.}(2020)\citenamefont {Chi},
  \citenamefont {Lau}, \citenamefont {Xu}, \citenamefont {Ohkubo},
  \citenamefont {Hono},\ and\ \citenamefont {Hayashi}}]{chi2020sciadv}%
  \BibitemOpen
  \bibfield  {author} {\bibinfo {author} {\bibfnamefont {Z.}~\bibnamefont
  {Chi}}, \bibinfo {author} {\bibfnamefont {Y.-C.}\ \bibnamefont {Lau}},
  \bibinfo {author} {\bibfnamefont {X.}~\bibnamefont {Xu}}, \bibinfo {author}
  {\bibfnamefont {T.}~\bibnamefont {Ohkubo}}, \bibinfo {author} {\bibfnamefont
  {K.}~\bibnamefont {Hono}}, \ and\ \bibinfo {author} {\bibfnamefont
  {M.}~\bibnamefont {Hayashi}},\ }\href@noop {} {\bibfield  {journal} {\bibinfo
   {journal} {Science Advances}\ }\textbf {\bibinfo {volume} {6}},\ \bibinfo
  {pages} {eaay2324} (\bibinfo {year} {2020})}\BibitemShut {NoStop}%
\bibitem [{\citenamefont {Clementi}\ \emph {et~al.}(1967)\citenamefont
  {Clementi}, \citenamefont {Raimondi},\ and\ \citenamefont
  {Reinhardt}}]{clementi1967jcp}%
  \BibitemOpen
  \bibfield  {author} {\bibinfo {author} {\bibfnamefont {E.}~\bibnamefont
  {Clementi}}, \bibinfo {author} {\bibfnamefont {D.~L.}\ \bibnamefont
  {Raimondi}}, \ and\ \bibinfo {author} {\bibfnamefont {W.~P.}\ \bibnamefont
  {Reinhardt}},\ }\href@noop {} {\bibfield  {journal} {\bibinfo  {journal}
  {Journal of Chemical Physics}\ }\textbf {\bibinfo {volume} {47}},\ \bibinfo
  {pages} {1300} (\bibinfo {year} {1967})}\BibitemShut {NoStop}%
\bibitem [{\citenamefont {Behnia}\ \emph {et~al.}(2007)\citenamefont {Behnia},
  \citenamefont {Balicas},\ and\ \citenamefont
  {Kopelevich}}]{behnia2007science}%
  \BibitemOpen
  \bibfield  {author} {\bibinfo {author} {\bibfnamefont {K.}~\bibnamefont
  {Behnia}}, \bibinfo {author} {\bibfnamefont {L.}~\bibnamefont {Balicas}}, \
  and\ \bibinfo {author} {\bibfnamefont {Y.}~\bibnamefont {Kopelevich}},\
  }\href@noop {} {\bibfield  {journal} {\bibinfo  {journal} {Science}\ }\textbf
  {\bibinfo {volume} {317}},\ \bibinfo {pages} {1729} (\bibinfo {year}
  {2007})}\BibitemShut {NoStop}%
\bibitem [{\citenamefont {Kim}\ \emph {et~al.}(2013)\citenamefont {Kim},
  \citenamefont {Sinha}, \citenamefont {Hayashi}, \citenamefont {Yamanouchi},
  \citenamefont {Fukami}, \citenamefont {Suzuki}, \citenamefont {Mitani},\ and\
  \citenamefont {Ohno}}]{kim2013nmat}%
  \BibitemOpen
  \bibfield  {author} {\bibinfo {author} {\bibfnamefont {J.}~\bibnamefont
  {Kim}}, \bibinfo {author} {\bibfnamefont {J.}~\bibnamefont {Sinha}}, \bibinfo
  {author} {\bibfnamefont {M.}~\bibnamefont {Hayashi}}, \bibinfo {author}
  {\bibfnamefont {M.}~\bibnamefont {Yamanouchi}}, \bibinfo {author}
  {\bibfnamefont {S.}~\bibnamefont {Fukami}}, \bibinfo {author} {\bibfnamefont
  {T.}~\bibnamefont {Suzuki}}, \bibinfo {author} {\bibfnamefont
  {S.}~\bibnamefont {Mitani}}, \ and\ \bibinfo {author} {\bibfnamefont
  {H.}~\bibnamefont {Ohno}},\ }\href@noop {} {\bibfield  {journal} {\bibinfo
  {journal} {Nat. Mater.}\ }\textbf {\bibinfo {volume} {12}},\ \bibinfo {pages}
  {240} (\bibinfo {year} {2013})}\BibitemShut {NoStop}%
\bibitem [{\citenamefont {Garello}\ \emph {et~al.}(2013)\citenamefont
  {Garello}, \citenamefont {Miron}, \citenamefont {Avci}, \citenamefont
  {Freimuth}, \citenamefont {Mokrousov}, \citenamefont {Blugel}, \citenamefont
  {Auffret}, \citenamefont {Boulle}, \citenamefont {Gaudin},\ and\
  \citenamefont {Gambardella}}]{garello2013nnano}%
  \BibitemOpen
  \bibfield  {author} {\bibinfo {author} {\bibfnamefont {K.}~\bibnamefont
  {Garello}}, \bibinfo {author} {\bibfnamefont {I.~M.}\ \bibnamefont {Miron}},
  \bibinfo {author} {\bibfnamefont {C.~O.}\ \bibnamefont {Avci}}, \bibinfo
  {author} {\bibfnamefont {F.}~\bibnamefont {Freimuth}}, \bibinfo {author}
  {\bibfnamefont {Y.}~\bibnamefont {Mokrousov}}, \bibinfo {author}
  {\bibfnamefont {S.}~\bibnamefont {Blugel}}, \bibinfo {author} {\bibfnamefont
  {S.}~\bibnamefont {Auffret}}, \bibinfo {author} {\bibfnamefont
  {O.}~\bibnamefont {Boulle}}, \bibinfo {author} {\bibfnamefont
  {G.}~\bibnamefont {Gaudin}}, \ and\ \bibinfo {author} {\bibfnamefont
  {P.}~\bibnamefont {Gambardella}},\ }\href@noop {} {\bibfield  {journal}
  {\bibinfo  {journal} {Nat. Nanotechnol.}\ }\textbf {\bibinfo {volume} {8}},\
  \bibinfo {pages} {587} (\bibinfo {year} {2013})}\BibitemShut {NoStop}%
\bibitem [{\citenamefont {Hayashi}\ \emph {et~al.}(2014)\citenamefont
  {Hayashi}, \citenamefont {Kim}, \citenamefont {Yamanouchi},\ and\
  \citenamefont {Ohno}}]{hayashi2014prb}%
  \BibitemOpen
  \bibfield  {author} {\bibinfo {author} {\bibfnamefont {M.}~\bibnamefont
  {Hayashi}}, \bibinfo {author} {\bibfnamefont {J.}~\bibnamefont {Kim}},
  \bibinfo {author} {\bibfnamefont {M.}~\bibnamefont {Yamanouchi}}, \ and\
  \bibinfo {author} {\bibfnamefont {H.}~\bibnamefont {Ohno}},\ }\href@noop {}
  {\bibfield  {journal} {\bibinfo  {journal} {Phys. Rev. B}\ }\textbf {\bibinfo
  {volume} {89}},\ \bibinfo {pages} {144425} (\bibinfo {year}
  {2014})}\BibitemShut {NoStop}%
\bibitem [{\citenamefont {Avci}\ \emph {et~al.}(2014)\citenamefont {Avci},
  \citenamefont {Garello}, \citenamefont {Gabureac}, \citenamefont {Ghosh},
  \citenamefont {Fuhrer}, \citenamefont {Alvarado},\ and\ \citenamefont
  {Gambardella}}]{avci2014prb}%
  \BibitemOpen
  \bibfield  {author} {\bibinfo {author} {\bibfnamefont {C.~O.}\ \bibnamefont
  {Avci}}, \bibinfo {author} {\bibfnamefont {K.}~\bibnamefont {Garello}},
  \bibinfo {author} {\bibfnamefont {M.}~\bibnamefont {Gabureac}}, \bibinfo
  {author} {\bibfnamefont {A.}~\bibnamefont {Ghosh}}, \bibinfo {author}
  {\bibfnamefont {A.}~\bibnamefont {Fuhrer}}, \bibinfo {author} {\bibfnamefont
  {S.~F.}\ \bibnamefont {Alvarado}}, \ and\ \bibinfo {author} {\bibfnamefont
  {P.}~\bibnamefont {Gambardella}},\ }\href@noop {} {\bibfield  {journal}
  {\bibinfo  {journal} {Phys. Rev. B}\ }\textbf {\bibinfo {volume} {90}},\
  \bibinfo {pages} {224427} (\bibinfo {year} {2014})}\BibitemShut {NoStop}%
\bibitem [{\citenamefont {Roschewsky}\ \emph {et~al.}(2019)\citenamefont
  {Roschewsky}, \citenamefont {Walker}, \citenamefont {Gowtham}, \citenamefont
  {Muschinske}, \citenamefont {Hellman}, \citenamefont {Bank},\ and\
  \citenamefont {Salahuddin}}]{roschewsky2019prb}%
  \BibitemOpen
  \bibfield  {author} {\bibinfo {author} {\bibfnamefont {N.}~\bibnamefont
  {Roschewsky}}, \bibinfo {author} {\bibfnamefont {E.~S.}\ \bibnamefont
  {Walker}}, \bibinfo {author} {\bibfnamefont {P.}~\bibnamefont {Gowtham}},
  \bibinfo {author} {\bibfnamefont {S.}~\bibnamefont {Muschinske}}, \bibinfo
  {author} {\bibfnamefont {F.}~\bibnamefont {Hellman}}, \bibinfo {author}
  {\bibfnamefont {S.~R.}\ \bibnamefont {Bank}}, \ and\ \bibinfo {author}
  {\bibfnamefont {S.}~\bibnamefont {Salahuddin}},\ }\href@noop {} {\bibfield
  {journal} {\bibinfo  {journal} {Phys. Rev. B}\ }\textbf {\bibinfo {volume}
  {99}},\ \bibinfo {pages} {195103} (\bibinfo {year} {2019})}\BibitemShut
  {NoStop}%
\bibitem [{\citenamefont {Pai}\ \emph {et~al.}(2015)\citenamefont {Pai},
  \citenamefont {Ou}, \citenamefont {Vilela-Leao}, \citenamefont {Ralph},\ and\
  \citenamefont {Buhrman}}]{pai2015prb}%
  \BibitemOpen
  \bibfield  {author} {\bibinfo {author} {\bibfnamefont {C.-F.}\ \bibnamefont
  {Pai}}, \bibinfo {author} {\bibfnamefont {Y.}~\bibnamefont {Ou}}, \bibinfo
  {author} {\bibfnamefont {L.~H.}\ \bibnamefont {Vilela-Leao}}, \bibinfo
  {author} {\bibfnamefont {D.~C.}\ \bibnamefont {Ralph}}, \ and\ \bibinfo
  {author} {\bibfnamefont {R.~A.}\ \bibnamefont {Buhrman}},\ }\href@noop {}
  {\bibfield  {journal} {\bibinfo  {journal} {Phys. Rev. B}\ }\textbf {\bibinfo
  {volume} {92}},\ \bibinfo {pages} {064426} (\bibinfo {year}
  {2015})}\BibitemShut {NoStop}%
\bibitem [{\citenamefont {Zhang}\ \emph {et~al.}(2015)\citenamefont {Zhang},
  \citenamefont {Han}, \citenamefont {Jiang}, \citenamefont {Yang},\ and\
  \citenamefont {Parkin}}]{zhang2015nphys}%
  \BibitemOpen
  \bibfield  {author} {\bibinfo {author} {\bibfnamefont {W.}~\bibnamefont
  {Zhang}}, \bibinfo {author} {\bibfnamefont {W.}~\bibnamefont {Han}}, \bibinfo
  {author} {\bibfnamefont {X.}~\bibnamefont {Jiang}}, \bibinfo {author}
  {\bibfnamefont {S.-H.}\ \bibnamefont {Yang}}, \ and\ \bibinfo {author}
  {\bibfnamefont {S.~S.~P.}\ \bibnamefont {Parkin}},\ }\href@noop {} {\bibfield
   {journal} {\bibinfo  {journal} {Nature Physics}\ }\textbf {\bibinfo {volume}
  {11}},\ \bibinfo {pages} {496} (\bibinfo {year} {2015})}\BibitemShut
  {NoStop}%
\bibitem [{\citenamefont {Rojas-Sanchez}\ \emph {et~al.}(2014)\citenamefont
  {Rojas-Sanchez}, \citenamefont {Reyren}, \citenamefont {Laczkowski},
  \citenamefont {Savero}, \citenamefont {Attane}, \citenamefont {Deranlot},
  \citenamefont {Jamet}, \citenamefont {George}, \citenamefont {Vila},\ and\
  \citenamefont {Jaffres}}]{rojassanchez2014prl}%
  \BibitemOpen
  \bibfield  {author} {\bibinfo {author} {\bibfnamefont {J.~C.}\ \bibnamefont
  {Rojas-Sanchez}}, \bibinfo {author} {\bibfnamefont {N.}~\bibnamefont
  {Reyren}}, \bibinfo {author} {\bibfnamefont {P.}~\bibnamefont {Laczkowski}},
  \bibinfo {author} {\bibfnamefont {W.}~\bibnamefont {Savero}}, \bibinfo
  {author} {\bibfnamefont {J.~P.}\ \bibnamefont {Attane}}, \bibinfo {author}
  {\bibfnamefont {C.}~\bibnamefont {Deranlot}}, \bibinfo {author}
  {\bibfnamefont {M.}~\bibnamefont {Jamet}}, \bibinfo {author} {\bibfnamefont
  {J.~M.}\ \bibnamefont {George}}, \bibinfo {author} {\bibfnamefont
  {L.}~\bibnamefont {Vila}}, \ and\ \bibinfo {author} {\bibfnamefont
  {H.}~\bibnamefont {Jaffres}},\ }\href@noop {} {\bibfield  {journal} {\bibinfo
   {journal} {Phys. Rev. Lett.}\ }\textbf {\bibinfo {volume} {112}},\ \bibinfo
  {pages} {106602} (\bibinfo {year} {2014})}\BibitemShut {NoStop}%
\bibitem [{\citenamefont {Sagasta}\ \emph {et~al.}(2016)\citenamefont
  {Sagasta}, \citenamefont {Omori}, \citenamefont {Isasa}, \citenamefont
  {Gradhand}, \citenamefont {Hueso}, \citenamefont {Niimi}, \citenamefont
  {Otani},\ and\ \citenamefont {Casanova}}]{sagasta2016prb}%
  \BibitemOpen
  \bibfield  {author} {\bibinfo {author} {\bibfnamefont {E.}~\bibnamefont
  {Sagasta}}, \bibinfo {author} {\bibfnamefont {Y.}~\bibnamefont {Omori}},
  \bibinfo {author} {\bibfnamefont {M.}~\bibnamefont {Isasa}}, \bibinfo
  {author} {\bibfnamefont {M.}~\bibnamefont {Gradhand}}, \bibinfo {author}
  {\bibfnamefont {L.~E.}\ \bibnamefont {Hueso}}, \bibinfo {author}
  {\bibfnamefont {Y.}~\bibnamefont {Niimi}}, \bibinfo {author} {\bibfnamefont
  {Y.}~\bibnamefont {Otani}}, \ and\ \bibinfo {author} {\bibfnamefont
  {F.}~\bibnamefont {Casanova}},\ }\href@noop {} {\bibfield  {journal}
  {\bibinfo  {journal} {Phys. Rev. B}\ }\textbf {\bibinfo {volume} {94}},\
  \bibinfo {pages} {060412} (\bibinfo {year} {2016})}\BibitemShut {NoStop}%
\bibitem [{\citenamefont {Hou}\ \emph {et~al.}(2012)\citenamefont {Hou},
  \citenamefont {Qiu}, \citenamefont {Harii}, \citenamefont {Kajiwara},
  \citenamefont {Uchida}, \citenamefont {Fujikawa}, \citenamefont {Nakayama},
  \citenamefont {Yoshino}, \citenamefont {An}, \citenamefont {Ando},
  \citenamefont {Jin},\ and\ \citenamefont {Saitoh}}]{hou2012apl}%
  \BibitemOpen
  \bibfield  {author} {\bibinfo {author} {\bibfnamefont {D.~Z.}\ \bibnamefont
  {Hou}}, \bibinfo {author} {\bibfnamefont {Z.}~\bibnamefont {Qiu}}, \bibinfo
  {author} {\bibfnamefont {K.}~\bibnamefont {Harii}}, \bibinfo {author}
  {\bibfnamefont {Y.}~\bibnamefont {Kajiwara}}, \bibinfo {author}
  {\bibfnamefont {K.}~\bibnamefont {Uchida}}, \bibinfo {author} {\bibfnamefont
  {Y.}~\bibnamefont {Fujikawa}}, \bibinfo {author} {\bibfnamefont
  {H.}~\bibnamefont {Nakayama}}, \bibinfo {author} {\bibfnamefont
  {T.}~\bibnamefont {Yoshino}}, \bibinfo {author} {\bibfnamefont
  {T.}~\bibnamefont {An}}, \bibinfo {author} {\bibfnamefont {K.}~\bibnamefont
  {Ando}}, \bibinfo {author} {\bibfnamefont {X.~F.}\ \bibnamefont {Jin}}, \
  and\ \bibinfo {author} {\bibfnamefont {E.}~\bibnamefont {Saitoh}},\
  }\href@noop {} {\bibfield  {journal} {\bibinfo  {journal} {Appl. Phys.
  Lett.}\ }\textbf {\bibinfo {volume} {101}},\ \bibinfo {pages} {042403}
  (\bibinfo {year} {2012})}\BibitemShut {NoStop}%
\bibitem [{\citenamefont {Emoto}\ \emph {et~al.}(2016)\citenamefont {Emoto},
  \citenamefont {Ando}, \citenamefont {Eguchi}, \citenamefont {Ohshima},
  \citenamefont {Shikoh}, \citenamefont {Fuseya}, \citenamefont {Shinjo},\ and\
  \citenamefont {Shiraishi}}]{emoto2016prb}%
  \BibitemOpen
  \bibfield  {author} {\bibinfo {author} {\bibfnamefont {H.}~\bibnamefont
  {Emoto}}, \bibinfo {author} {\bibfnamefont {Y.}~\bibnamefont {Ando}},
  \bibinfo {author} {\bibfnamefont {G.}~\bibnamefont {Eguchi}}, \bibinfo
  {author} {\bibfnamefont {R.}~\bibnamefont {Ohshima}}, \bibinfo {author}
  {\bibfnamefont {E.}~\bibnamefont {Shikoh}}, \bibinfo {author} {\bibfnamefont
  {Y.}~\bibnamefont {Fuseya}}, \bibinfo {author} {\bibfnamefont
  {T.}~\bibnamefont {Shinjo}}, \ and\ \bibinfo {author} {\bibfnamefont
  {M.}~\bibnamefont {Shiraishi}},\ }\href@noop {} {\bibfield  {journal}
  {\bibinfo  {journal} {Phys. Rev. B}\ }\textbf {\bibinfo {volume} {93}},\
  \bibinfo {pages} {174428} (\bibinfo {year} {2016})}\BibitemShut {NoStop}%
\bibitem [{\citenamefont {Yue}\ \emph {et~al.}(2018)\citenamefont {Yue},
  \citenamefont {Lin}, \citenamefont {Li}, \citenamefont {Jin},\ and\
  \citenamefont {Chien}}]{yue2018prl}%
  \BibitemOpen
  \bibfield  {author} {\bibinfo {author} {\bibfnamefont {D.}~\bibnamefont
  {Yue}}, \bibinfo {author} {\bibfnamefont {W.~W.}\ \bibnamefont {Lin}},
  \bibinfo {author} {\bibfnamefont {J.~J.}\ \bibnamefont {Li}}, \bibinfo
  {author} {\bibfnamefont {X.~F.}\ \bibnamefont {Jin}}, \ and\ \bibinfo
  {author} {\bibfnamefont {C.~L.}\ \bibnamefont {Chien}},\ }\href@noop {}
  {\bibfield  {journal} {\bibinfo  {journal} {Phys. Rev. Lett.}\ }\textbf
  {\bibinfo {volume} {121}},\ \bibinfo {pages} {037201} (\bibinfo {year}
  {2018})}\BibitemShut {NoStop}%
\bibitem [{\citenamefont {Hirose}\ \emph {et~al.}(2018)\citenamefont {Hirose},
  \citenamefont {Ito}, \citenamefont {Kawaguchi}, \citenamefont {Lau},\ and\
  \citenamefont {Hayashi}}]{hirose2021prb}%
  \BibitemOpen
  \bibfield  {author} {\bibinfo {author} {\bibfnamefont {H.}~\bibnamefont
  {Hirose}}, \bibinfo {author} {\bibfnamefont {N.}~\bibnamefont {Ito}},
  \bibinfo {author} {\bibfnamefont {M.}~\bibnamefont {Kawaguchi}}, \bibinfo
  {author} {\bibfnamefont {Y.~C.}\ \bibnamefont {Lau}}, \ and\ \bibinfo
  {author} {\bibfnamefont {M.}~\bibnamefont {Hayashi}},\ }\href@noop {}
  {\bibfield  {journal} {\bibinfo  {journal} {Appl. Phys. Lett.}\ }\textbf
  {\bibinfo {volume} {113}},\ \bibinfo {pages} {222404} (\bibinfo {year}
  {2018})}\BibitemShut {NoStop}%
\bibitem [{\citenamefont {Smith}\ \emph {et~al.}(1964)\citenamefont {Smith},
  \citenamefont {Rowell},\ and\ \citenamefont {Baraff}}]{smith1964pr}%
  \BibitemOpen
  \bibfield  {author} {\bibinfo {author} {\bibfnamefont {G.~E.}\ \bibnamefont
  {Smith}}, \bibinfo {author} {\bibfnamefont {J.~M.}\ \bibnamefont {Rowell}}, \
  and\ \bibinfo {author} {\bibfnamefont {G.~A.}\ \bibnamefont {Baraff}},\
  }\href@noop {} {\bibfield  {journal} {\bibinfo  {journal} {Phys. Rev.}\
  }\textbf {\bibinfo {volume} {135}},\ \bibinfo {pages} {1118} (\bibinfo {year}
  {1964})}\BibitemShut {NoStop}%
\bibitem [{\citenamefont {Fuseya}\ \emph
  {et~al.}(2015{\natexlab{b}})\citenamefont {Fuseya}, \citenamefont {Zhu},
  \citenamefont {Fauque}, \citenamefont {Kang}, \citenamefont {Lenoir},\ and\
  \citenamefont {Behnia}}]{fuseya2015prl}%
  \BibitemOpen
  \bibfield  {author} {\bibinfo {author} {\bibfnamefont {Y.}~\bibnamefont
  {Fuseya}}, \bibinfo {author} {\bibfnamefont {Z.}~\bibnamefont {Zhu}},
  \bibinfo {author} {\bibfnamefont {B.}~\bibnamefont {Fauque}}, \bibinfo
  {author} {\bibfnamefont {W.}~\bibnamefont {Kang}}, \bibinfo {author}
  {\bibfnamefont {B.}~\bibnamefont {Lenoir}}, \ and\ \bibinfo {author}
  {\bibfnamefont {K.}~\bibnamefont {Behnia}},\ }\href@noop {} {\bibfield
  {journal} {\bibinfo  {journal} {Phys. Rev. Lett.}\ }\textbf {\bibinfo
  {volume} {115}},\ \bibinfo {pages} {216401} (\bibinfo {year}
  {2015}{\natexlab{b}})}\BibitemShut {NoStop}%
\bibitem [{\citenamefont {Fujimoto}\ and\ \citenamefont
  {Ogata}(2021)}]{fujimoto2021condmat}%
  \BibitemOpen
  \bibfield  {author} {\bibinfo {author} {\bibfnamefont {J.}~\bibnamefont
  {Fujimoto}}\ and\ \bibinfo {author} {\bibfnamefont {M.}~\bibnamefont
  {Ogata}},\ }\href@noop {} {\bibfield  {journal} {\bibinfo  {journal}
  {arXiv:2106.04050}\ } (\bibinfo {year} {2021})}\BibitemShut {NoStop}%
\bibitem [{\citenamefont {Mellnik}\ \emph {et~al.}(2014)\citenamefont
  {Mellnik}, \citenamefont {Lee}, \citenamefont {Richardella}, \citenamefont
  {Grab}, \citenamefont {Mintun}, \citenamefont {Fischer}, \citenamefont
  {Vaezi}, \citenamefont {Manchon}, \citenamefont {Kim}, \citenamefont
  {Samarth},\ and\ \citenamefont {Ralph}}]{mellnik2014nature}%
  \BibitemOpen
  \bibfield  {author} {\bibinfo {author} {\bibfnamefont {A.~R.}\ \bibnamefont
  {Mellnik}}, \bibinfo {author} {\bibfnamefont {J.~S.}\ \bibnamefont {Lee}},
  \bibinfo {author} {\bibfnamefont {A.}~\bibnamefont {Richardella}}, \bibinfo
  {author} {\bibfnamefont {J.~L.}\ \bibnamefont {Grab}}, \bibinfo {author}
  {\bibfnamefont {P.~J.}\ \bibnamefont {Mintun}}, \bibinfo {author}
  {\bibfnamefont {M.~H.}\ \bibnamefont {Fischer}}, \bibinfo {author}
  {\bibfnamefont {A.}~\bibnamefont {Vaezi}}, \bibinfo {author} {\bibfnamefont
  {A.}~\bibnamefont {Manchon}}, \bibinfo {author} {\bibfnamefont {E.~A.}\
  \bibnamefont {Kim}}, \bibinfo {author} {\bibfnamefont {N.}~\bibnamefont
  {Samarth}}, \ and\ \bibinfo {author} {\bibfnamefont {D.~C.}\ \bibnamefont
  {Ralph}},\ }\href@noop {} {\bibfield  {journal} {\bibinfo  {journal}
  {Nature}\ }\textbf {\bibinfo {volume} {511}},\ \bibinfo {pages} {449}
  (\bibinfo {year} {2014})}\BibitemShut {NoStop}%
\bibitem [{\citenamefont {Liu}\ \emph {et~al.}(2011)\citenamefont {Liu},
  \citenamefont {Moriyama}, \citenamefont {Ralph},\ and\ \citenamefont
  {Buhrman}}]{liu2011prl}%
  \BibitemOpen
  \bibfield  {author} {\bibinfo {author} {\bibfnamefont {L.}~\bibnamefont
  {Liu}}, \bibinfo {author} {\bibfnamefont {T.}~\bibnamefont {Moriyama}},
  \bibinfo {author} {\bibfnamefont {D.~C.}\ \bibnamefont {Ralph}}, \ and\
  \bibinfo {author} {\bibfnamefont {R.~A.}\ \bibnamefont {Buhrman}},\
  }\href@noop {} {\bibfield  {journal} {\bibinfo  {journal} {Phys. Rev. Lett.}\
  }\textbf {\bibinfo {volume} {106}},\ \bibinfo {pages} {036601} (\bibinfo
  {year} {2011})}\BibitemShut {NoStop}%
\end{thebibliography}%

\end{document}